\documentclass[%
 reprint,
 amsmath,amssymb,
 aps,
]{revtex4-2}

\usepackage{graphicx}
\usepackage{dcolumn}
\usepackage{bm}
\usepackage{subfigure}
\usepackage{lineno}

\begin{document}

\preprint{RAL-P-2022-001}

\title{Multiple Coulomb Scattering of muons in Lithium Hydride}

\collaboration{MICE Collaboration}

\author{M.~Bogomilov,  R.~Tsenov, G.~Vankova-Kirilova}
\affiliation{
Department of Atomic Physics, St.~Kliment Ohridski University of Sofia, 5 James Bourchier Blvd, Sofia, Bulgaria
}

\author{Y.~P.~Song, J.~Y.~Tang}
\affiliation{
Institute of High Energy Physics, Chinese Academy of Sciences, 19 Yuquan Rd, Shijingshan District, Beijing, China
}

\author{Z.~H.~Li}
\affiliation{
Sichuan University, 252 Shuncheng St, Chengdu, China
}

\author{R.~Bertoni, M.~Bonesini, F.~Chignoli, R.~Mazza}
\affiliation{
Sezione INFN Milano Bicocca and Dipartimento di Fisica G.~Occhialini, Piazza della Scienza 3, Milano, Italy
}

\author{V.~Palladino}
\affiliation{
Sezione INFN Napoli and Dipartimento di Fisica, Universit\`{a} Federico II, Complesso Universitario di Monte S.~Angelo, via Cintia, Napoli, Italy
}

\author{A.~de Bari}
\affiliation{ 
Sezione INFN Pavia and Dipartimento di Fisica, Universit\`{a} di Pavia, Via Agostino Bassi 6, Pavia, Italy
}

\author{D.~Orestano, L.~Tortora}
\affiliation{
Sezione INFN Roma Tre and Dipartimento di Matematica e Fisica, Universit\`{a} Roma Tre, Via della Vasca Navale 84, Roma, Italy
}

\author{Y.~Kuno, H.~Sakamoto\footnote{Current address RIKEN 2-1 Hirosawa, Wako, Saitama, Japan}, A.~Sato}
\affiliation{
Osaka University, Graduate School of Science, Department of Physics, 1-1 Machikaneyamacho, Toyonaka, Osaka, Japan
}

\author{S.~Ishimoto}
\affiliation{
High Energy Accelerator Research Organization (KEK), Institute of Particle and Nuclear Studies, Tsukuba, Ibaraki, Japan
}

\author{M.~Chung, C.~K.~Sung}
\affiliation{ 
Department of Physics, UNIST, 50 UNIST-gil, Ulsan, South Korea
}

\author{F.~Filthaut\footnote{Also at Radboud University, Houtlaan 4, Nijmegen, Netherlands}}
\affiliation{
Nikhef, Science Park 105, Amsterdam, Netherlands
}

\author{M.~Fedorov}
\affiliation{
Radboud University, Houtlaan 4, Nijmegen, Netherlands
}

\author{D.~Jokovic, D.~Maletic, M.~Savic}
\affiliation{
Institute of Physics, University of Belgrade, Serbia
}


\author{N.~Jovancevic, J.~Nikolov}
\affiliation{
Faculty of Sciences, University of Novi Sad, Trg Dositeja Obradovi\'{c}a 3, Novi Sad, Serbia
}

\author{M.~Vretenar, S.~Ramberger}
\affiliation{
CERN, Esplanade des Particules 1, Geneva, Switzerland
}

\author{R.~Asfandiyarov, A.~Blondel, F.~Drielsma\footnote{Current address SLAC National Accelerator Laboratory, 2575 Sand Hill Road, Menlo Park, California, USA}, Y.~Karadzhov }
\affiliation{
DPNC, Section de Physique, Universit\'e de Gen\`eve, 24 Quai Ernest-Ansermet, Geneva, Switzerland
}

\author{G.~Charnley, N.~Collomb,  K.~Dumbell, A.~Gallagher, A.~Grant, S.~Griffiths,  T.~Hartnett, B.~Martlew, 
A.~Moss, A.~Muir, I.~Mullacrane, A.~Oates, P.~Owens, G.~Stokes, P.~Warburton, C.~White}
\affiliation{
STFC Daresbury Laboratory, Keckwick Ln, Daresbury, Cheshire, UK
}

\author{D.~Adams,   V.~Bayliss, J.~Boehm, T.~W.~Bradshaw, C.~Brown\footnote{Also at College of Engineering, Design and Physical Sciences, Brunel University, Kingston Lane, Uxbridge, UK}, M.~Courthold,  J.~Govans, M.~Hills, J.-B.~Lagrange, C.~Macwaters, A.~Nichols, R.~Preece, S.~Ricciardi, C.~Rogers, T.~Stanley, J.~Tarrant,  
M.~Tucker, S.~Watson\footnote{Current address ATC, Royal Observatory Edinburgh, Blackford Hill, Edinburgh, UK}, A.~Wilson}
\affiliation{
STFC Rutherford Appleton Laboratory, Harwell Campus, Didcot, UK
}

\author{R.~Bayes\footnote{Current address Laurentian University, 935 Ramsey Lake Road, Sudbury, Ontario, Canada},  J.~C.~Nugent, F.~J.~P.~Soler}
\affiliation{
School of Physics and Astronomy, Kelvin Building, University of Glasgow, Glasgow, UK
}

\author{R.~Gamet, P.~Cooke}
\affiliation{
Department of Physics, University of Liverpool, Oxford St, Liverpool, UK
}

\author{V.~J.~Blackmore, D.~Colling, A.~Dobbs\footnote{Current address OPERA Simulation Software, Network House, Langford Locks, Kidlington, UK}, P.~Dornan, P.~Franchini\footnote{Current address Lancaster University, Physics avenue, Lancaster, UK}, C.~Hunt\footnote{Current address CERN, Esplanade des Particules 1, Geneva, Switzerland}, P.~B.~Jurj, A.~Kurup, K.~Long, J.~Martyniak,  S.~Middleton\footnote{Current address School of Physics and Astronomy, University of Manchester, Oxford Road, Manchester, UK}, J.~Pasternak, M.~A.~Uchida\footnote{Current address Rutherford Building, Cavendish Laboratory, JJ Thomson Avenue, Cambridge, UK}}
\affiliation{
Department of Physics, Blackett Laboratory, Imperial College London, Exhibition Road, London, UK
}

\author{J.~H.~Cobb}
\affiliation{
Department of Physics, University of Oxford, Denys Wilkinson Building, Keble Rd, Oxford, UK
}

\author{C.~N.~Booth, P.~Hodgson, J.~Langlands, E.~Overton\footnote{Current address Arm, City Gate, 8 St Mary's Gate, Sheffield, UK}, V.~Pec,  P.~J.~Smith, S.~Wilbur}
\affiliation{
Department of Physics and Astronomy, University of Sheffield, Hounsfield Rd, Sheffield, UK
}

\author{G.~T.~Chatzitheodoridis\footnote{Also at School of Physics and Astronomy, Kelvin Building, University of Glasgow, Glasgow, UK}$^,$\footnotemark, A.~J.~Dick\footnotemark[\value{footnote}],  K.~Ronald\footnotemark[\value{footnote}], C.~G.~Whyte\footnotemark[\value{footnote}], A.~R.~Young\footnotemark[\value{footnote}]
\footnotetext{Also at Cockcroft Institute, Daresbury Laboratory, Sci-Tech Daresbury, Keckwick Ln, Daresbury, Warrington, UK}}
\affiliation{
SUPA and the Department of Physics, University of Strathclyde, 107 Rottenrow, Glasgow, UK
}

\author{S.~Boyd,  J.~R.~Greis\footnote{Current address TNG Technology Consulting, Beta-Strasse 13A, Unterf\"{o}hring, Germany}, T.~Lord, C.~Pidcott\footnote{Current address Department of Physics and Astronomy, University of Sheffield, Hounsfield Rd, Sheffield, UK}, I.~Taylor\footnote{Current address Defence Science and Technology Laboratory, Porton Down, Salisbury, UK}}
\affiliation{
Department of Physics, University of Warwick, Gibbet Hill Road, Coventry, UK
}

\author{M.~Ellis\footnote{Current address Macquarie Group, 50 Martin Place, Sydney, Australia}, R.B.S.~Gardener\footnote{Current address Inawisdom, Columba House, Adastral park, Martlesham, Ipswich, UK}, P.~Kyberd, J.~J.~Nebrensky\footnote{Current address UK Atomic Energy Authority, Culham Science Centre, Abingdon, UK}}
\affiliation{
College of Engineering, Design and Physical Sciences, Brunel University, Kingston Lane, Uxbridge, UK
}

\author{M.~Palmer, H.~Witte}
\affiliation{
Brookhaven National Laboratory, 98 Rochester St, Upton, New York, USA
}

\newcounter{FNEuclid}
\author{D.~Adey\footnote{Current address Institute of High Energy Physics, Chinese Academy of Sciences, 19 Yuquan Rd, Shijingshan District, Beijing, China}, A.~D.~Bross, D.~Bowring, P.~Hanlet, A.~Liu\footnotemark\footnotetext{Current address Euclid Techlabs, 367 Remington Blvd, Bolingbrook, Illinois, USA}\setcounter{FNEuclid}{\value{footnote}}, D.~Neuffer, M.~Popovic, P.~Rubinov}
\affiliation{
Fermilab, Kirk Rd and Pine St, Batavia, Illinois, USA
}

\author{A.~DeMello, S.~Gourlay, A.~Lambert, D.~Li, T.~Luo, S.~Prestemon,  S.~Virostek}
\affiliation{
Lawrence Berkeley National Laboratory, 1 Cyclotron Rd, Berkeley, California, USA
}

\author{B.~Freemire\footnotemark[\value{FNEuclid}], D.~M.~Kaplan, T.~A.~Mohayai\footnote{Current address Fermilab, Kirk Rd and Pine St, Batavia, Illinois, USA}, D.~Rajaram\footnote{Current address KLA, 2350 Green Rd, Ann Arbor, Michigan, USA}, P.~Snopok, Y.~Torun}
\affiliation{
Illinois Institute of Technology, 10 West 35th St, Chicago, Illinois, USA
}

\author{L.~M.~Cremaldi, D.~A.~Sanders, D.~J.~Summers\footnote{Deceased}}
\affiliation{
University of Mississippi, University Ave, Oxford, Mississippi, USA
}

\author{L.~R.~Coney\footnote{Current address European Spallation Source ERIC, Partikelgatan 2, Lund, Sweden}, G.~G.~Hanson, C.~Heidt\footnote{Current address Swish Analytics, Oakland, California, USA}}
\affiliation{
University of California, 900 University Ave, Riverside, California, USA
}

\date{\today}

\begin{abstract}
Multiple Coulomb Scattering (MCS) is a well known phenomenon occurring when charged particles traverse materials. Measurements of muons traversing low $Z$ materials made in the MuScat experiment showed that theoretical models and simulation codes, such as GEANT4 (v7.0), over-estimated the scattering. The Muon Ionization Cooling Experiment (MICE)  measured the cooling of a muon beam traversing a liquid hydrogen or lithium hydride (LiH) energy absorber as part of a programme to develop muon accelerator facilities, such as a Neutrino Factory or a Muon Collider. The energy loss and MCS that occur in the absorber material are competing effects that alter the performance of the cooling channel. Therefore measurements of MCS are required in order to validate the simulations used to predict the cooling performance in future accelerator facilities. We report measurements made in the MICE apparatus of MCS using a LiH absorber and muons within the momentum range 160 to 245\,MeV/$c$. The measured RMS scattering width is about 9\% smaller than that predicted by the approximate formula proposed by the Particle Data Group. Data at 172, 200 and 240\,MeV/$c$ are compared to the GEANT4 (v9.6) default scattering model. These measurements show agreement with this more recent GEANT4 (v9.6) version over the range of incident muon momenta. 
\end{abstract}

\maketitle


\section{\label{Sect:Intro}INTRODUCTION}

Multiple Coulomb Scattering (MCS) describes the multiple interactions of charged particles in the Coulomb field of the nuclei and electrons of a material. Rossi and Greisen derived a simple expression for the root-mean-square (RMS) scattering angle in the small angle approximation \cite{Rossi:1941zza} by integrating the Rutherford cross section \cite{Rutherford:1911zz}.  The mean square scattering angle $\left\langle  \theta^2\right\rangle$ after multiple collisions traversing a thickness $dz$ of material can be expressed as a function of radiation length $X_{0}$
\begin{equation}\label{eq:Rossi}
\frac{d\left\langle \theta^2 \right\rangle}{dz} = \frac{E_s^2}{p^2\beta^2}\frac{1}{X_{0}},
\end{equation}
where $E_s=21.2$\,MeV/$c$, $p$ is the momentum of the charged particle and $\beta$ its speed in units of the speed of light, $c$. The projection of the scattering angle onto a plane containing the incident track gives the RMS projected scattering angle $\theta_0 = \sqrt{\left\langle \theta^2/2 \right\rangle}$ \cite{Zyla:2020zbs}
\begin{equation}
\theta_0  = \frac{14.85\textrm{ MeV/$c$}}{p\beta}\sqrt{\frac{\Delta z}{X_{0}}}.
\end{equation}

Moli\`ere \cite{Moliere:1947zza,Moliere:1948zz} developed a theory of MCS based on the scattering of fast charged particles from atomic nuclei that showed good agreement with data.  Bethe \cite{Bethe:1953va} improved the treatment by taking into account interactions with electrons within the atom. The theory was subsequently improved by Fano \cite{Fano:1954zz} to account for elastic and inelastic scattering. 

Most of the models of MCS mentioned above reproduce data very well \cite{Attwood:2005zz} for small angle scatters and when the atomic number, $Z$, of the target nuclei is large. Highland \cite{Highland:1975pq} compared the Moli\`ere theory with the simple formula by Rossi and Greisen (Eq. \ref{eq:Rossi}), and found a distinct $Z$ dependence of the value of $E_s$. As a consequence, Highland recommended that a logarithmic term be added to the Rossi-Greisen formula to improve the agreement with Moli\`ere's theory, especially at low $Z$ such as for liquid hydrogen or lithium hydride. The formula for $\theta_0$, the RMS width of the Gaussian approximation for the central 98\% of the  projected scattering angle distribution on a plane, was reviewed by Lynch and Dahl \cite{Lynch:1990sq} and is now recommended by the Particle Data Group \cite{Zyla:2020zbs} as
\begin{equation}\label{eq:pdg}
\theta_0 = \frac{13.6\textrm{ MeV/$c$}}{p\beta}\sqrt{\frac{\Delta z}{X_{0}}}\left(1 + 0.038\ln{\frac{\Delta z}{X_{0}\beta^2 }}\right),
	\end{equation}
claimed to be accurate to 11\% over the full range of values of $Z$. 

Multiple scattering has not been well modelled for low $Z$ materials in standard simulations. Data collected by the MuScat experiment \cite{Attwood:2005zz} indicate that GEANT4 v7.0 \cite{Agostinelli:2002hh} and the Moli\`ere model overestimate MCS for these materials. However, a simple Monte Carlo method, which samples the Wentzel scattering cross section \cite{Wentzel:1926} to generate the MCS distributions, was shown by Carlisle and Cobb in \cite{Carlisle:2013dda} to agree very well with muon scattering data from the MuScat experiment. Since the time of MuScat, GEANT4 has evolved through several versions and the comparison to data made in this analysis uses GEANT4 v9.6.

Emittance is a measure of the average spread of particle coordinates in position and momentum phase space and has dimensions of length times angle, e.g., mm$\cdot$radians, usually written as just mm. The Muon Ionization Cooling Experiment (MICE) made measurements of emittance reduction in low $Z$ absorbers, i.e., those materials that can be used to reduce muon-beam emittance via ionization cooling \cite{Neuffer:1983jr}, thus providing the first observation of the ionization cooling process \cite{Bogomilov:2019kfj} that can be used to cool beams of muons for a Neutrino Factory \cite{Bogomilov:2014koa} or a Muon Collider \cite{Neuffer:2017bnx,PhysRevSTAB.16.040104,PhysRevSTAB.18.031003,Neuffer_2017}. The normalized transverse emittance of the MICE muon beam \cite{Bogomilov:2012sr} is  reduced due to energy loss and increased by the scattering in the absorber material. The rate of change in the normalized emittance, $\epsilon_n$, \cite{Neuffer:1983jr} is given by
	\begin{equation}\label{eq:cce}
	\frac{d\epsilon_n}{dz} \approx -\frac{\epsilon_n}{p_{\mu}\beta}\Bigg\langle\frac{dE_{\mu}}{dz}\Bigg\rangle
		+ \frac{\beta_{\perp}p_{\mu}}{2m_\mu}\frac{d\theta_0^{2}}{dz}\,,
	\end{equation}
where $\frac{dE_\mu}{dz}$ is the energy loss of muons per unit distance, $m_{\mu}$ the muon mass, $p_{\mu}$ the muon momentum and $\beta_{\perp}$ the betatron function. 

To make accurate predictions of the emittance in the absorber materials,
the model in the simulation must be validated. This is particularly important for the prediction of the
equilibrium emittance, the case when $d\epsilon_{n}/dz = 0$ and
\begin{equation}
\epsilon_n = \frac{\beta_{\perp}p_{\mu}^{2}\beta}{2m_\mu}\frac{d\theta_0^{2}}{dz}\left\langle\frac{dE_{\mu}}{dz}\right\rangle^{-1}.
\end{equation}
This provides the minimum emittance for which cooling is effective and is lowest for low $Z$ absorbers.
There is thus great interest in performing a detailed measurement of MCS of muons traversing low $Z$ absorbers, such as liquid hydrogen or lithium hydride (LiH). Here, we report the first measurement of MCS of muons in lithium hydride in the muon momentum range 160 to 245\,MeV/$c$, using the MICE apparatus. Accurate MCS modelling will ensure  design studies for future facilities are as informative as possible \cite{Long:2020wfp}. This paper is divided as follows: Section \ref{Sect:Method} outlines the MICE experiment, describes the analysis method and defines the relevant measurement angles, Section \ref{Sect:Data} describes the data collected and the event selection and Section \ref{Sect:Results} describes the data deconvolution method and the multiple scattering results, with a final short conclusion in Section \ref{Sect:Conclusions}.

\section{\label{Sect:Method}METHOD}

The MICE configuration for the MCS measurements presented here consisted of two scintillating fiber trackers, one upstream (US) and one downstream (DS) of a lithium hydride absorber. Each tracker contained five stations, each composed of three planes of scintillating fiber employing 120$^\circ$ stereo views, immersed in helium gas \cite{Ellis:2010bb}. Thin aluminum windows separated the helium volume from the vacuum containing the absorber. The tracker position resolution was determined to be 470\,$\mu$m \cite{Mice:2021lqu}. 
The solenoid magnets surrounding the trackers were turned off for these measurements 
to allow straight-track reconstruction of the muons before and after the absorber. 

The muon beam was generated by protons with a kinetic energy of 700\,MeV at the STFC Rutherford Appleton Laboratory ISIS synchrotron facility \cite{Bogomilov:2012sr,Adams:2013lba} impinging on a titanium target \cite{Booth:2012qz,Booth:2016lno}. The beam line is described in \cite{Bogomilov:2012sr}.

\begin{figure*}[htbp]
\begin{center}
\includegraphics[width=\textwidth]{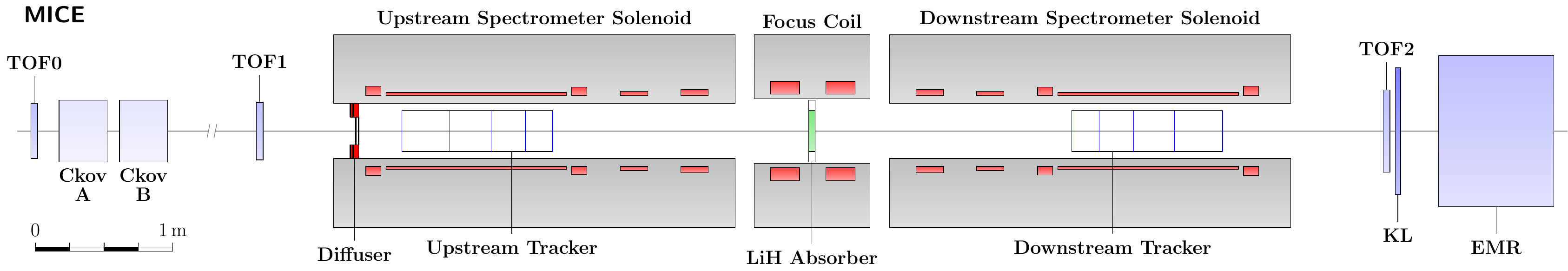}
\end{center}
\caption{Schematic of the MICE cooling channel. The spectrometer solenoids and focus coils were not powered during the measurements described here. A variable thickness diffuser upstream of the trackers was fully retracted during the measurements. Acronyms are defined in the text.}
\label{fig:micecc}
\end{figure*}

A schematic diagram of the MICE cooling channel and detectors is shown in Fig. \ref{fig:micecc}. A time of flight (TOF) system, consisting of three detectors (TOF0 and TOF1 upstream and TOF2 downstream of the apparatus), was used to measure the momentum of reconstructed muons \cite{BERTONI201014}. The Cherenkov detector, pre-shower system (KL) and Electron-Muon Ranger (EMR) were used to confirm the TOF's particle identification performance \cite{Bogomilov:2012sr,Adams:2015wxp,Adams:2015eva}.  The MICE coordinate system is defined with $+z$ pointing along the beam direction towards the downstream region,  $+y$ pointing upwards and $+x$ defined to be consistent with a right-handed coordinate system.

The MICE LiH absorber was a disk, 65.37$\pm$0.02~mm thick (along the $z$-axis) and 550~mm in diameter. The absorber was coated with a thin parylene layer to prevent the ingress of water or oxygen. The composition of the LiH disk by weight was 81\% $^{6}$Li, 4\% $^{7}$Li and 14\% $^{1}$H with some trace amounts of carbon, oxygen and calcium. The density of the disk was measured to be 0.6957$\pm$0.0006 g/cm$^{3}$, and the radiation length was calculated to be 70.38~g/cm$^2$.

Multiple scattering is characterized using either the three-dimensional (3D) angle between
the initial and final momentum vectors, $\theta_{Scatt}$, or the 2D projected angle of scattering. The projected angles between the track vectors in the $x$-$z$ ($\theta_Y$) and $y$-$z$ ($\theta_X$) planes of the experimental coordinate system can be used, but these are only the true projected angles if the incident muon has no component of momentum in a direction perpendicular to these planes, i.e., the $y$ or $x$ direction respectively. To obtain the correct projected angle, a plane of projection must be defined for each incoming muon. The rotation calculated about an axis in the plane defined for each incoming muon is, to a very good approximation, the rotation around the specified axis. The precise definitions of $\theta_X$ and $\theta_Y$ are given in the Appendix. 

Table \ref{tab:RMSpred} shows the expected RMS projected scattering angles, $\theta_0$, obtained using 
Eq. \ref{eq:pdg}, for the LiH absorber and the material in each of the trackers. The number of radiation lengths traversed by a muon as it passes through the absorber was larger than that which it traversed as it passed through the trackers hence the majority of the scattering occurs in the absorber. Nevertheless the scattering in the trackers is significant and must be corrected for.

\begin{table*}
\caption{Material budget affecting particles passing through the MICE LiH
  absorber. The material thickness normalized by the radiation length
  is given with the RMS width of the scattering distribution calculated from
  the full PDG formula \cite{Zyla:2020zbs} in eq. \ref{eq:pdg}. Note that the 
  thickness shown for the tracker materials (He, Al windows and Scintillating Fibres) includes 
  both trackers.}
\label{tab:RMSpred}
\vspace{3mm}
\centering
\begin{ruledtabular}
\begin{tabular}{c|ccc|ccc}
& & & & \multicolumn{3}{c}{$\theta_0$ (mrad)} \\ 
Material & $z$ (cm) & $z/X_{0}$ & $\rho$ (g cm$^{-3}$) & 172\,MeV/$c$ & 200\,MeV/$c$ & 240\, MeV/$c$ \\ 
\hline
Tracker He & 226 & 0.00030 & 1.663$\times$10$^{-4}$ & 1.09 & 0.91 & 0.73 \\ 
Al Window & 0.032 & 0.0036 & 2.699 & 4.31 & 3.58 & 2.89 \\ 
Scintillating Fibres & 1.48 & 0.036 & 1.06 & 14.9 & 12.4 & 10.0 \\
Total Tracker &  & 0.038 & & 15.8 & 13.2 & 10.6 \\
LiH & 6.5 &  0.0641 & 0.6957      & 21.3 & 17.7 & 14.3 \\ 
\hline
Total with LiH & & 0.1058 & & 29.9 & 24.8 & 20.0 \\
\end{tabular}
\end{ruledtabular}
\end{table*}

\section{\label{Sect:Data}DATA SELECTION AND RECONSTRUCTION}

A coincidence of two PMTs firing in TOF1 was used to trigger readout of the detector system including the trackers. The muon rate was such that only a single incident particle was observed in the apparatus per readout. Data reconstruction and simulation were carried out using MAUS (MICE Analysis and User Software) v3.3.2 \cite{Asfandiyarov:2018yds} (which uses GEANT4 v9.6.p02). Position and angle reconstruction was performed using data from the MICE trackers while momentum reconstruction was performed using data from the TOF detectors.

\subsection{Position and Angle Reconstruction}
\label{sec:Recon}
Space points were created from the signals generated in the three scintillating fiber planes contained in a tracker station. Multiple space points that formed a straight line through the tracker were associated together. Space points that did not match a possible track were rejected. A Kalman filter \cite{Kalma:1960} was used to provide an improved estimate of the track position and angle in each tracker at the plane nearest to the absorber. 

An upstream track was required for the event to be considered for analysis, with a minimum of three space points among the five stations of the upstream tracker. No requirement was made on the presence of a downstream track. All scattering distributions were normalized to the number of upstream tracks selected in the analysis. The efficiency of the trackers has been shown to be very close to $100$\% \cite{Dobbs:2016ejn}. 

A residual misalignment between the upstream and downstream trackers was corrected by rotating all upstream tracks by a fixed angle in the range 1--7\,mrad. The final uncertainty in the rotation angles following the alignment procedure was 0.07\,mrad. 
\subsection{Momentum Reconstruction}

\begin{figure}[htp]
\begin{center}
\subfigure{\label{sfig:recvtrue}\includegraphics[width=0.5\textwidth]{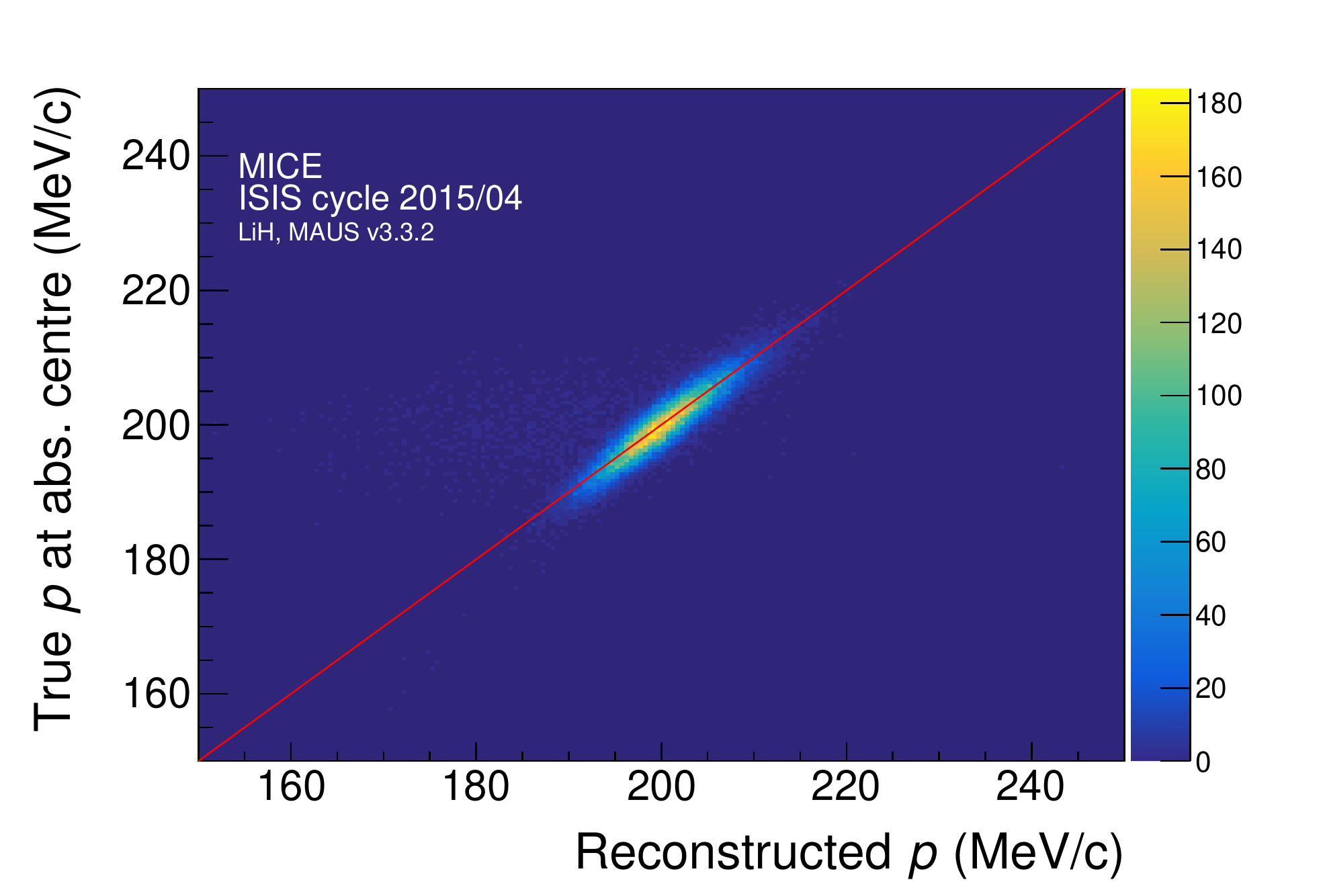}}
\hspace{.5cm}
\subfigure{\label{sfig:recvtruepfl}\includegraphics[width=0.5\textwidth]{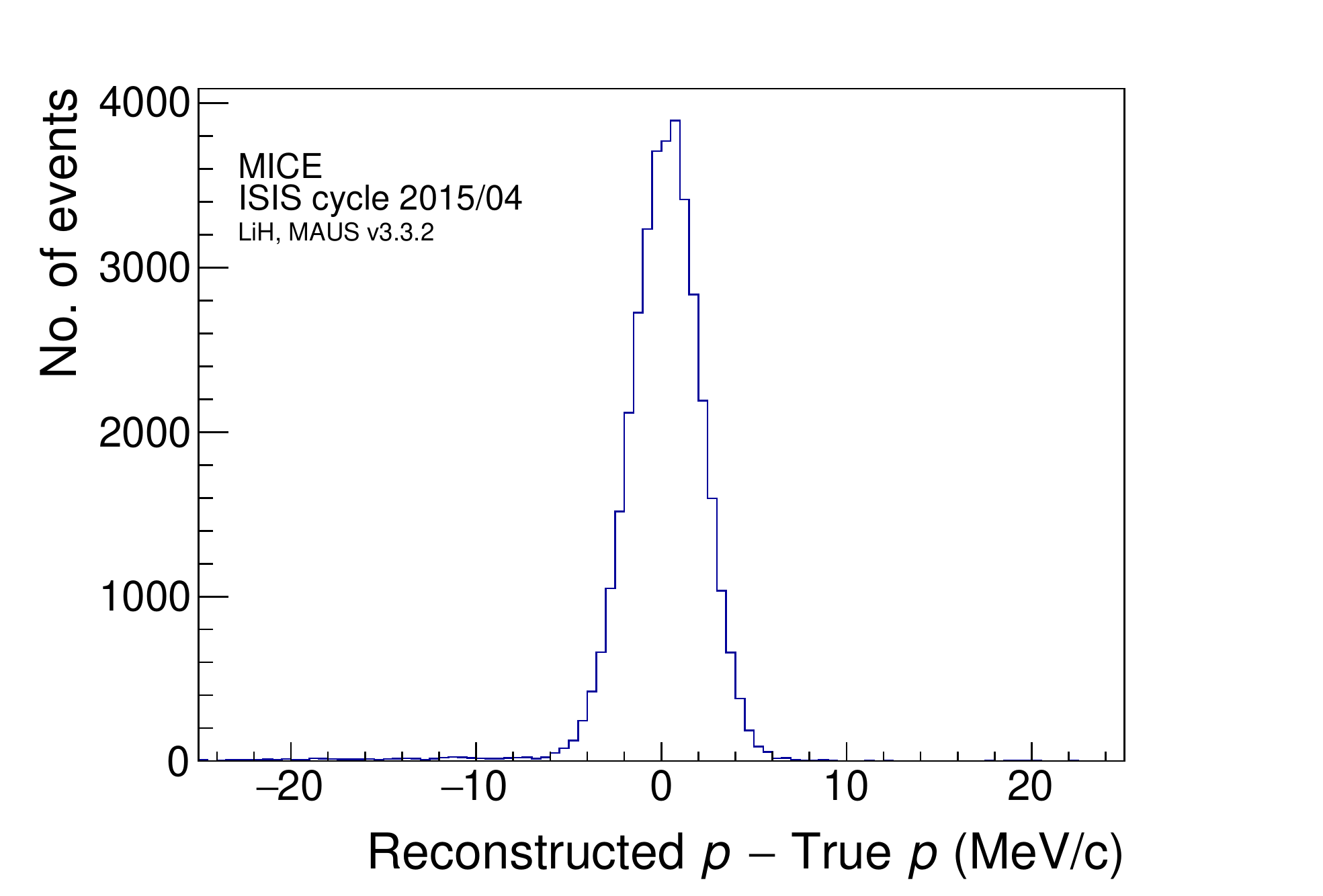}}
\caption{Top: comparison of the reconstructed and true momentum for the MC sample, for the bin with average momentum 200 MeV/c. Bottom: residual between reconstructed and true momentum for  the MC sample. The systematic error associated with the momentum reconstruction is discussed in section \ref{subsec:systematics}}
\label{fig:beampdep}
\end{center}
\end{figure}

Time of flight was used to measure the momentum of the muon at the absorber. Two time of flight measurements were used, designated as TOF01, the time of flight between TOF0 and TOF1, and TOF12, the time of flight between TOF1 and TOF2. The average momentum between time of flight detectors was calculated by evaluating 
\begin{equation}
p=\frac{m_\mu c}{\sqrt{\frac{t_{\mu}^2}{t_{e}^2}-1}}-\Delta p_{BB}-p_{MC},
\label{momexp}
\end{equation}
which assumes the mass of the electron to be $\approx$\,0 and where $t_{\mu}$ is the time of flight of the muon and $t_e$ is the average 
time of flight of positrons ($t_e=25.40$\,ns for TOF01 and $27.38$\,ns for TOF12). $\Delta p_{BB}$ was an additional term which accounted for the Bethe-Bloch most probable energy loss of the muon as it passes through matter and was chosen to yield an optimal reconstructed momentum at the center of the absorber. When measuring the momentum using TOF01, accounting for the material upstream of the LiH absorber, $\Delta p_{BB}$ was of order $\sim$25\,MeV/$c$ (the correction varied as a function of muon momentum and was calculated separately for each selected sample of muons). $p_{MC}$ accounted for the bias between the reconstructed and true momentum observed in the Monte Carlo (MC) simulation; this arises primarily due to the simplifying assumptions intrinsic to Eq. \ref{momexp}, e.g., that the path length between the TOF detectors can be approximated to the straight line on-axis distance between the two detectors when in fact the particle's trajectory may have curved through various magnetic fields or scattered in material. $p_{MC}$ was used when calculating the momentum with both TOF01 and TOF12 and the correction, $p_{MC}$, was $\sim$2--6\,MeV/$c$. After correction, the reconstructed data were well described by the MC as shown in Fig. \ref{fig:beampdep}. 

For muons reaching the end of the channel, the momentum measurement was made using TOF1 and TOF2. In this case the absorber sits near the midpoint between the detectors and the distance between them was larger than the distance between TOF0 and TOF1 which results in a slightly smaller uncertainty. In the selected samples, $\sim$\,90\% of muons reach TOF2. If no hit was recorded in TOF2, the momentum measurement was made using TOF0 and TOF1. The TOF01 distribution is shown in Fig. \ref{fig:tofsel_plots}.

Characteristics of the time-of-flight samples selected using TOF01 are shown in Table~\ref{tab:tofsel}. The resolution of the TOF system was $\approx$70 ps which corresponds to $\sim$4--10\,MeV/$c$ depending on the momentum setting. The agreement between the reconstructed momentum and the simulated true muon momentum at the centre of the absorber is shown in Fig. \ref{sfig:recvtrue} and a residual plot ($p_{\mathrm{Reconstructed}}-p_{\mathrm{Truth}}$) is shown in Fig. \ref{sfig:recvtruepfl}. 

\begin{table*}[htp]
\caption{Characteristics of the samples selected for model comparison; the standard deviation of the reconstructed momenta are compared with the spread of true momenta of equivalent samples selected from the simulation.}
\begin{center}
\begin{ruledtabular}
\begin{tabular}{c|cc|cc|c}
	Desired Momentum & Lower TOF & Upper TOF & Measured & Standard deviation & True MC Momentum\\
	(MeV/$c$) & limit (ns) & limit (ns) & $\langle p\rangle$ (MeV/$c$) & (MeV/$c$) & Spread (MeV/$c$)\\
	\hline
	172 & 28.60 & 28.80 & 171.55$\pm$0.06 & 4.37$\pm$0.06 & 4.82\\
	200 & 27.89 & 28.09 & 199.93$\pm$0.07 & 5.92$\pm$0.05 & 5.97\\
	240 & 27.16 & 27.36 & 239.76$\pm$0.13 & 8.95$\pm$0.09 & 8.21\\ 
\end{tabular}
\end{ruledtabular}
\end{center}
\label{tab:tofsel}
\end{table*}

\subsection{Data Collection}
Six data sets were collected during the ISIS user cycle 2015/04 using muon beams with a nominal 3\,mm emittance, at three nominal momenta (172, 200 and 240\,MeV/$c$). The three data sets collected with the LiH absorber in place are referred to as `LiH' data while the three data sets with no absorber in place are referred to as `No Absorber' data. The beams typically had RMS widths of $30$--$36$\,mm and divergences of $9.0$--$9.4$\,mrad, after the selection described in section  \ref{sec:selection}. The No Absorber data sets were used to determine the scattering attributable to the tracking detectors and thus to extract the true MCS distribution due to the LiH  absorber. Two methods, described in Section \ref{Sect:Results}, were used. Positively charged muon beams were used to minimize pion contamination, which was measured to be less than 1.4\% \cite{Adams:2015wxp}. Positron contamination was identified and rejected using the time-of-flight system. 

\subsection{Event Selection}
\label{sec:selection}

The data from the three nominal muon beams were merged into one sample and all muons in the sample were treated identically. Unbiased scattering distributions were selected 
from the data samples using the cuts listed in Table \ref{tab:selection}. The fraction of events selected by each cut is also shown. Events that produced one space point in TOF0 and one space point in TOF1 were selected.  A beam diffuser, otherwise used to increase the beam emittance, was fully retracted for all of the runs used in this analysis. A fraction of the muon beam traversed the diffuser ring in its retracted position, adding additional energy loss. Any upstream tracks that traversed the outer ring of the diffuser were removed.  

\begin{table*}[!htb]
\begin{center}
	\caption{Particle selection criteria and survival rates for the muon sample with a LiH absorber.}
		\label{tab:selection}
		\vspace*{3mm}
		\begin{ruledtabular}
		\begin{tabular}{l|p{8.4cm}|c}
	Selection & Description & Fraction events surviving each cut \\
	\hline
	Upstream track selection & Exactly one TOF0 space point, exactly one TOF1 space point and one upstream track. &  100.0\%\\
Diffuser cut & Upstream tracks were projected to the diffuser position. Any track outside the radius of the diffuser aperture was rejected. & 81.7\%\\
Fiducial selection & Upstream tracks, when projected to the far end of the downstream tracker, have a projected distance from axis less than 90 mm. & 3.7\%\\
	\end{tabular}
	\end{ruledtabular}
\end{center}
\end{table*}

A fiducial selection to ensure that the unscattered downstream track was likely to have been within the volume of the downstream tracker was also applied. If the upstream track, when projected to the downstream end of the downstream tracker, passed outside of the fiducial radius $r_0=90$~mm the track was rejected.

\begin{figure}[htp]
\begin{center}
\includegraphics[width=0.5\textwidth]{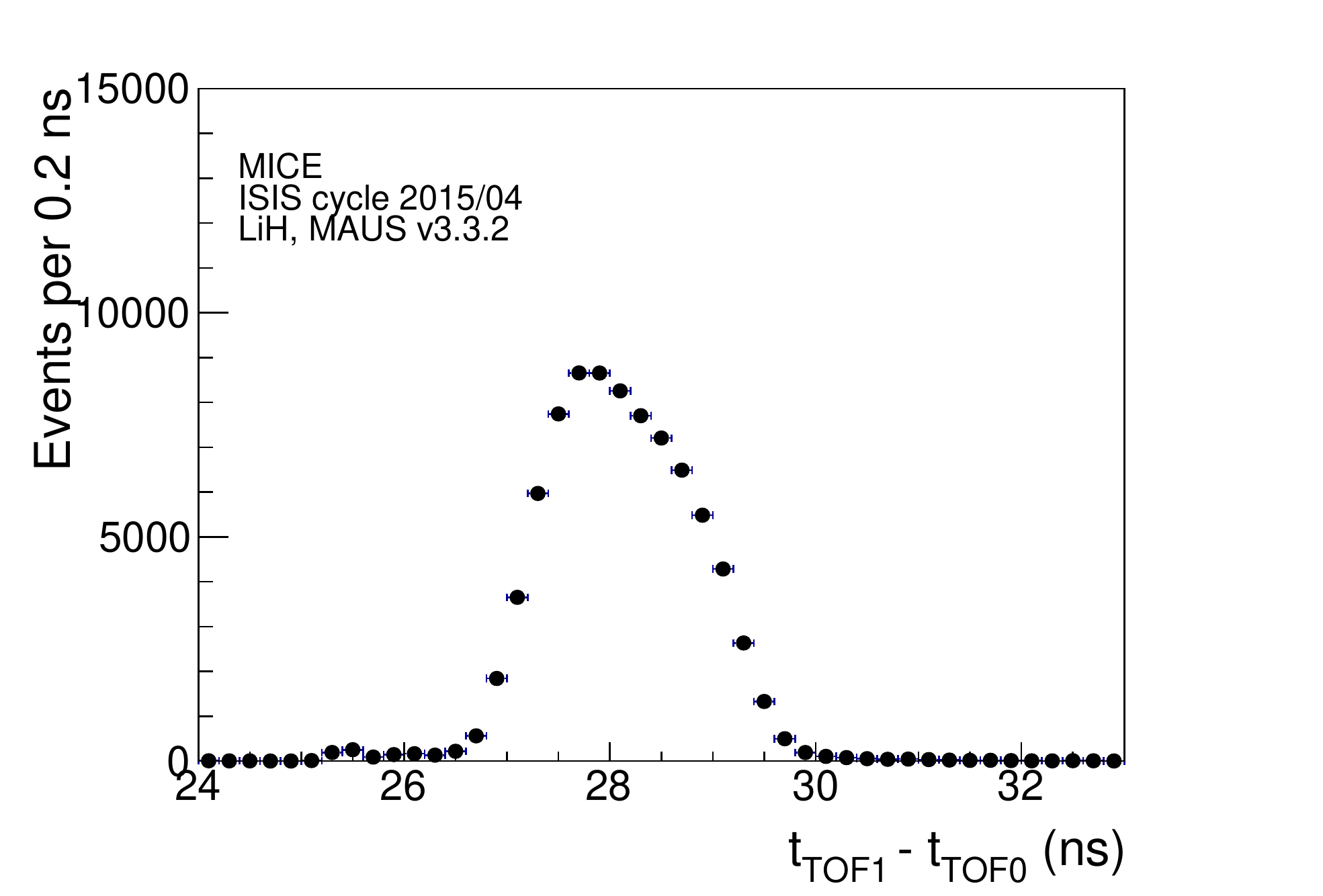}
\caption{Combined time-of-flight between TOF0 and TOF1 distribution of LiH data for all beam line settings after all selection cuts.}
\label{fig:tofsel_plots}
\end{center}
\end{figure}

Finally, particles with a time of flight between stations TOF0 and TOF1 compatible with the passage of a muon (above 26\,ns) were selected. The data were then binned in 200~ps $\Delta t_{01}$ bins (Fig. \ref{fig:tofsel_plots}) to yield eleven quasi-monochromatic samples. Most positrons, which had a TOF between 25 and 26\,ns, were excluded by this binning. Three of these samples, with mean momentum of 172, 200 and 240\,MeV/$c$ and containing 0.19, 0.25 and 0.19\% of the total number of events respectively, were compared to the GEANT4 and Moli\`ere models. The sample at 172\,MeV/$c$ enabled comparison with MuScat while samples at 200 and 240\,MeV/$c$ were of interest for the MICE experiment. The selected sample sizes are shown in Table \ref{tab:samplesizesel}.

\begin{table}[htp]
\caption{Sample size after selection.}
\begin{center}
\begin{ruledtabular}
\begin{tabular}{l|c|cc}
Absorber &	$p$ (MeV/$c$) & No. of events US & No. of events DS\\
	\hline
    & 172 & 6479 & 5906 \\
LiH	& 200 & 8589 & 8112 \\
	& 240 & 5612 & 5445 \\
	\hline
	& 172 & 1500 & 1469 \\
No Absorber	& 200 & 2025 & 1995 \\
	& 240 & 1394 & 1378 \\ 
\end{tabular}
\end{ruledtabular}
\end{center}
\label{tab:samplesizesel}
\end{table}

\subsection{Acceptance Correction}
\begin{figure}[htbp]
\begin{center}
\subfigure{\label{sfig:accx}\includegraphics[width=0.5\textwidth]{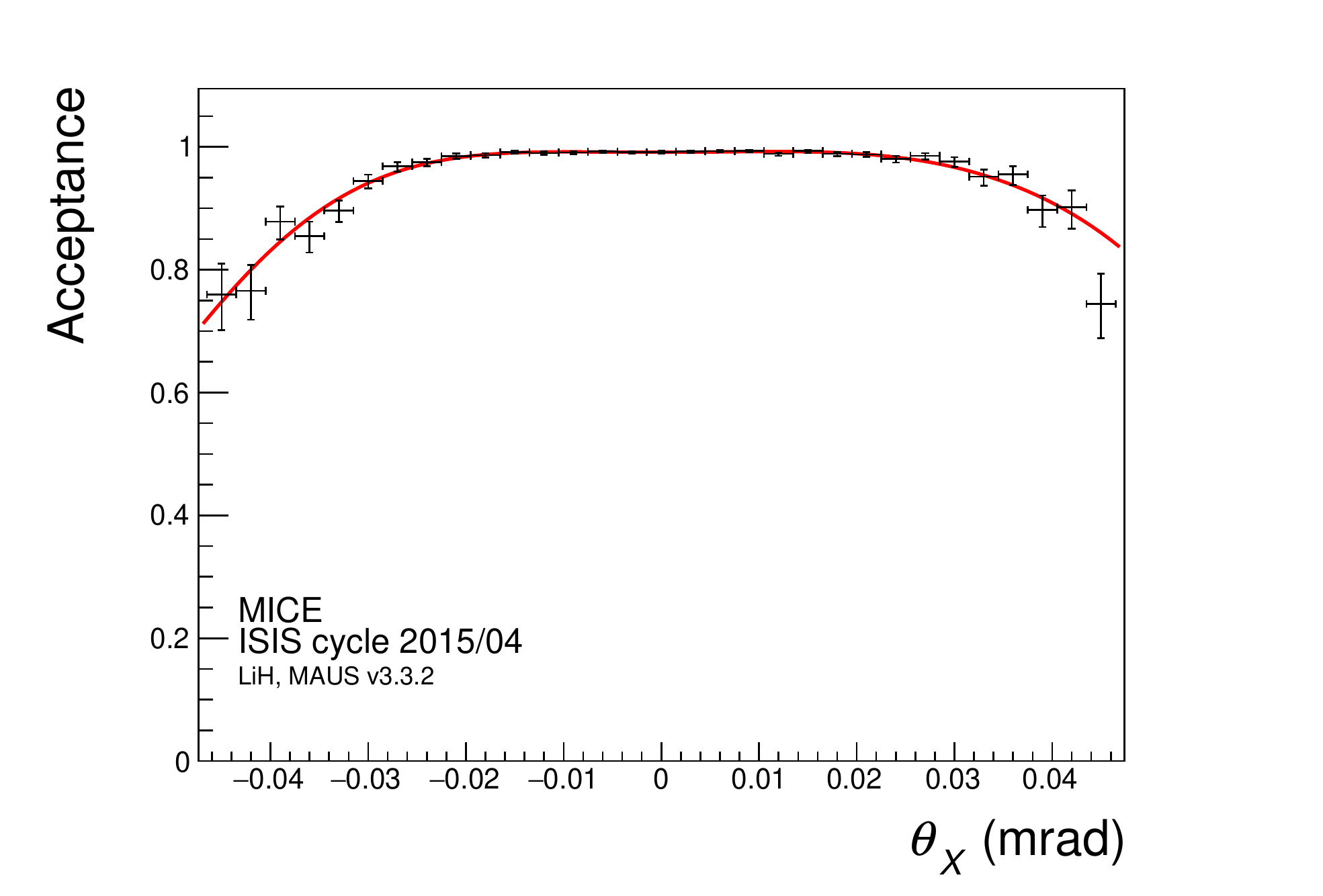}}
\subfigure{\label{sfig:accy}\includegraphics[width=0.5\textwidth]{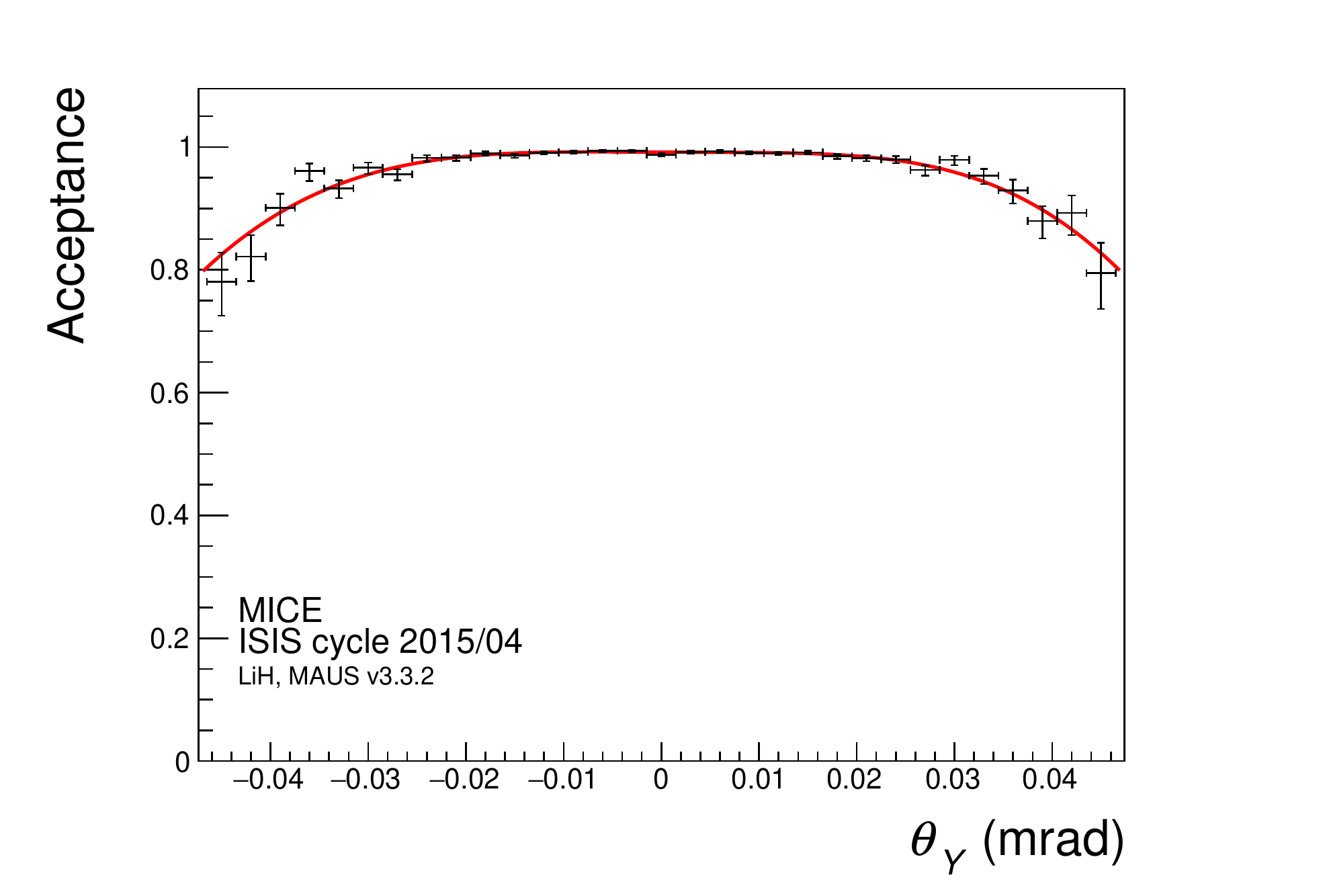}}
\caption{The simulated fraction of events reconstructed by the trackers as a
  function of scattering angle after event selection. The red curve is an asymmetric seventh order polynomial  fitted to the points and used for the acceptance correction.}
\label{fig:acc}
\end{center}
\end{figure}

The simulated geometric acceptance of the downstream tracker as a function of the
projected scattering angles $\theta_{X}$ and $\theta_{Y}$ is shown in
Fig. \ref{fig:acc}. The acceptance depends on the scattering
angle so the scattering angle distributions must be corrected by the acceptance determined from
simulation. The acceptance data were fitted by a seventh order polynomial, 
\[\epsilon = a + b\theta_i^2 + c\theta_i^4+d\theta_i^6+e\theta_i^7,\]
where $i$ is the bin number and $a, b, c$, $d$ and $e$ are fit parameters. This smoothed fluctuations in the tails of the acceptance function. 

\subsection{Comparison to simulation}
The MICE MC simulation models particles arising from protons incident on the target. G4beamline \cite{Roberts:2008zzc} was used to simulate particles from immediately after the target to just upstream of TOF0. The remainder of MICE, including the downstream portion of the beam line and cooling channel, was simulated using MAUS \cite{Asfandiyarov:2018yds}. The simulation is handled in this way to reduce the computing resources required, as only a small subset of particles at the target is transported to the end of the cooling channel. 

\begin{figure*}[htp]
\begin{center}
\subfigure{\label{sfig:MCDataPz172}\includegraphics[width=0.48\textwidth]{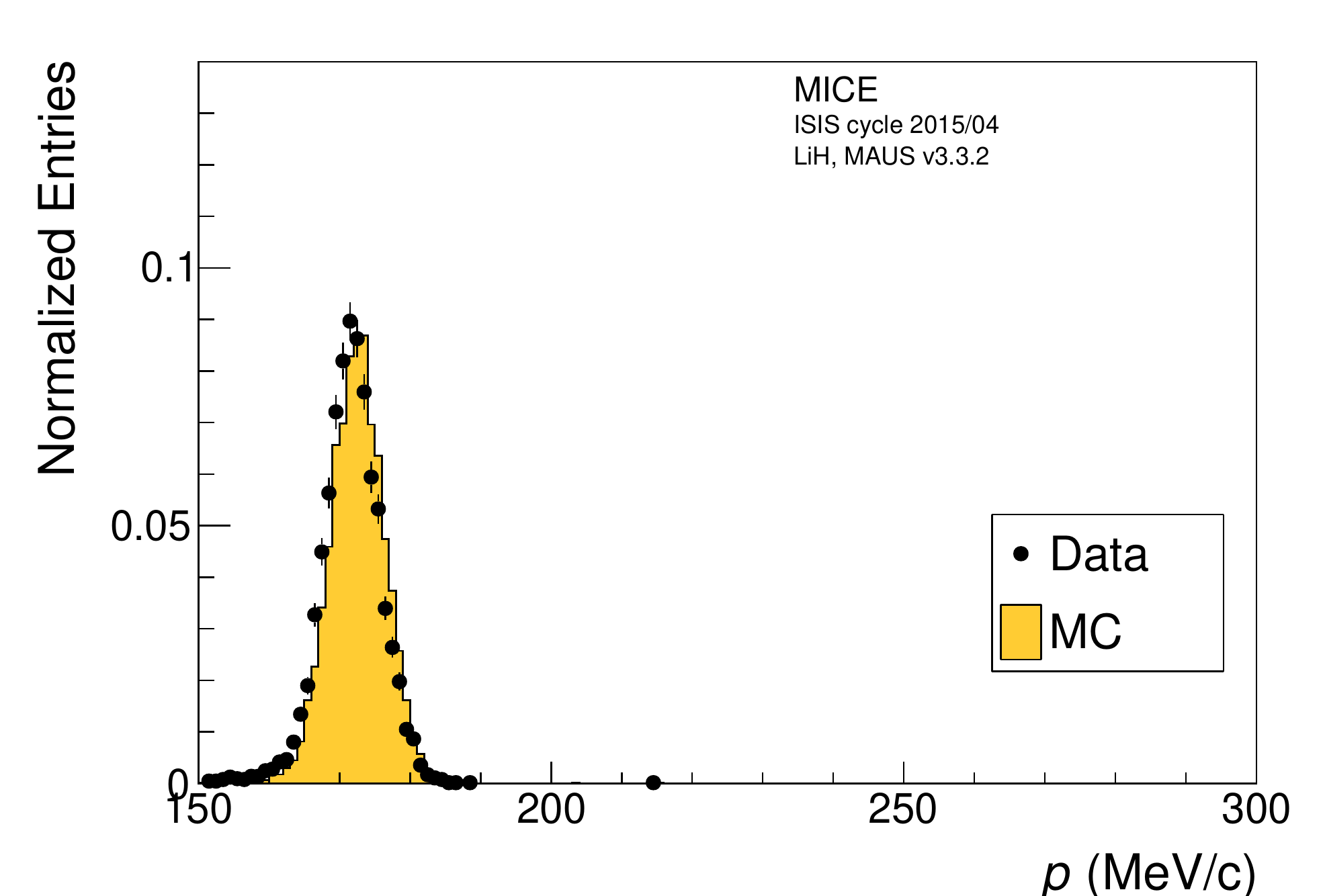}}\hspace{3mm}
\subfigure{\label{sfig:MCDataPz200}\includegraphics[width=0.48\textwidth]{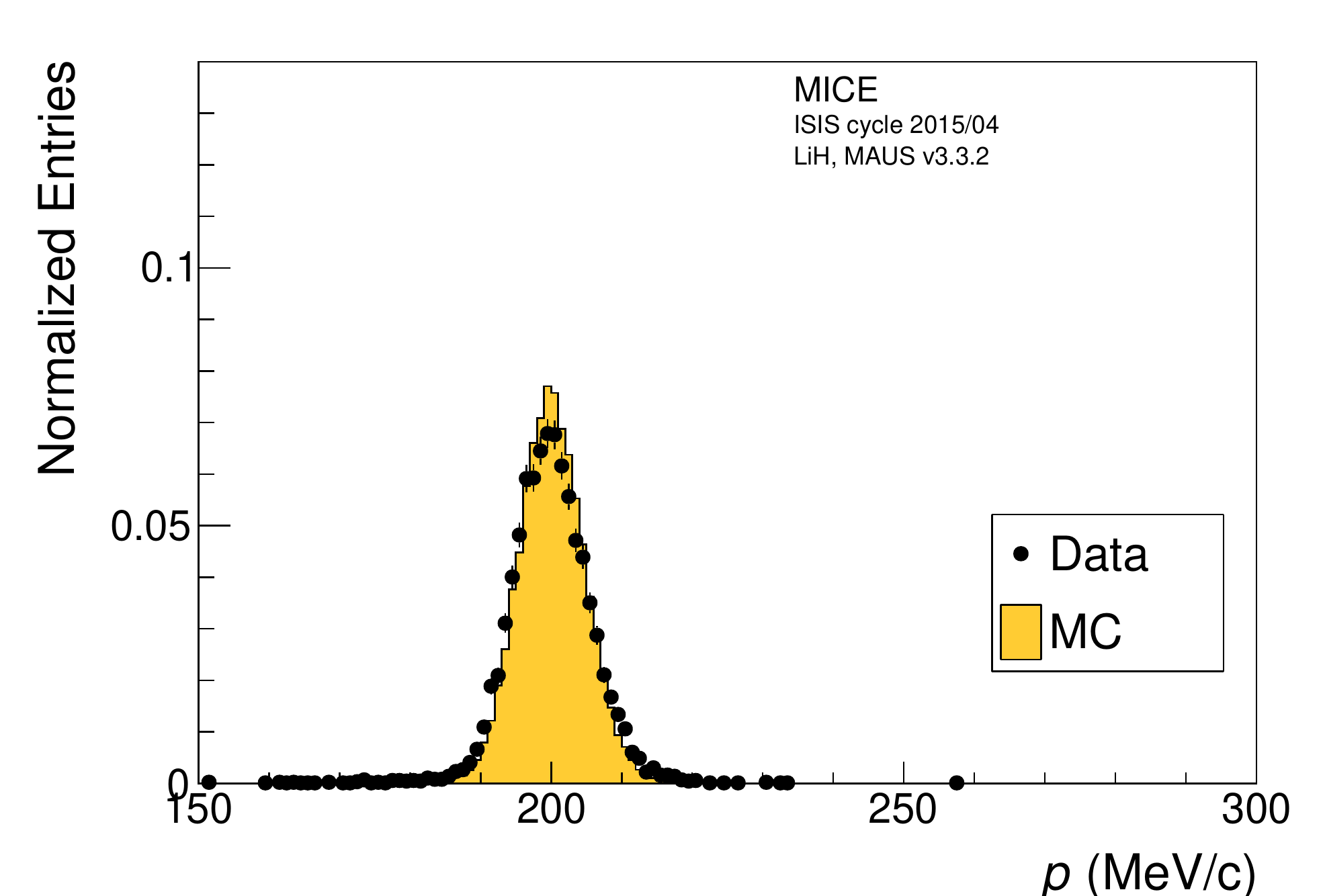}}
    \subfigure{\label{sfig:MCDataPz240}\includegraphics[width=0.48\textwidth]{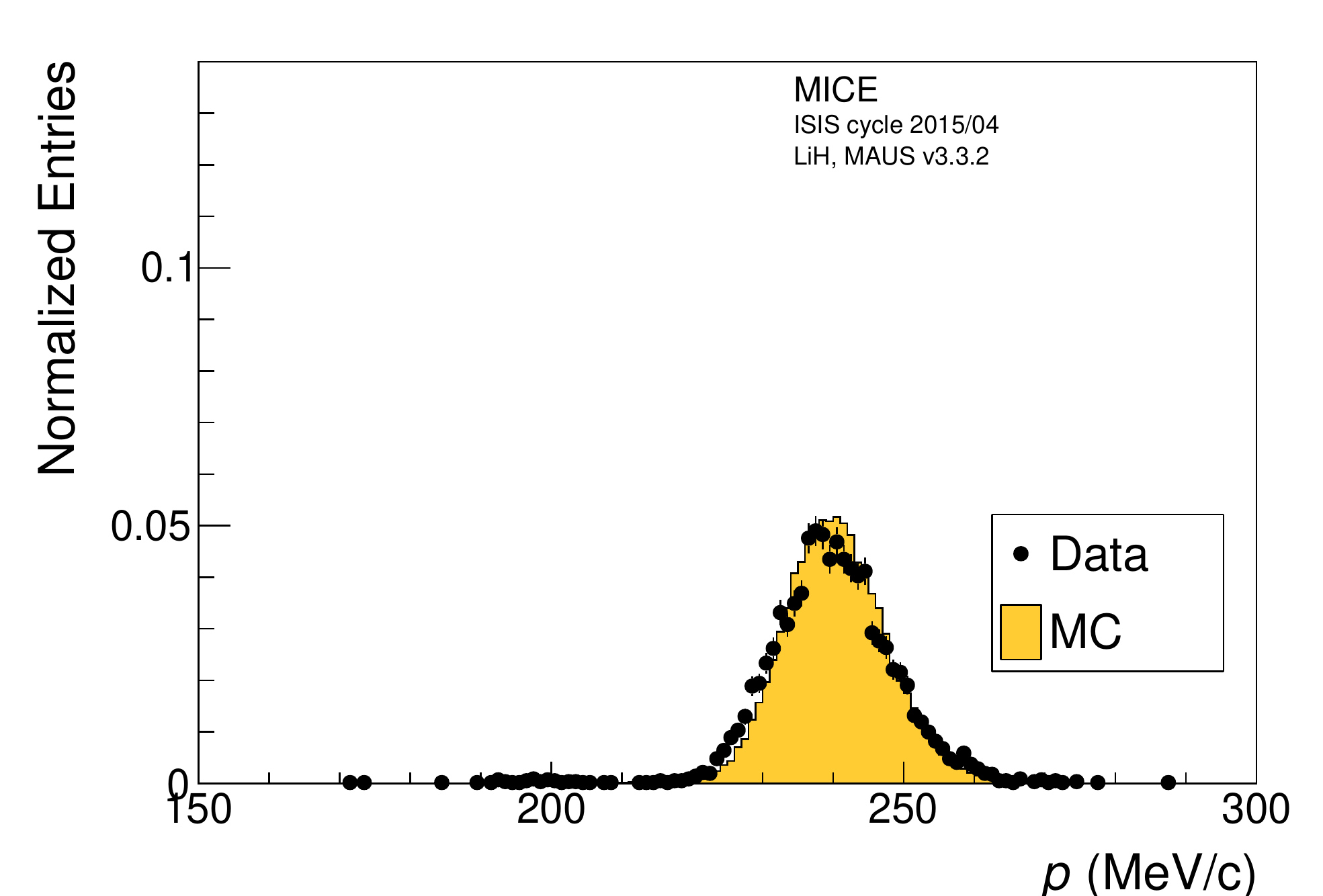}}\hspace{3mm}
\caption{Comparison of reconstructed muon momentum at the center of the absorber for the 172, 200 and 240\,MeV/$c$ samples for data and simulation.}
\label{fig:MCDataPz}
\end{center}
\end{figure*}

\begin{figure*}[htp]
\begin{center}
\subfigure{\label{sfig:MCDataX}\includegraphics[width=0.48\textwidth]{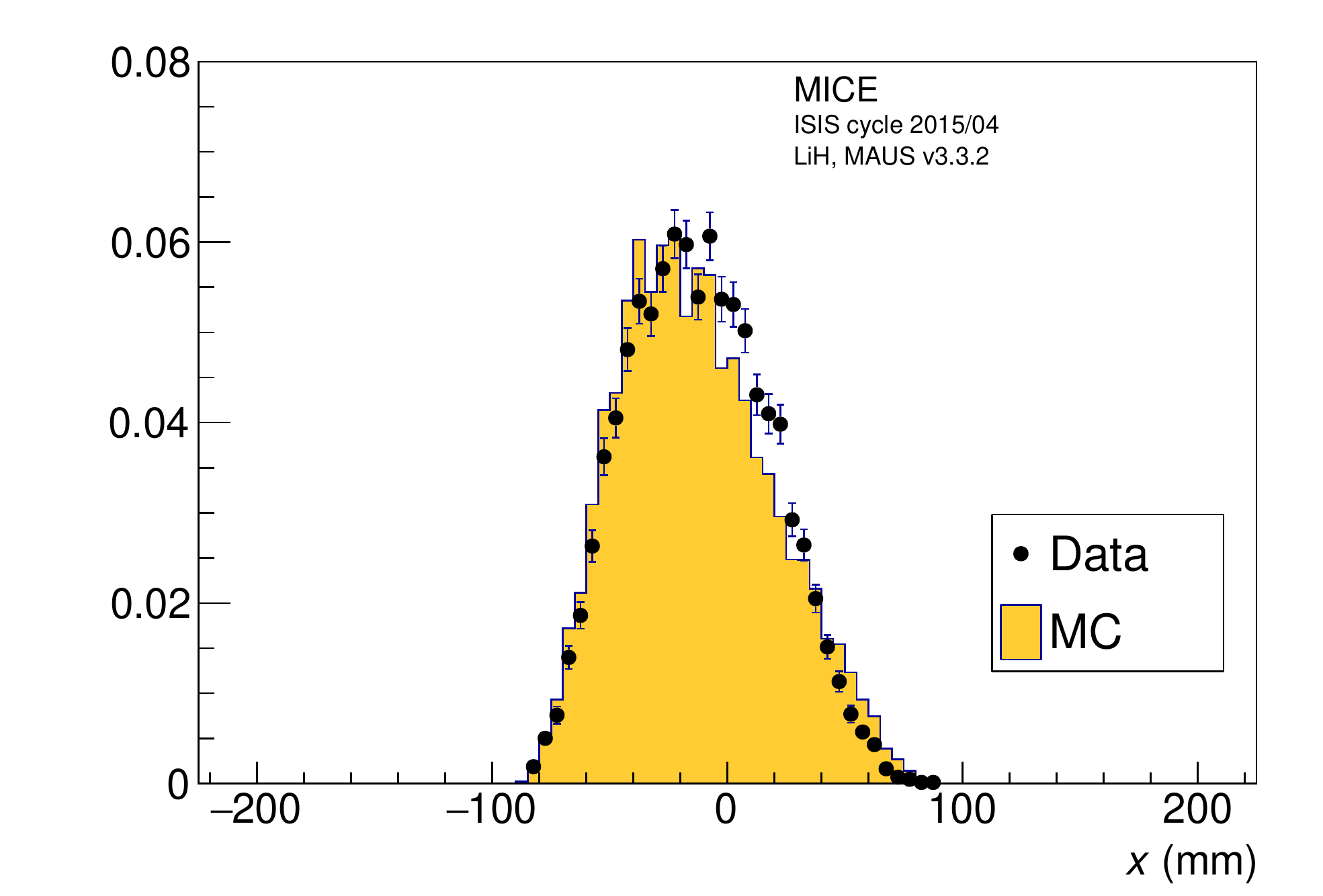}}
\subfigure{\label{sfig:MCDataY}\includegraphics[width=0.48\textwidth]{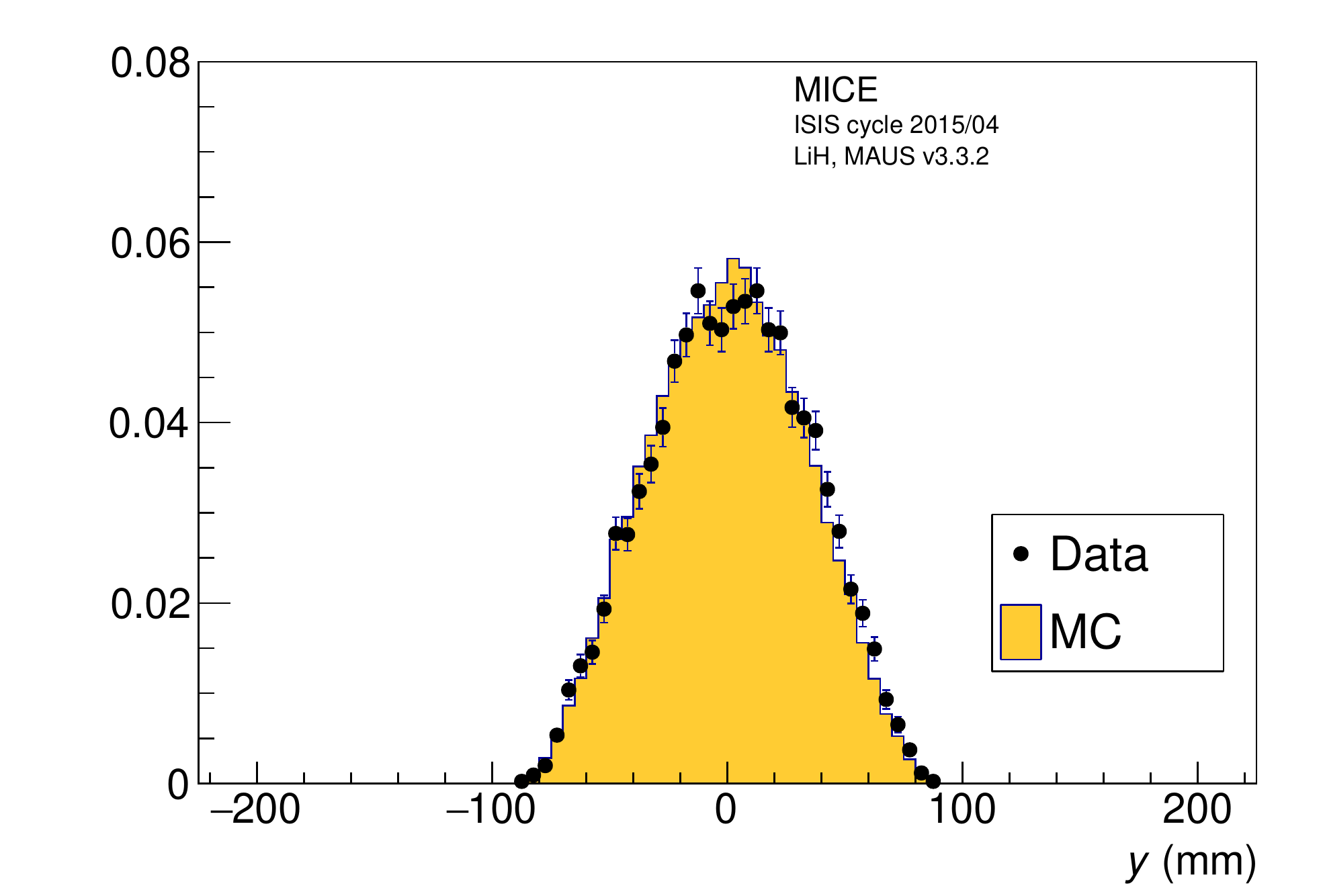}}
\subfigure{\label{sfig:MCDatadXdz}\includegraphics[width=0.48\textwidth]{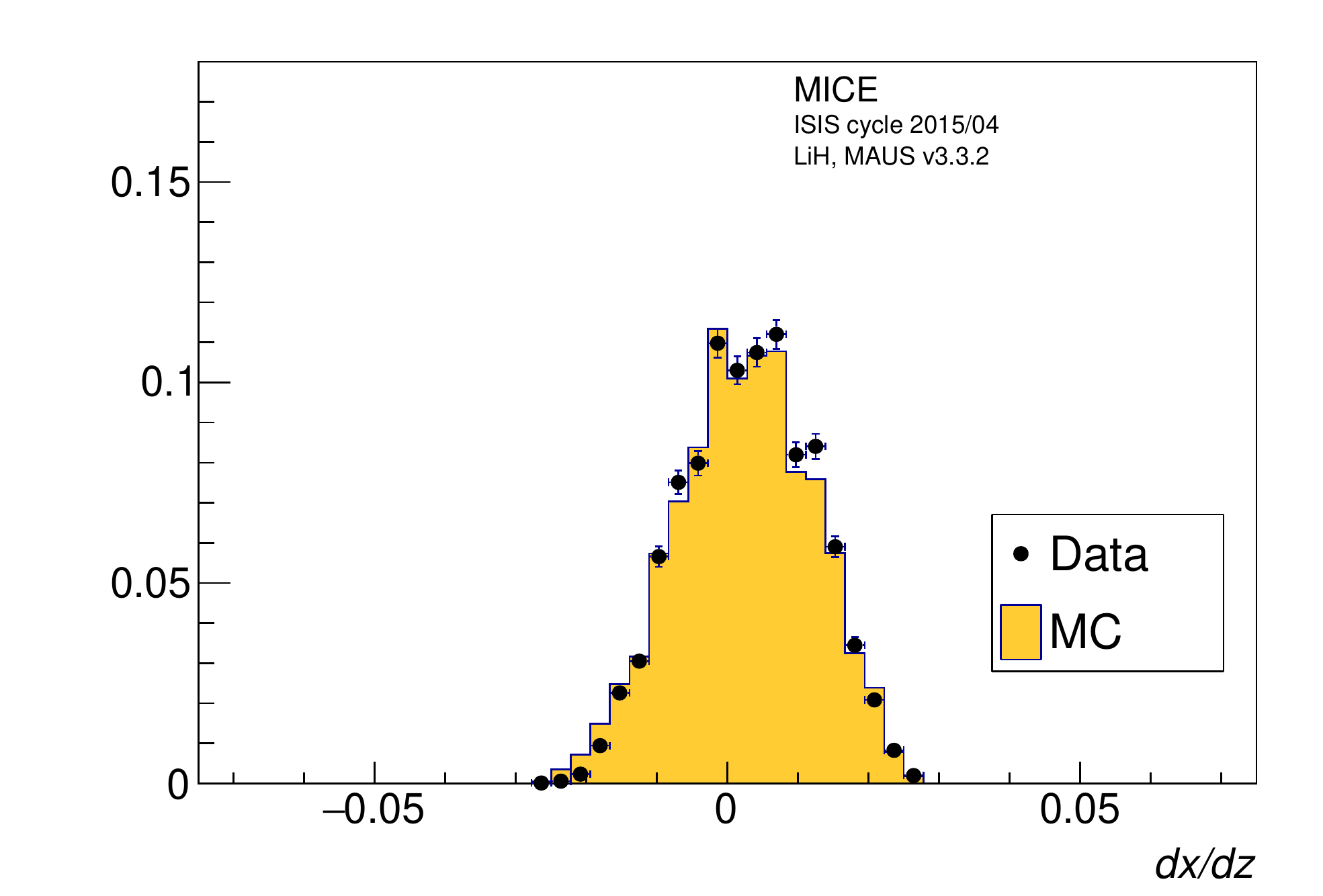}}
\subfigure{\label{sfig:MCDatadYdz}\includegraphics[width=0.48\textwidth]{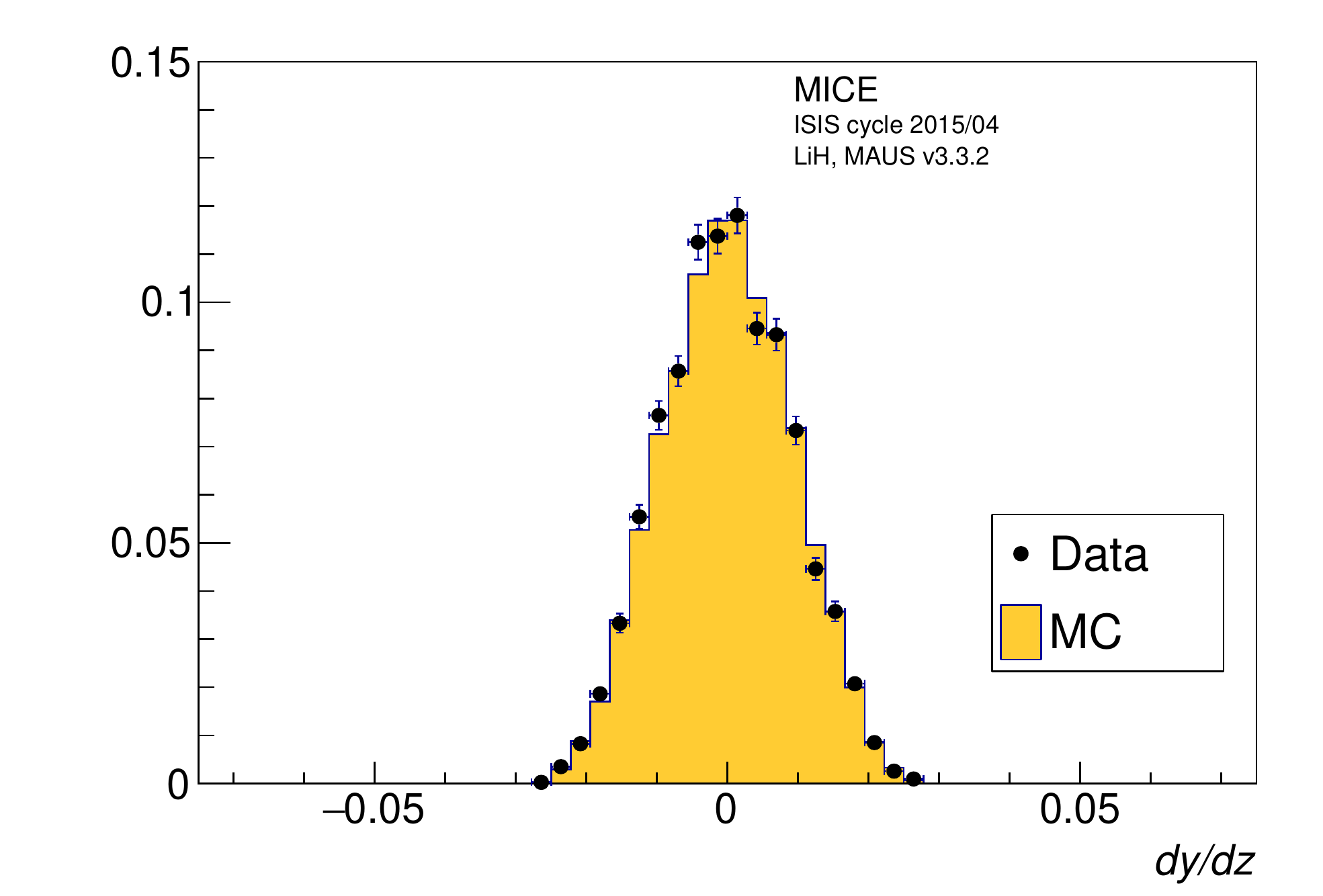}}
\caption{Comparison between Monte Carlo simulations and data for muons in the 200\,MeV/$c$ sample with the LiH absorber installed. All distributions are for the selected muons at the upstream reference plane. Top Left: $x$ distribution, top right: $y$ distribution, bottom left: $dx/dz$ distribution and bottom right: $dy/dz$ distribution }
\label{fig:MCData}
\end{center}
\end{figure*}

A comparison between the momentum distributions for reconstructed MC and data for the selected samples at three momenta (172, 200 and 240\,MeV/$c$) is shown in Fig. \ref{fig:MCDataPz}. The measured distributions of $x$ and $y$ positions and slopes for the selected upstream muon samples are well described by the GEANT4 (v9.6) MC, as illustrated in Fig. \ref{fig:MCData}. 

\section{\label{Sect:Results}RESULTS}

\subsection{Raw Data MC Comparison}
\begin{table*}
	\caption{Distribution widths of multiple scattering in lithium hydride and the $\chi^{2}$ comparisons between data and the GEANT4 simulation. The $\chi^{2}/N\!D\!F$ were calculated using the number of bins as the number of degrees of freedom. Statistical and systematic uncertainties are given for the data distributions. Only statistical uncertainties are given for the model.}
\vspace{3mm}
\begin{center}
\begin{ruledtabular}
	\begin{tabular}{cc|c|ccc}
	$p$ (MeV/$c$) & Angle & $\theta_{\mathrm{Data}}$ (mrad)& $\theta_{G4}$ (mrad) & $\chi^{2}/N\!D\!F$ & P-value \\
\hline
171.55& $\theta_X$ & 21.16$\pm$0.28$\pm$0.48 & 21.87$\pm$0.25 & 23.67 / 31 & 0.79\\
171.55& $\theta_Y$ & 20.97$\pm$0.27$\pm$0.48 & 21.51$\pm$0.25 & 37.86 / 31 & 0.15\\
\hline
199.93& $\theta_X$ & 18.38$\pm$0.18$\pm$0.33 & 18.76$\pm$0.09 & 17.75 / 31 & 0.96\\
199.93& $\theta_Y$ & 18.35$\pm$0.18$\pm$0.33 & 18.89$\pm$0.09 & 27.93 / 31 & 0.57\\
\hline
239.76& $\theta_X$ & 15.05$\pm$0.17$\pm$0.21 & 15.69$\pm$0.06 & 8.07 / 31 & 1.00\\
239.76& $\theta_Y$ & 15.03$\pm$0.16$\pm$0.21 & 15.55$\pm$0.06 & 8.23 / 31 & 1.00\\
	\end{tabular}
	\end{ruledtabular}
	\label{tab:dataMC}
\end{center}
\end{table*}

\begin{figure*}
\centering
\subfigure{\label{fig:MCdata_172_compX}
\includegraphics[width=0.48\textwidth]{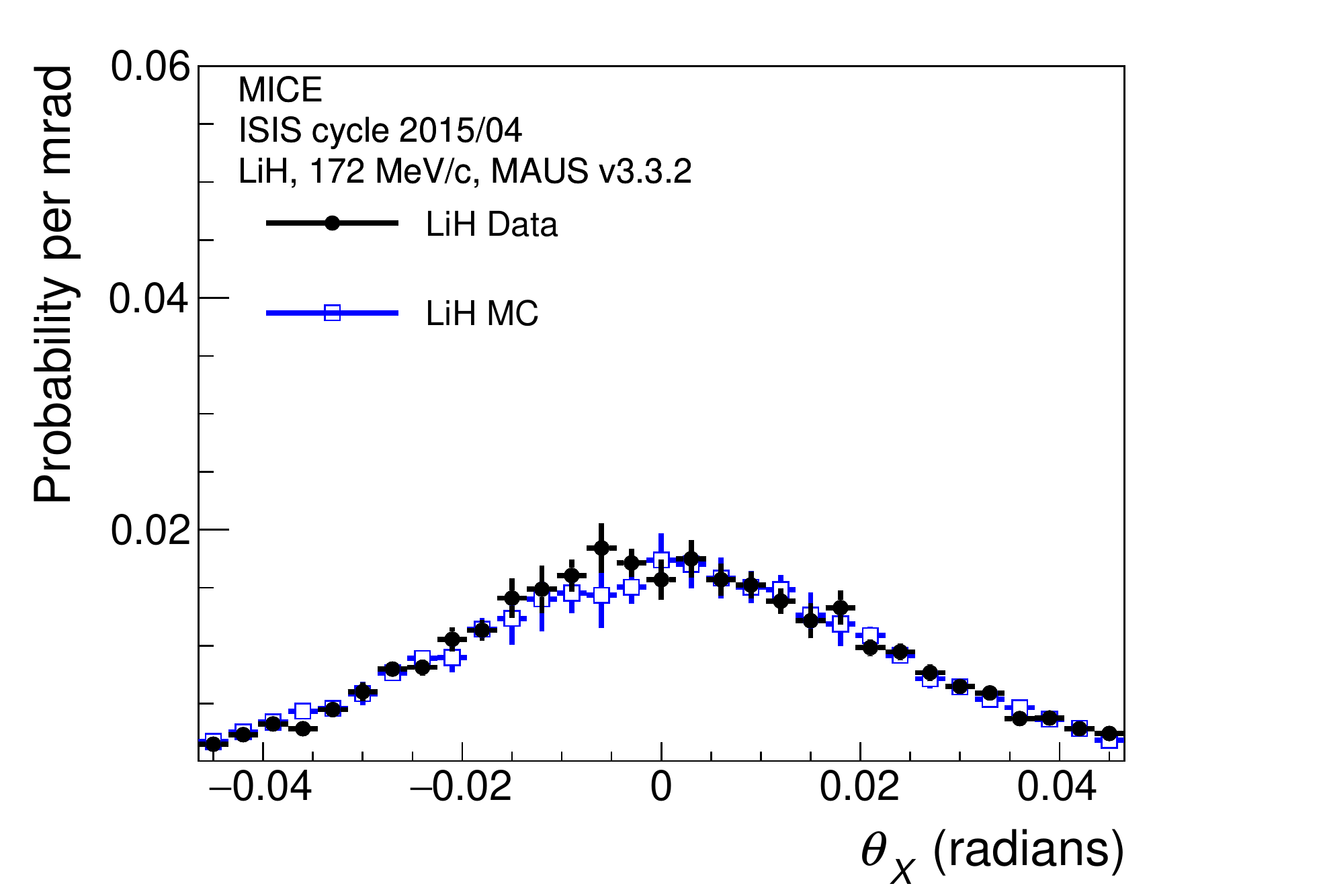}}
\subfigure{\label{fig:MCdata_172_compY}
\includegraphics[width=0.48\textwidth]{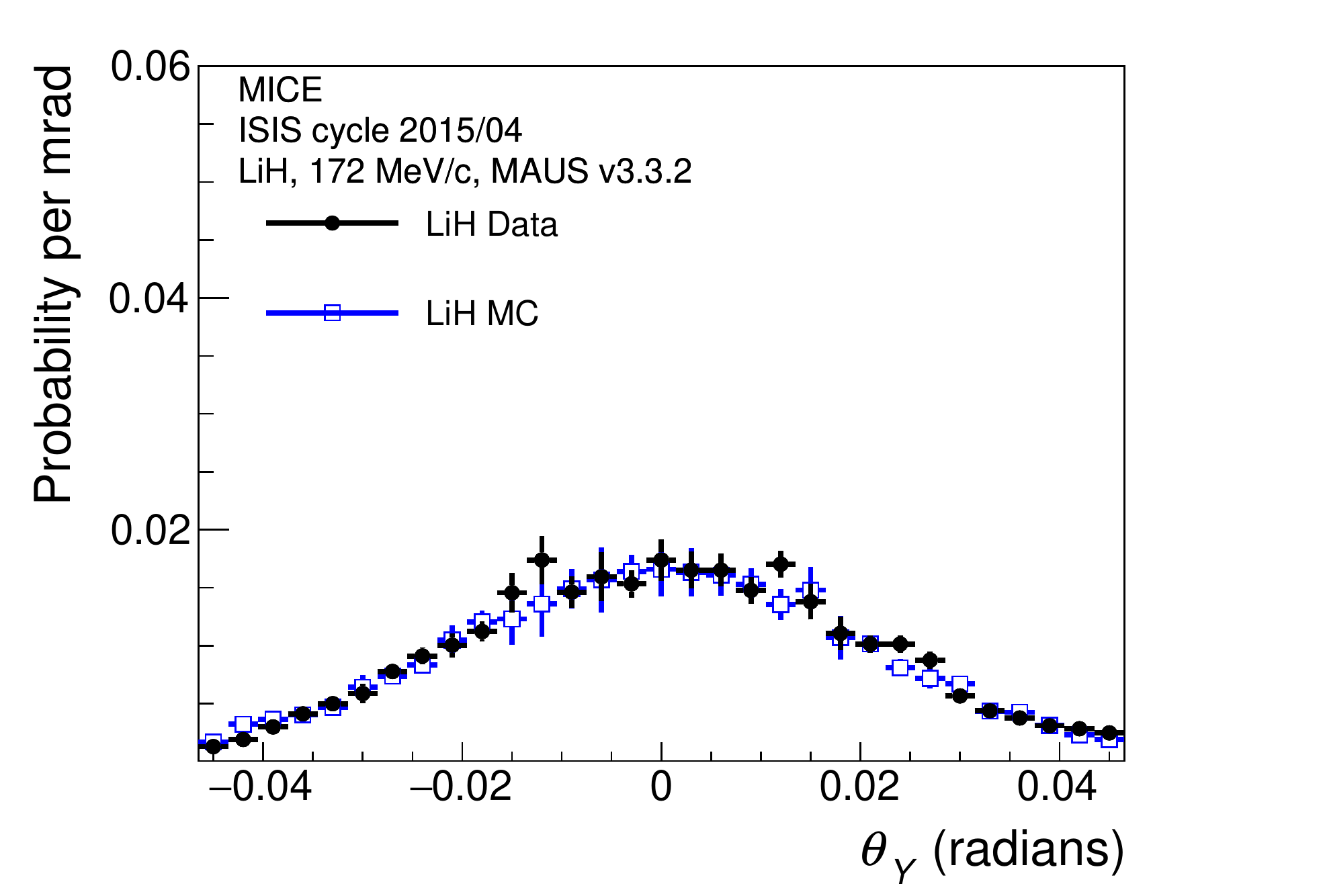}}
\subfigure{\label{fig:MCdata_172_compXref}
\includegraphics[width=0.48\textwidth]{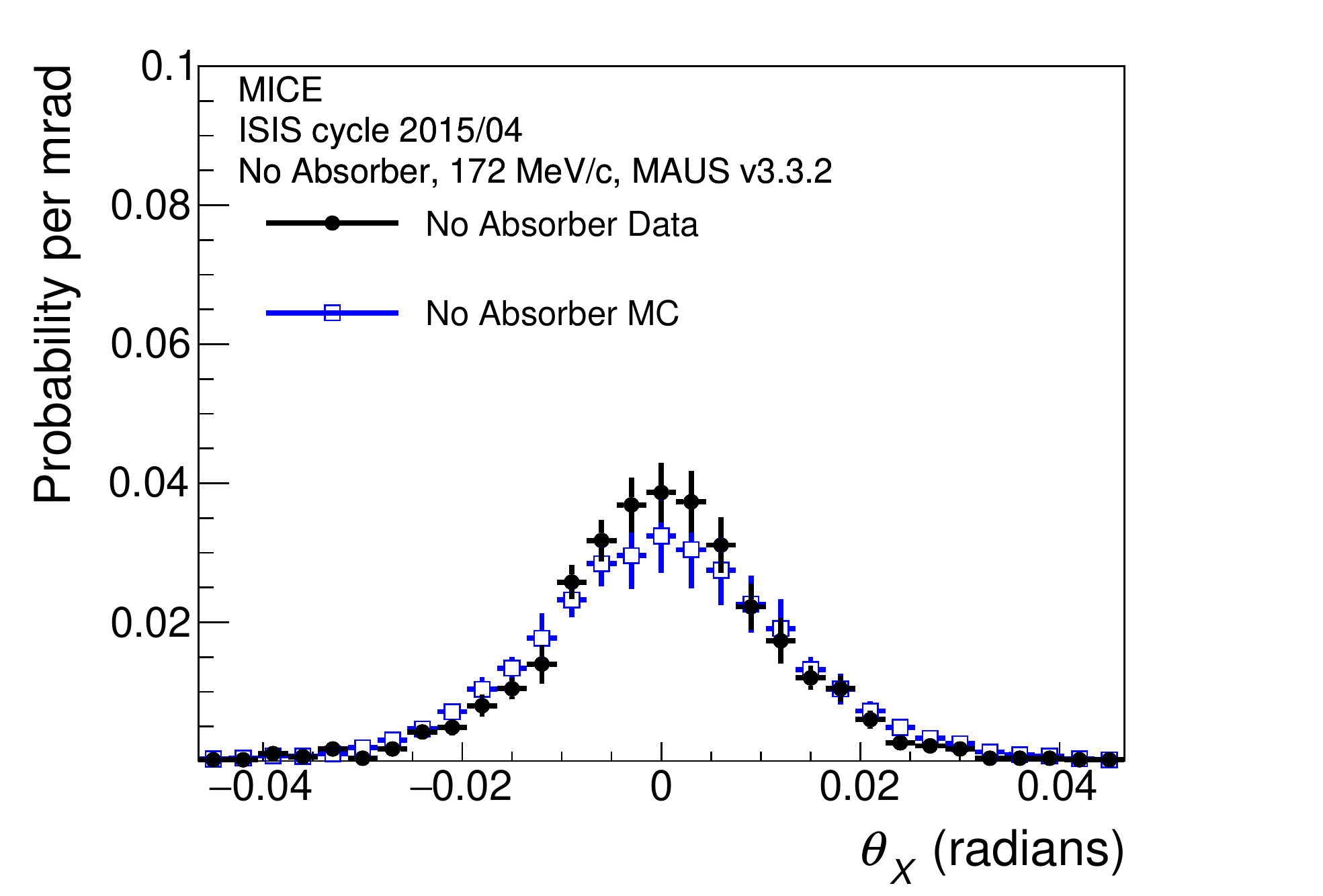}}
\subfigure{\label{fig:MCdata_172_compYref}
\includegraphics[width=0.48\textwidth]{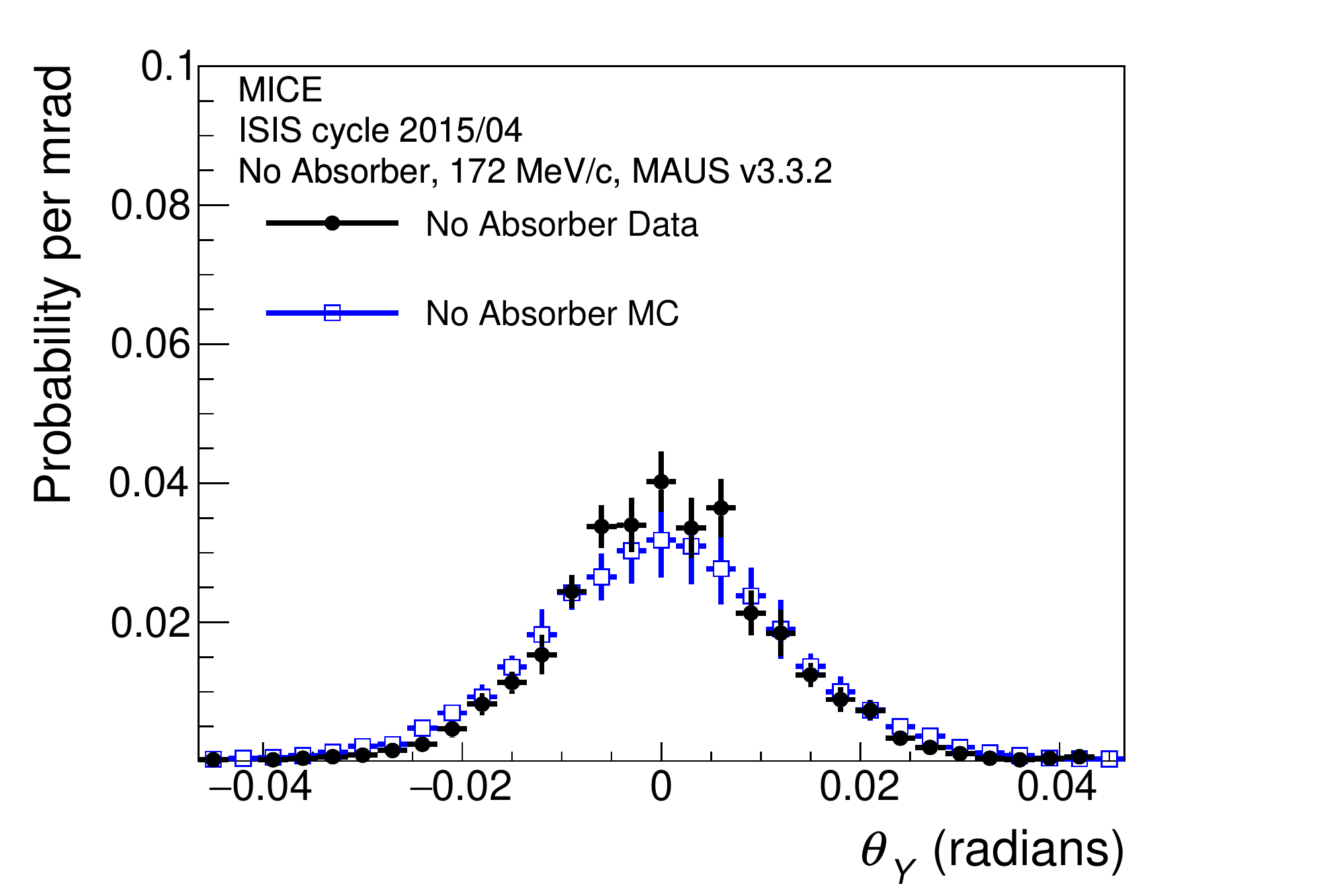}}
\caption{Scattering probability functions $\theta_X$ and $\theta_Y$ reconstructed from the 172\,MeV/$c$ muon beam with (top) and without (bottom) the LiH absorber in place compared to reconstructed MC scattering distributions. The black points are the real data and the blue open squares are the simulated data.}
\label{fig:MCdata_comp172}
\end{figure*}

\begin{figure*}
\centering
\subfigure{\label{fig:MCdata_200_compX}
\includegraphics[width=0.48\textwidth]{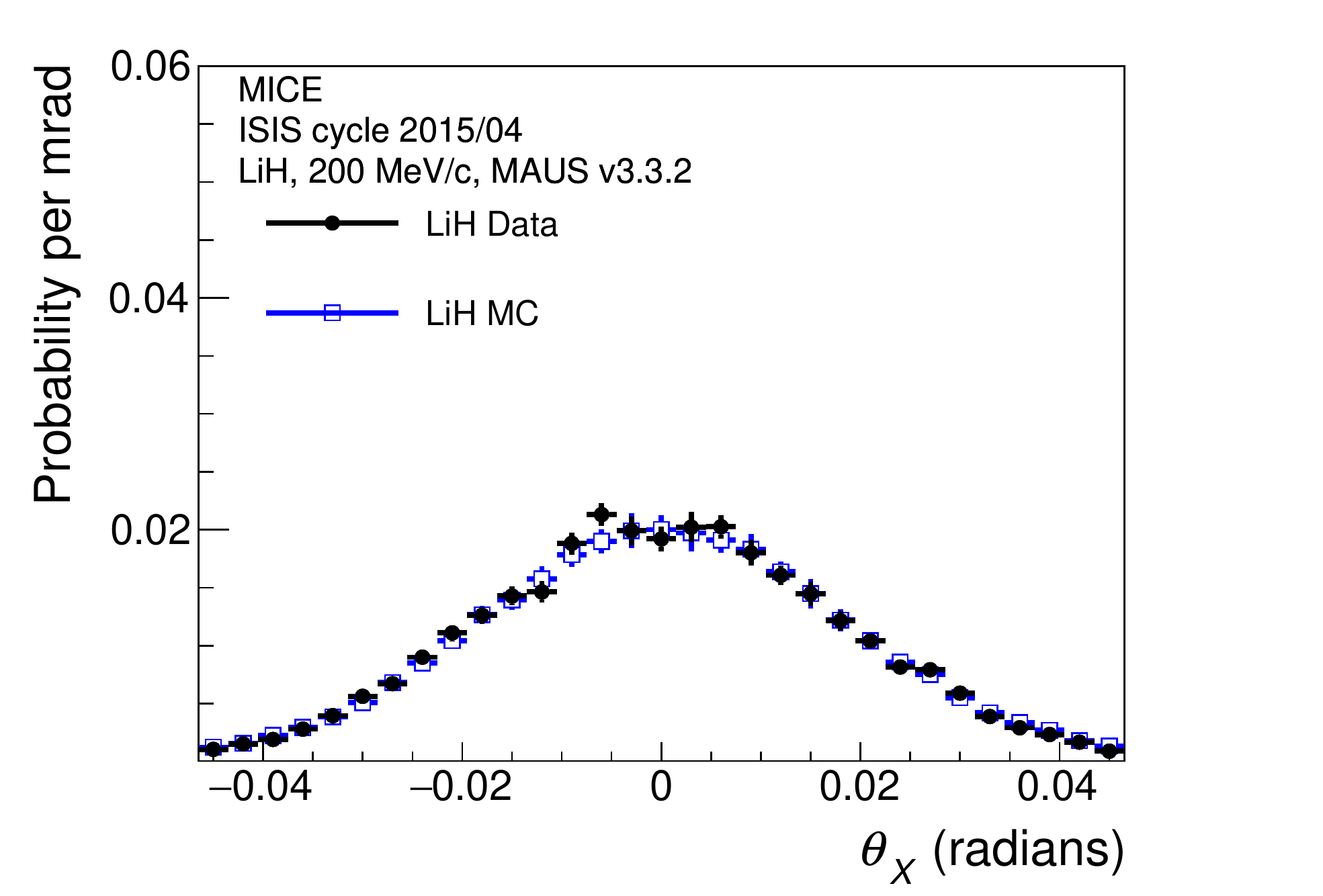}}
\subfigure{\label{fig:MCdata_200_compY}
\includegraphics[width=0.48\textwidth]{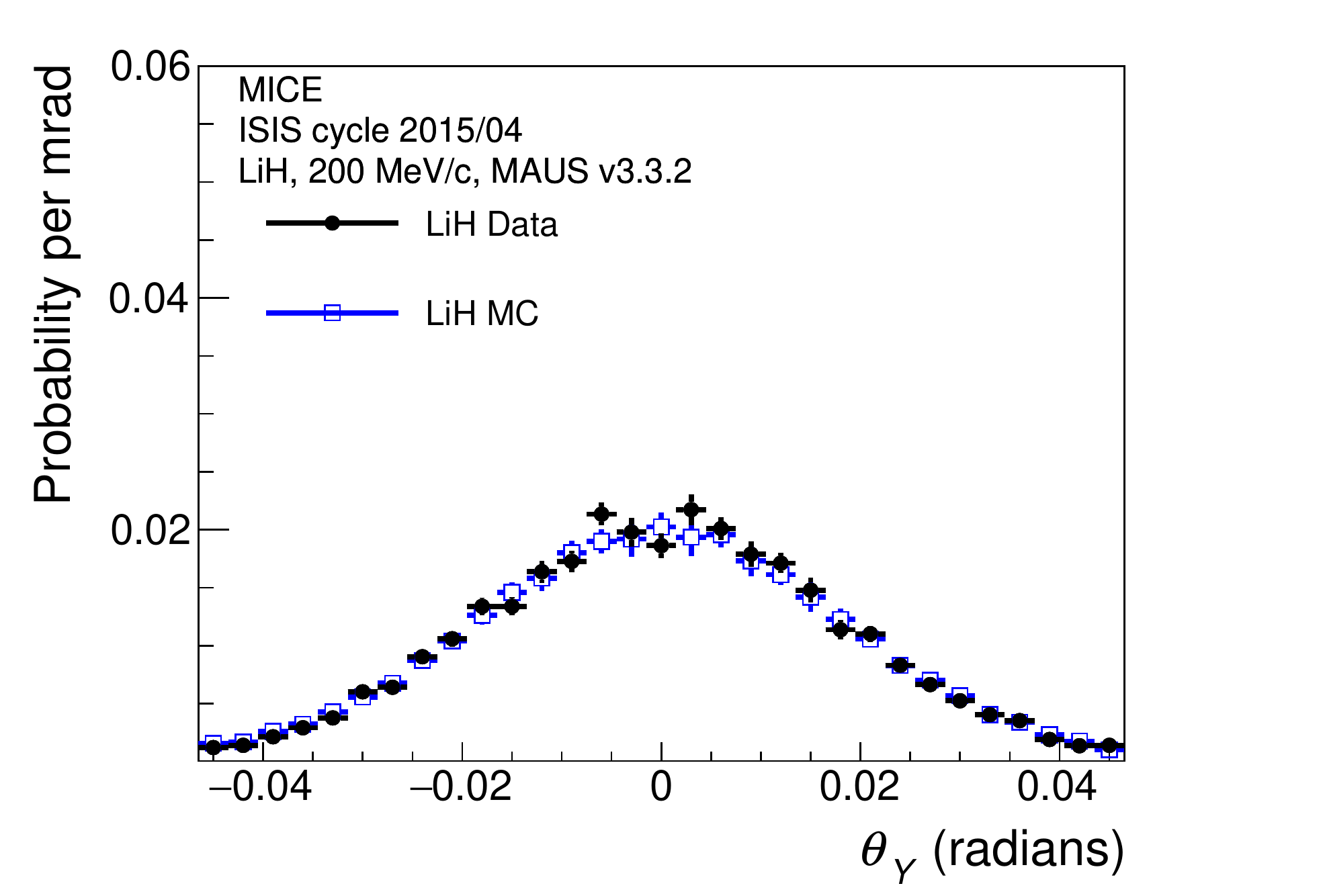}}
\subfigure{\label{fig:MCdata_200_compXref}
\includegraphics[width=0.48\textwidth]{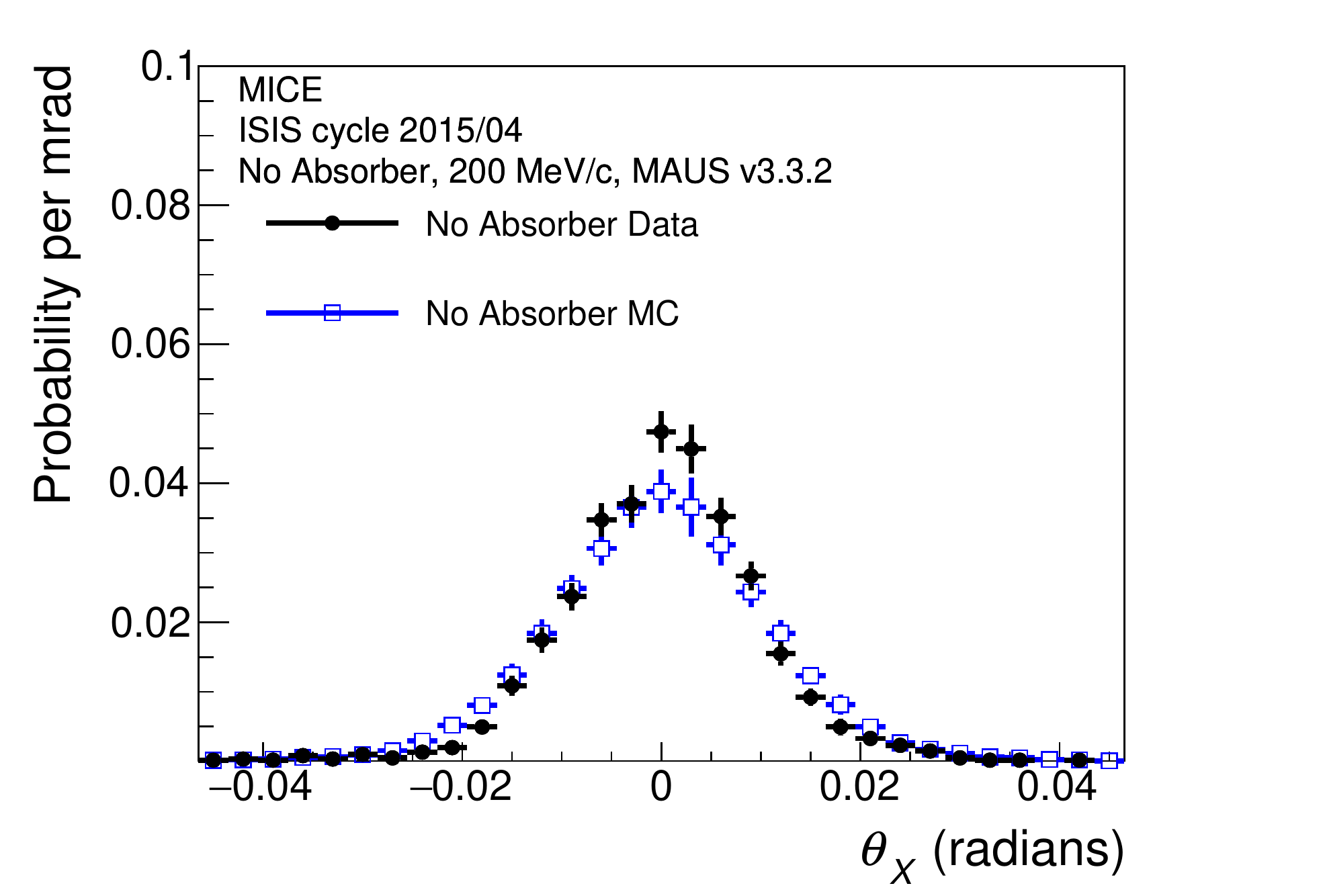}}
\subfigure{\label{fig:MCdata_200_compYref}
\includegraphics[width=0.48\textwidth]{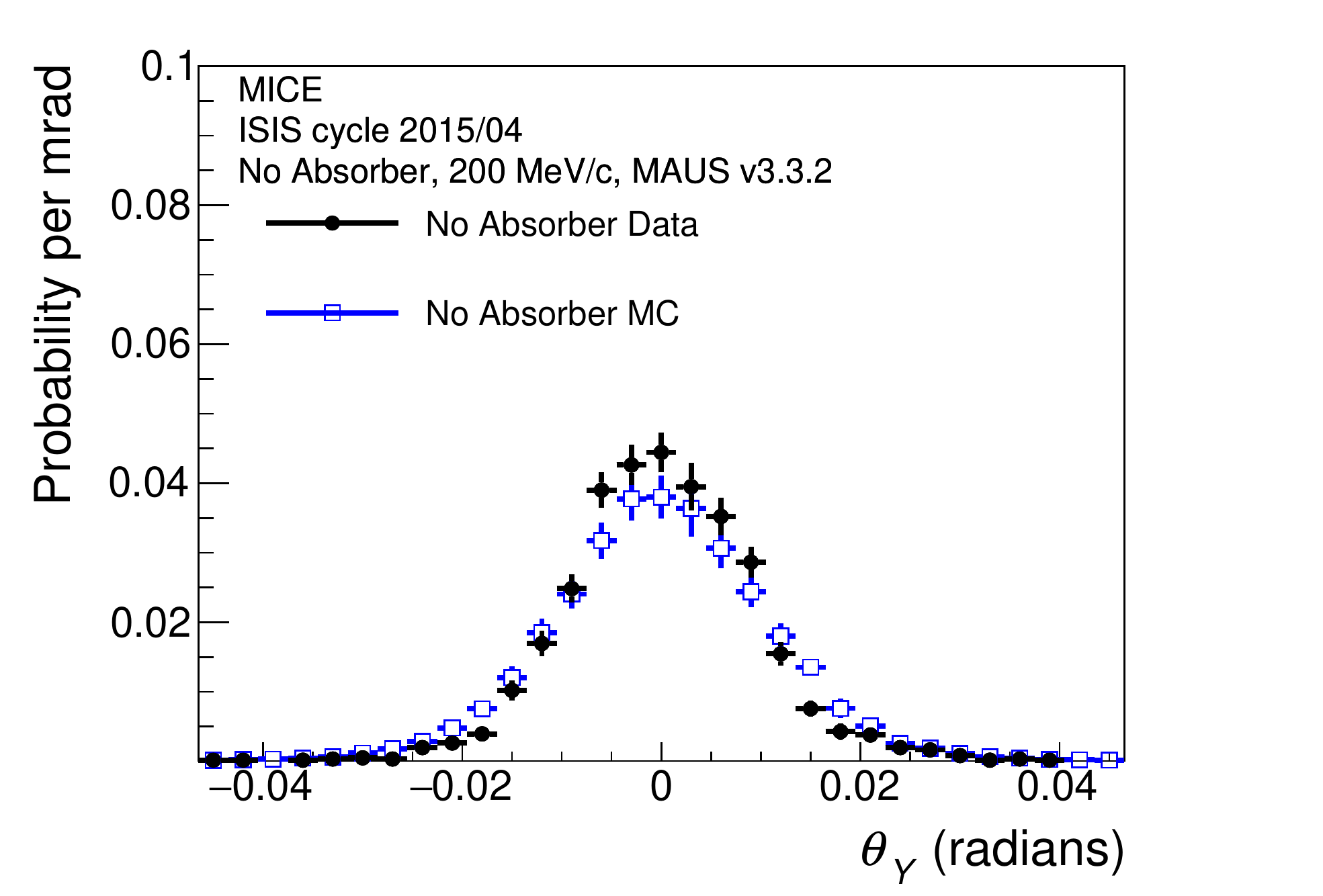}}
\caption{Scattering probability functions $\theta_X$ and $\theta_Y$ reconstructed from the 200\,MeV/$c$ muon beam with (top) and without (bottom) the LiH absorber in place compared to reconstructed MC scattering distributions. The black points are the real data and the blue open squares are the simulated data.}
\label{fig:MCdata_comp200}
\end{figure*}

\begin{figure*}
\centering
\subfigure{\label{fig:MCdata_240_compX}
\includegraphics[width=0.48\textwidth]{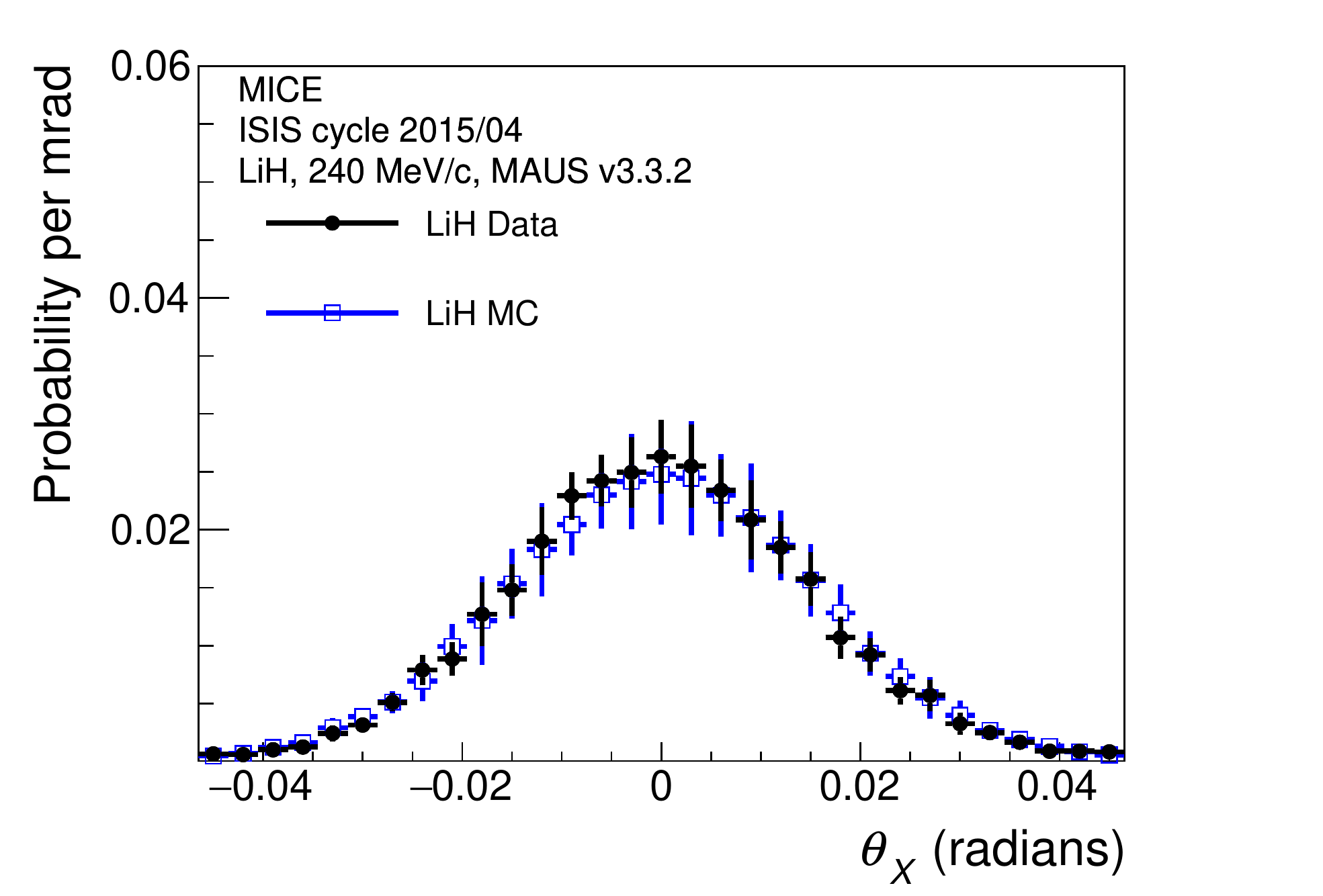}}
\subfigure{\label{fig:MCdata_240_compY}
\includegraphics[width=0.48\textwidth]{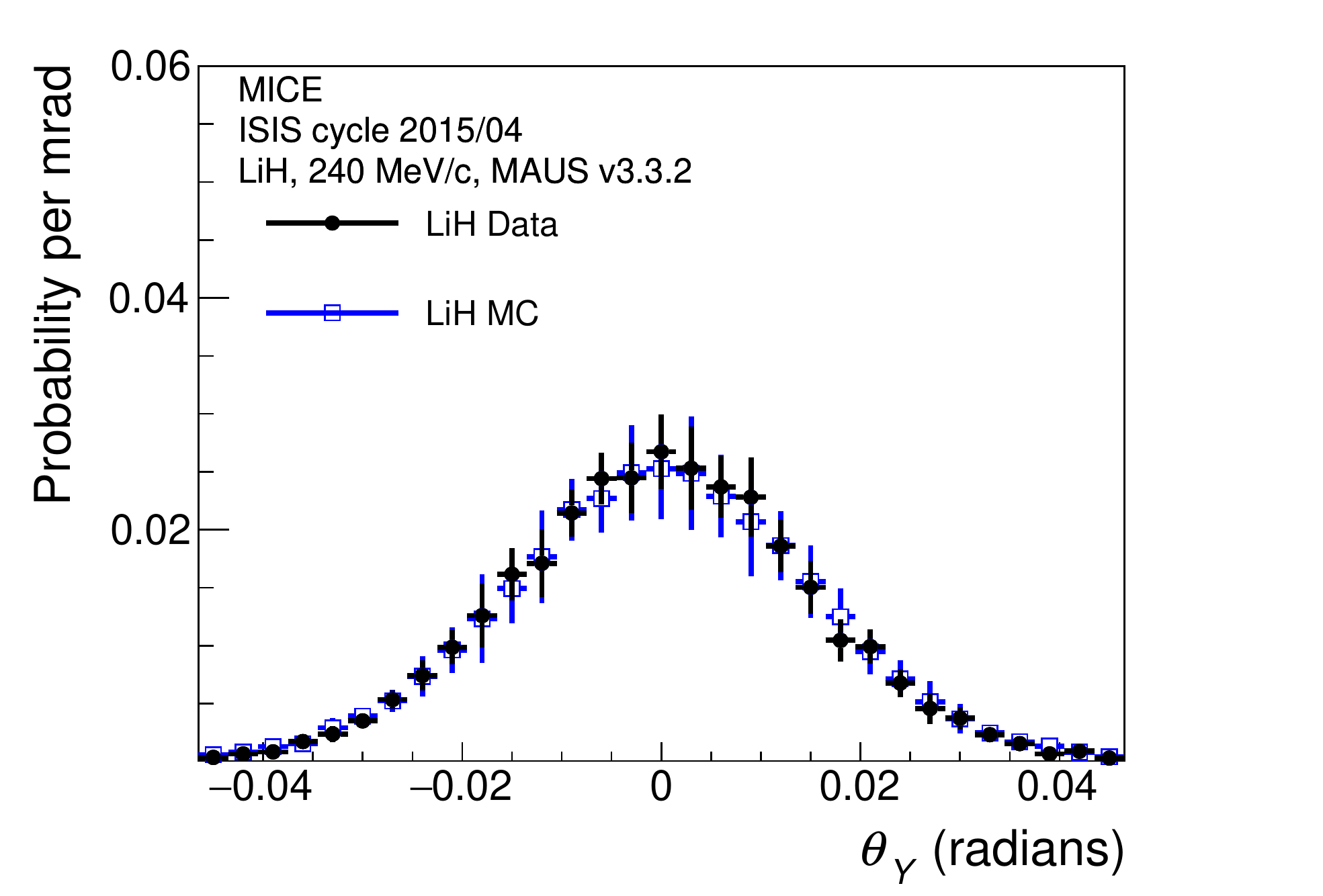}}
\subfigure{\label{fig:MCdata_240_compXref}
\includegraphics[width=0.48\textwidth]{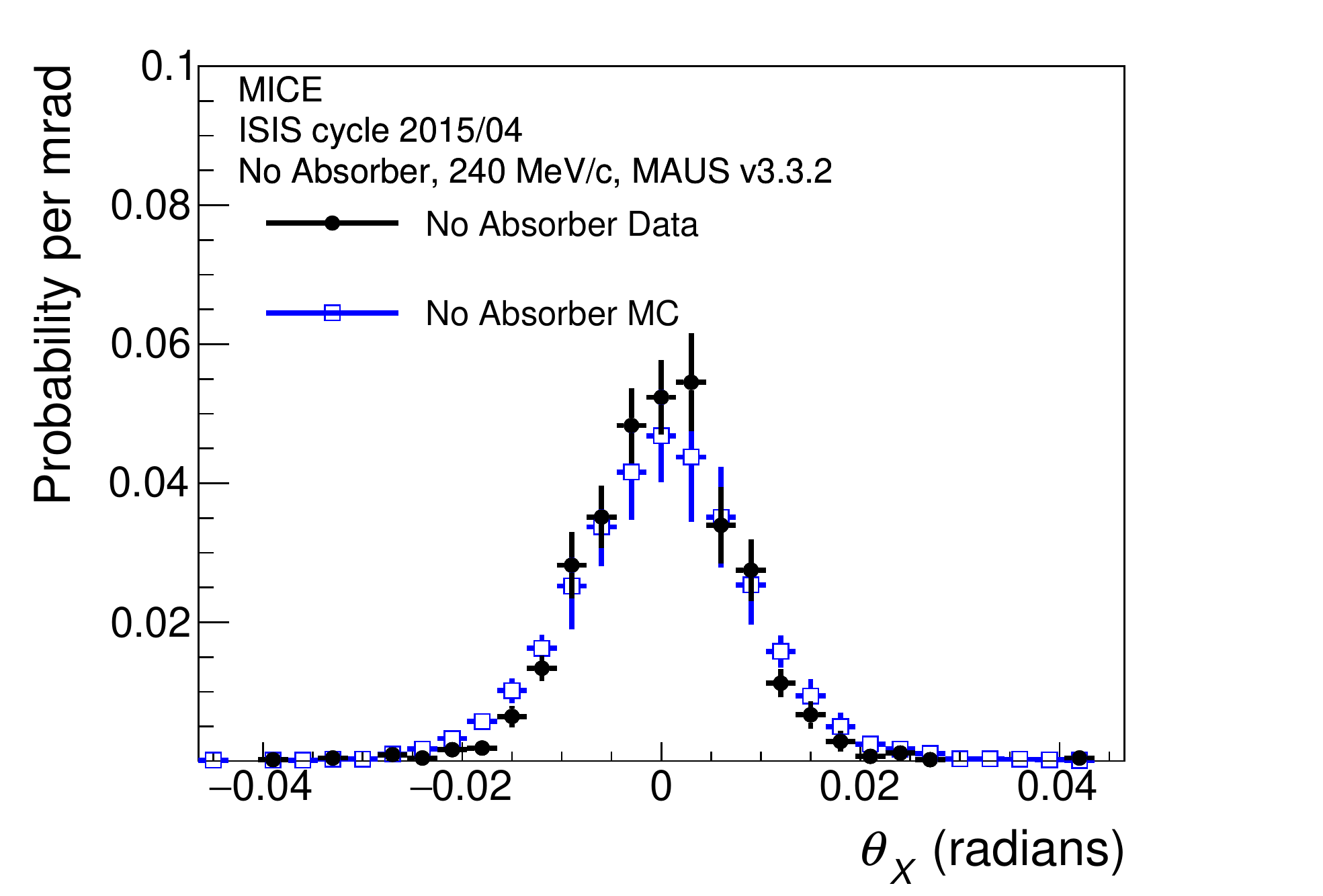}}
\subfigure{\label{fig:MCdata_240_compYref}
\includegraphics[width=0.48\textwidth]{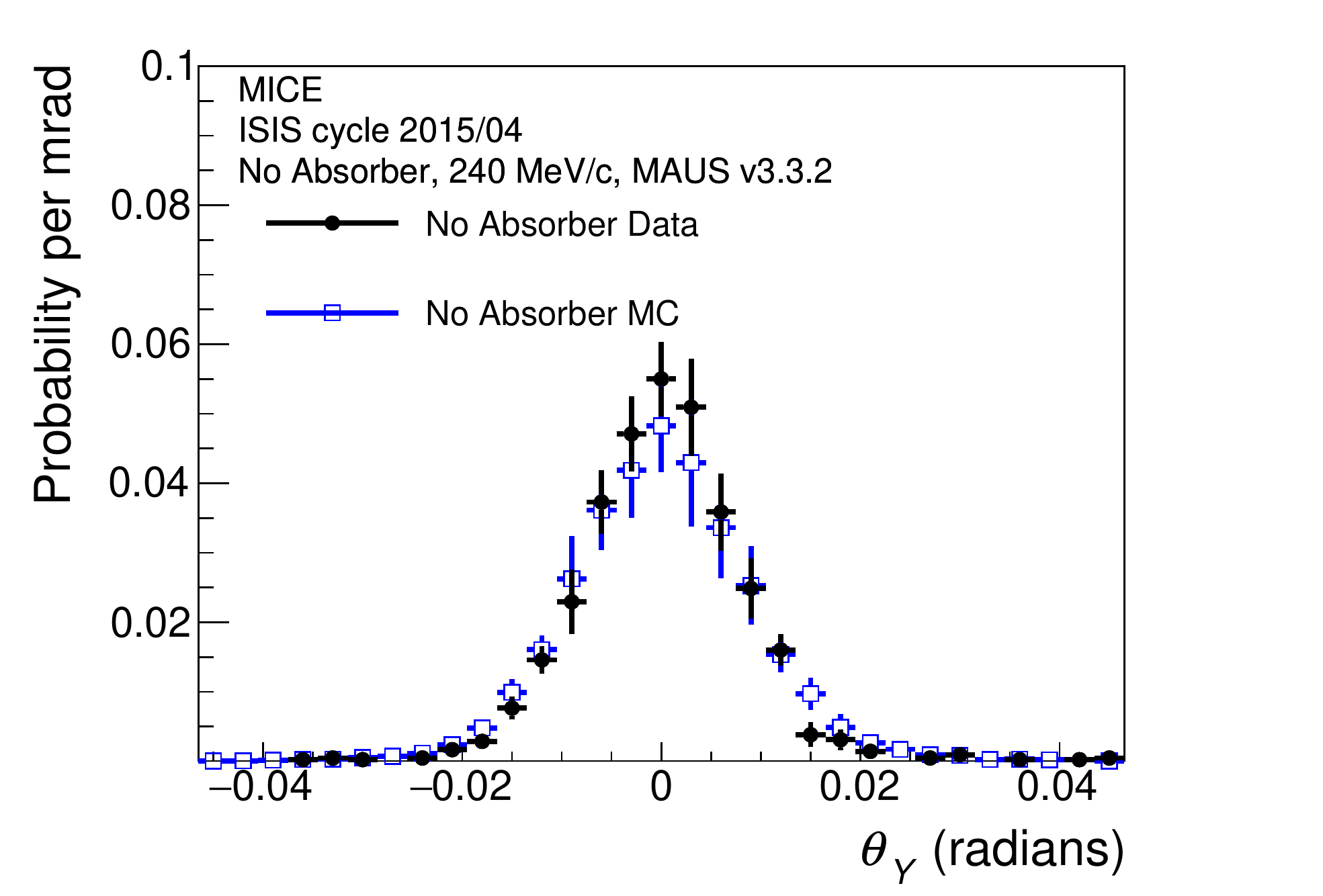}}
\caption{Scattering probability functions $\theta_X$ and $\theta_Y$  reconstructed from the 240\,MeV/$c$ muon beam with (top) and without (bottom) the LiH absorber in place compared to reconstructed MC scattering distributions. The black points are the real data and the blue open squares are the simulated data.}
\label{fig:MCdata_comp240}
\end{figure*}

\begin{figure*}
\centering
\subfigure{\label{fig:MCdata_172_scatt2}
\includegraphics[width=0.48\textwidth]{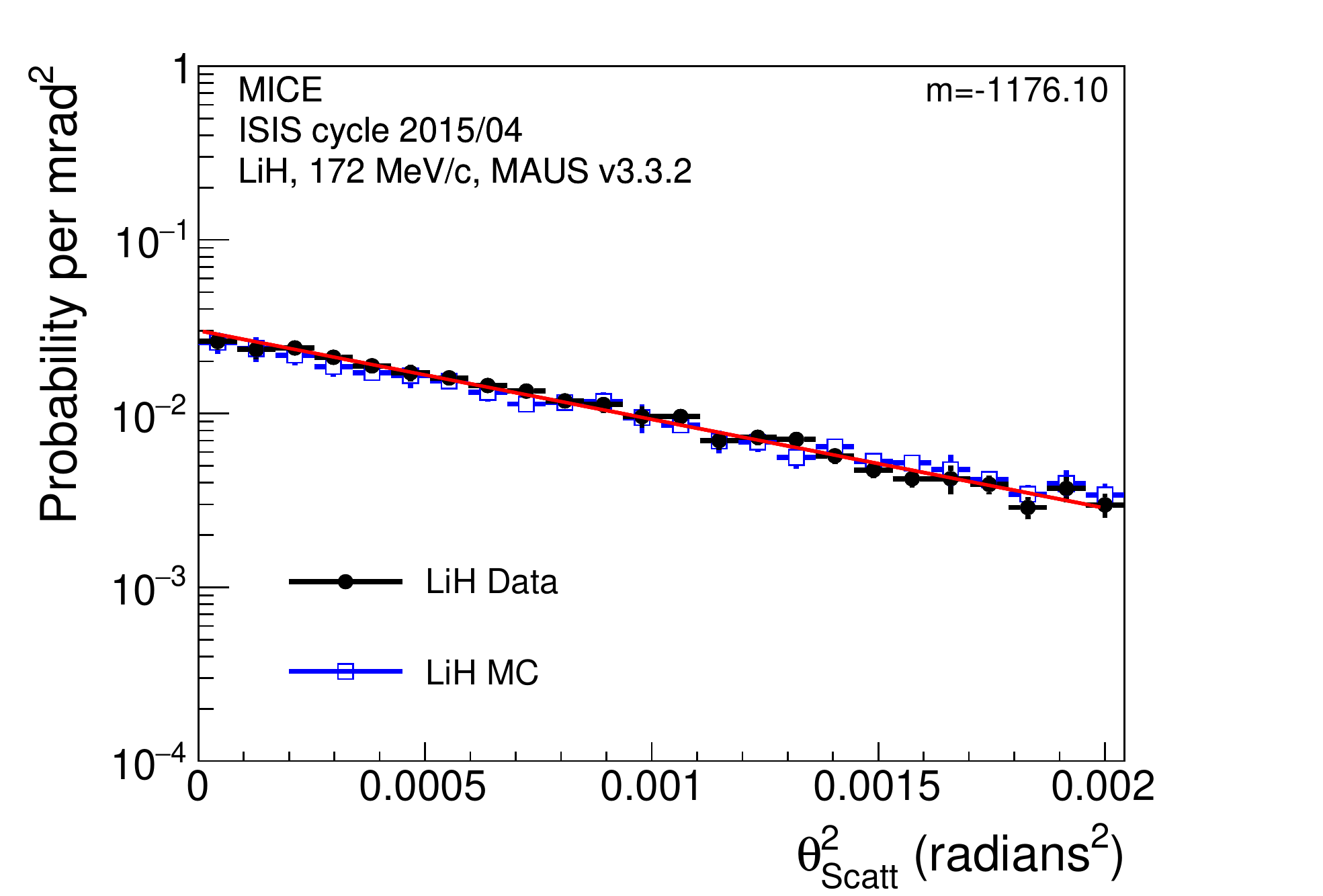}}
\subfigure{\label{fig:MCdata_200_scatt2}
\includegraphics[width=0.48\textwidth]{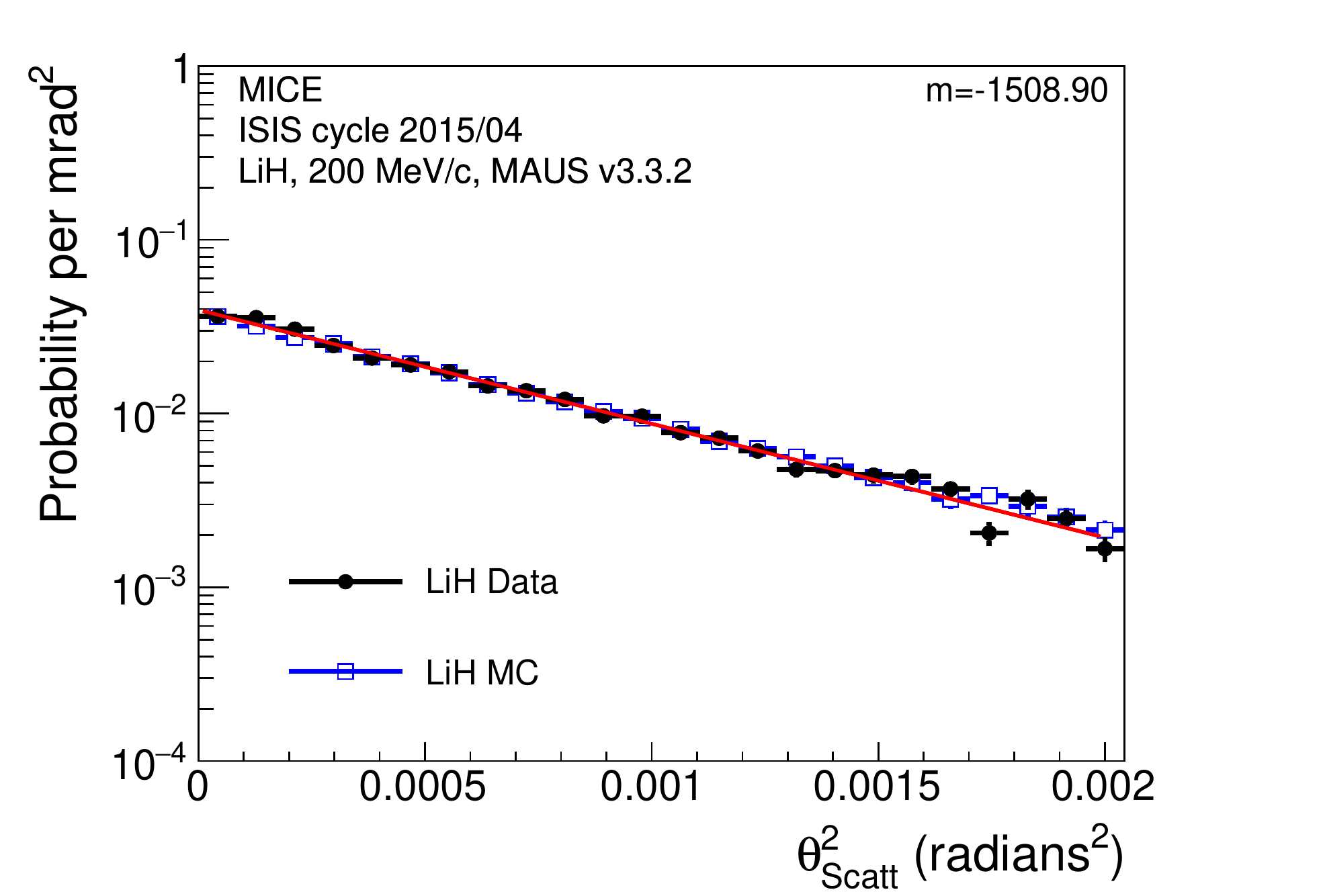}}
\subfigure{\label{fig:MCdata_240_scatt2}
\includegraphics[width=0.48\textwidth]{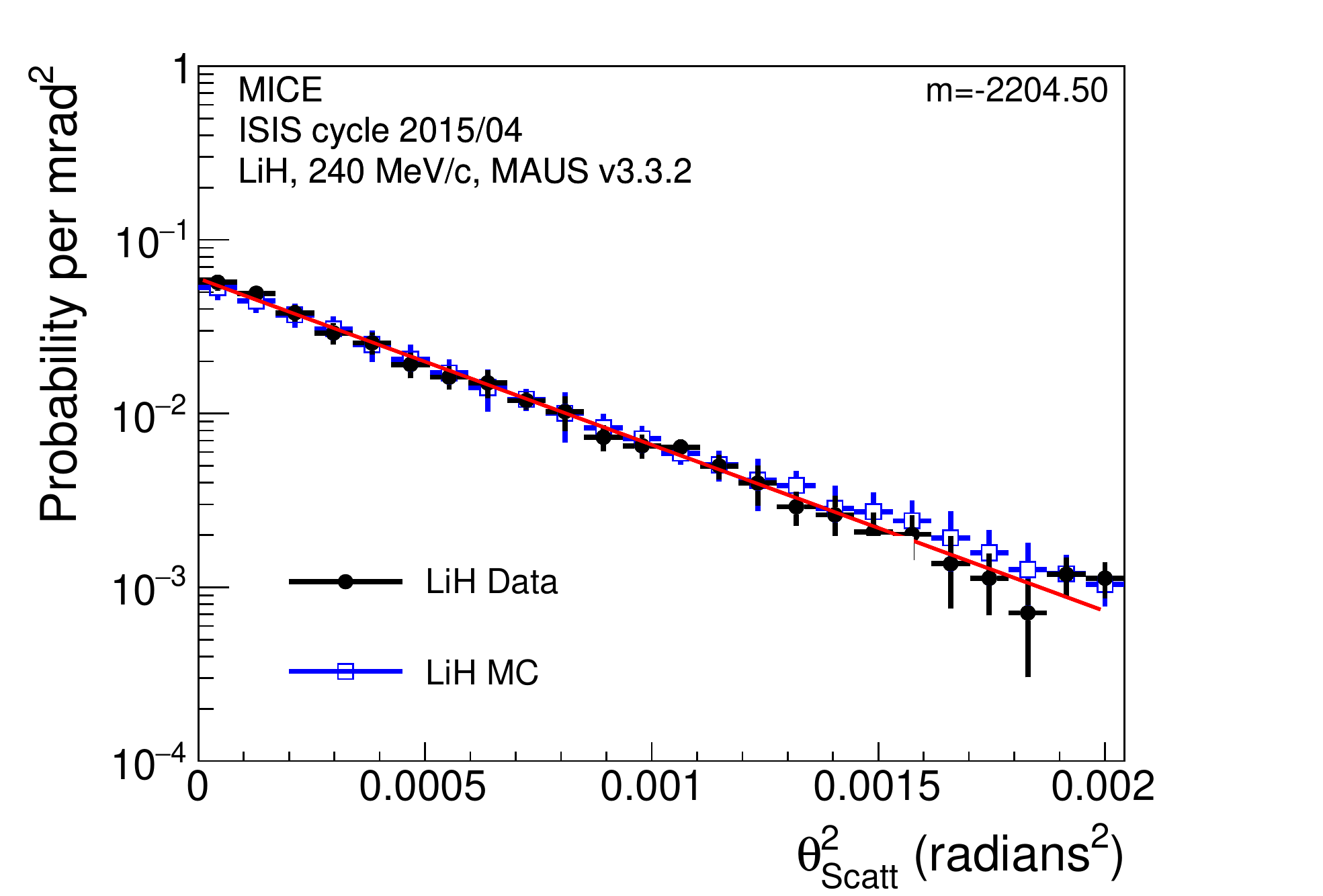}}
\caption{$\theta_{Scatt}^2$ distributions reconstructed from the 172, 200 and 240\,MeV/$c$ muon samples. The LiH absorber was in the beamline in these samples. The black points are the real data and the blue open squares are the reconstructed simulated data. A function $Ae^{-m\theta_{Scatt}^{2}}$ was fitted to the data distribution and is shown by the red line with $m$ displayed on the plot.}
\label{fig:MCdata_comp2Scatt}
\end{figure*}

The $\theta_X$ and $\theta_Y$ distributions from the LiH and No Absorber data are compared to GEANT4 (v9.6) simulations in Figs. \ref{fig:MCdata_comp172}, \ref{fig:MCdata_comp200} and \ref{fig:MCdata_comp240} and the $\theta^2_{Scatt}$ distribution in Fig. \ref{fig:MCdata_comp2Scatt}, at three momenta: 172, 200 and 240\,MeV/$c$. The simulation gives an adequate description of the data; a summary of the comparison given in Table \ref{tab:dataMC}. The integrals of these distributions are between 88\% and 96\% demonstrating that the selection criteria ensure high transmission for the selected sample. 
In this analysis GEANT4 (v9.6) is used with the QGSP\_BERT (v4.0) physics list. In this configuration, multiple Coulomb scattering is modelled by the G4WentzelVI model \cite{Allison:2016lfl, Ivanchenko_2010}. The G4WentzelVI model is a mixed algorithm simulating both the hard collisions one by one and using a multiple scattering theory to treat the effects of the soft collisions at the end of a given step; this prevents the number of steps in the simulation from becoming too large and also reduces the dependence on the step length. This model is expected to provide results similar in accuracy to single scattering but in a computationally efficient manner. Single scattering is based on the assumption that the effect of multiple scattering can be modelled as if the hard scatters are the sum of many individual scatters while soft scatters are sampled from a distribution. `Hard' scatters are inelastic and result in large-angle deflections and large energy transfers. `Soft' scatters are elastic and result in small-angle deflections with small energy transfers.

\subsection{Convolution with Scattering Models}
\label{subsec:convolution}

\begin{figure*}
\begin{center}
\subfigure{\label{fig:raw_compX_172}\includegraphics[width=0.48\textwidth]{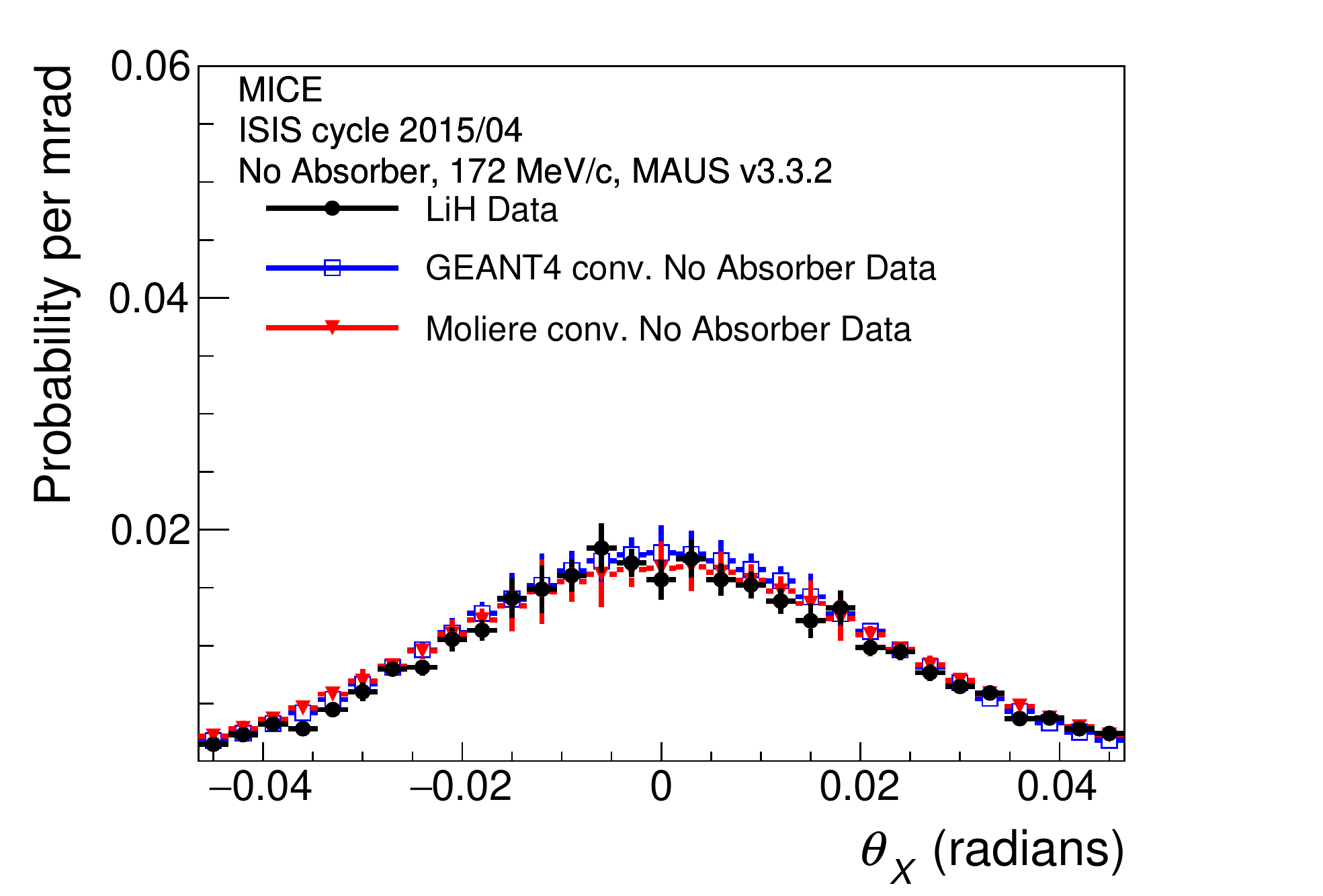}}\hspace{3mm}
\subfigure{\label{fig:raw_compY_172}\includegraphics[width=0.48\textwidth]{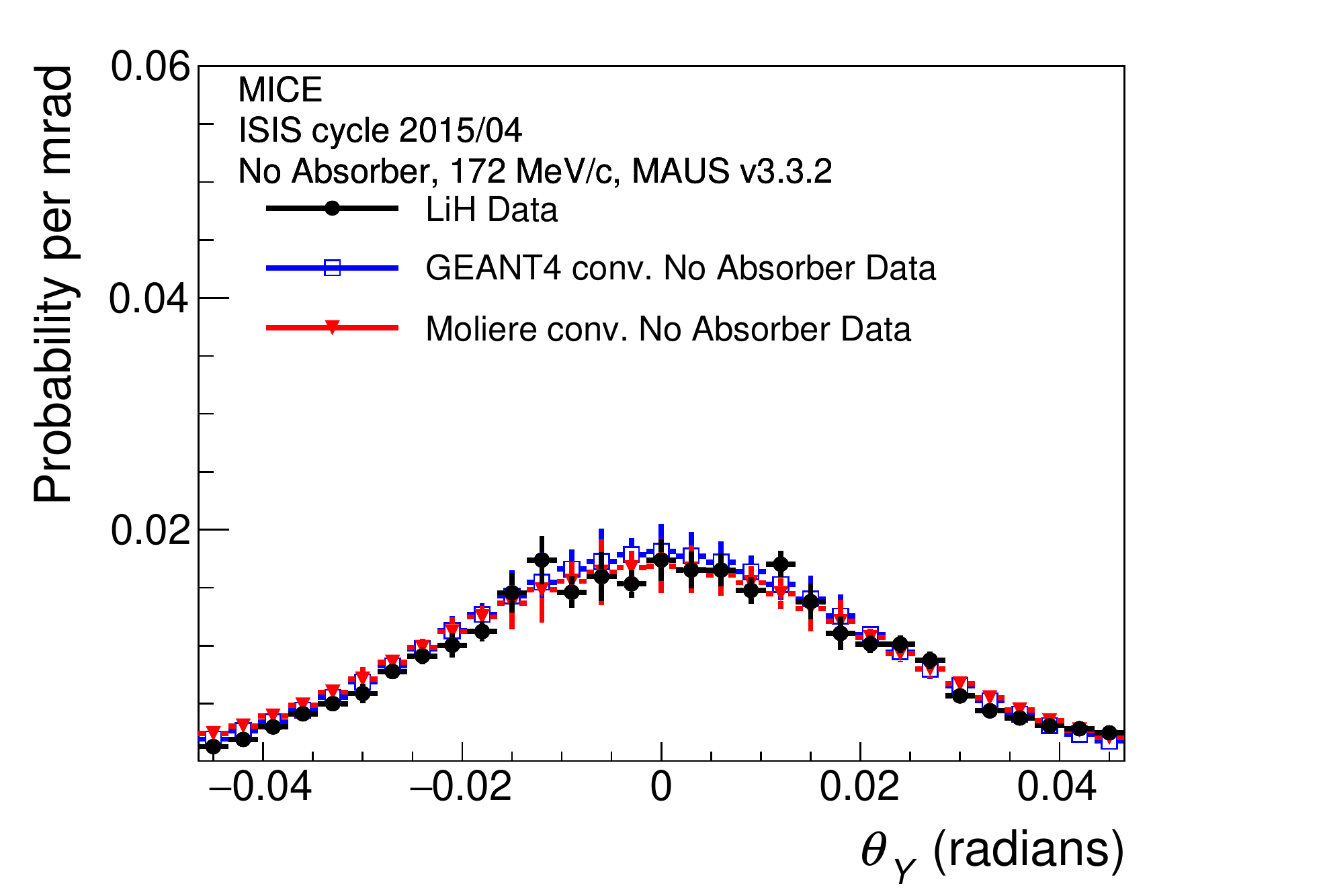}}
\subfigure{\label{fig:raw_compX_200}\includegraphics[width=0.48\textwidth]{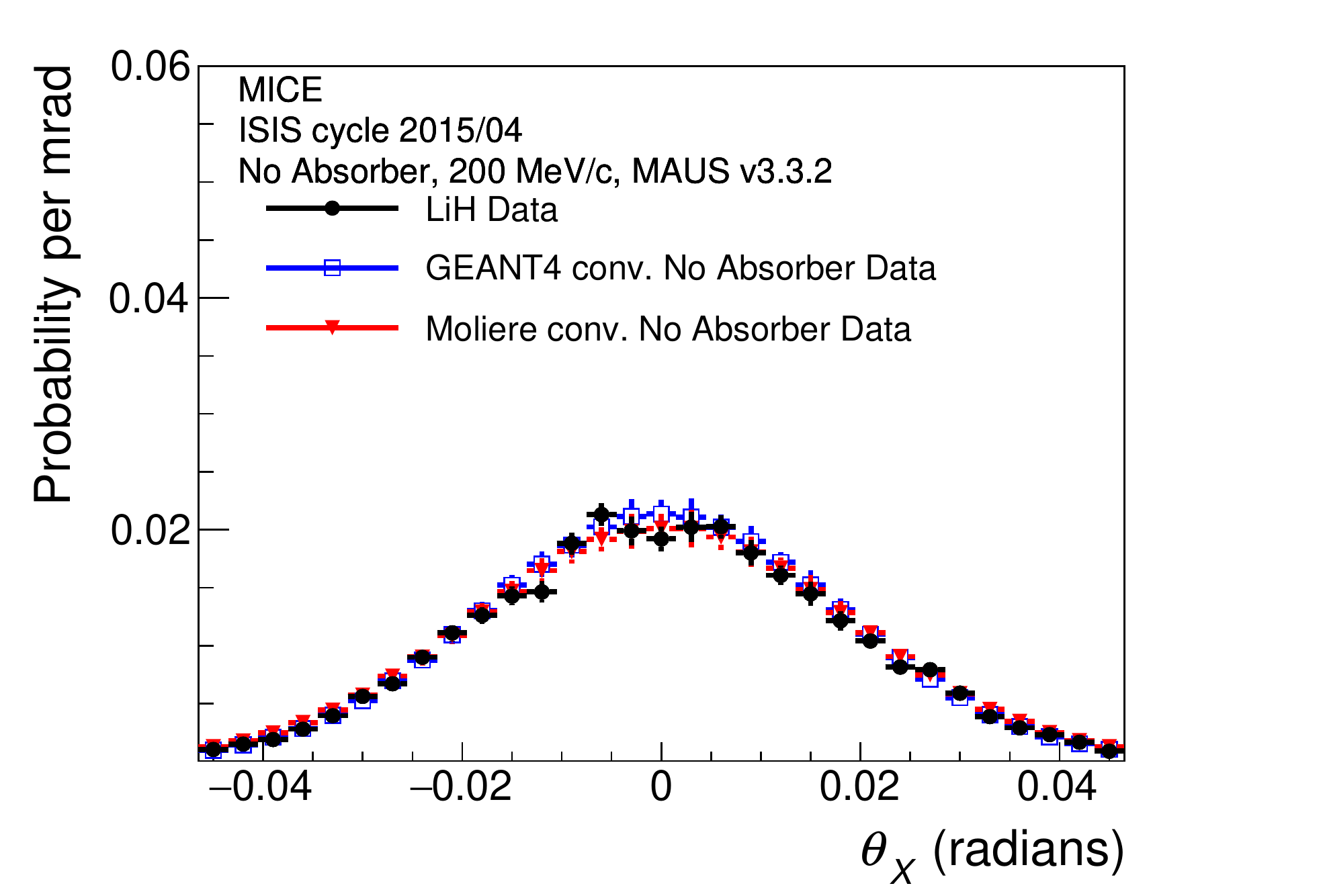}}\hspace{3mm}
\subfigure{\label{fig:raw_compY_200}\includegraphics[width=0.48\textwidth]{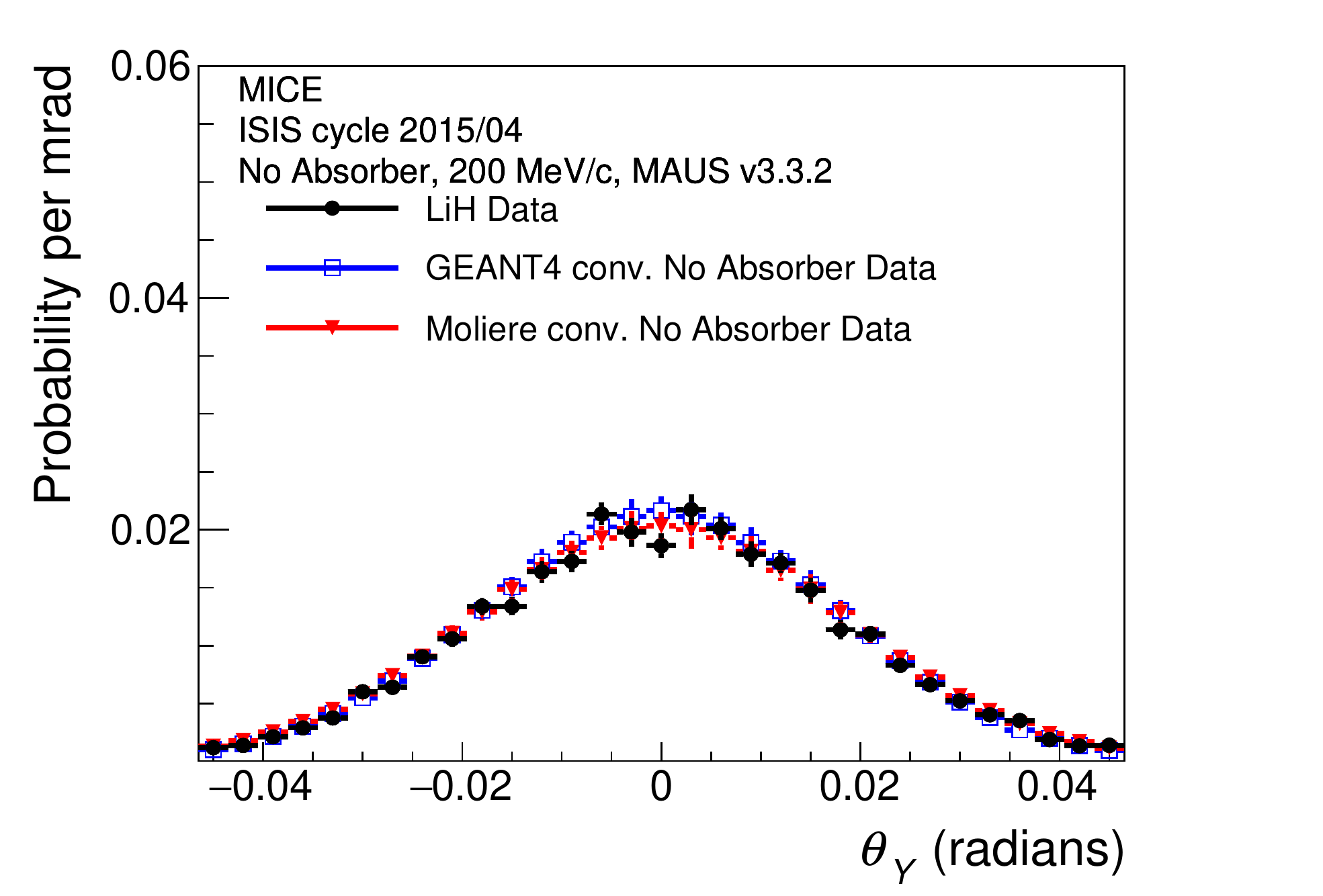}}
\subfigure{\label{fig:raw_compX_240}\includegraphics[width=0.48\textwidth]{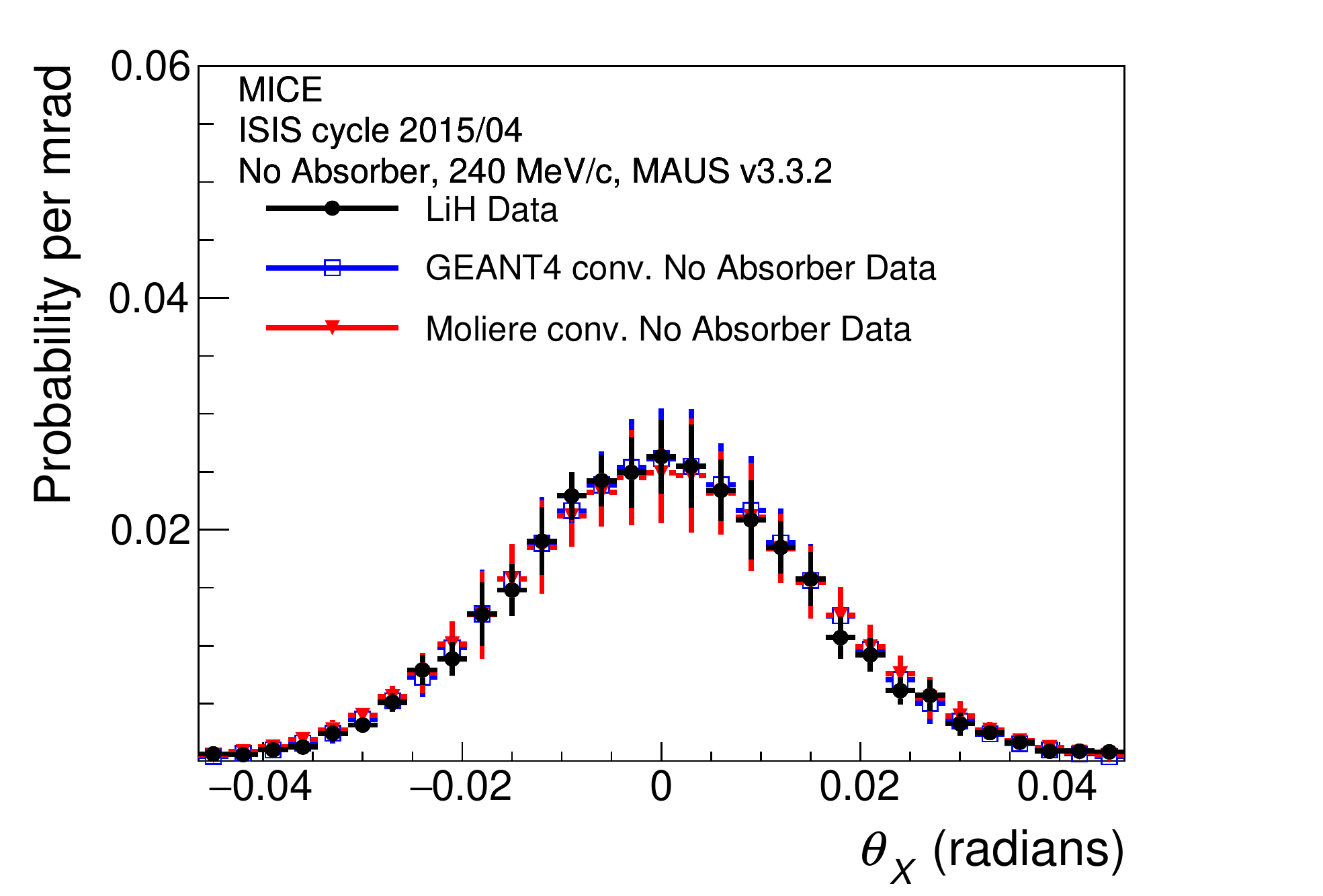}}\hspace{3mm}
\subfigure{\label{fig:raw_compY_240}\includegraphics[width=0.48\textwidth]{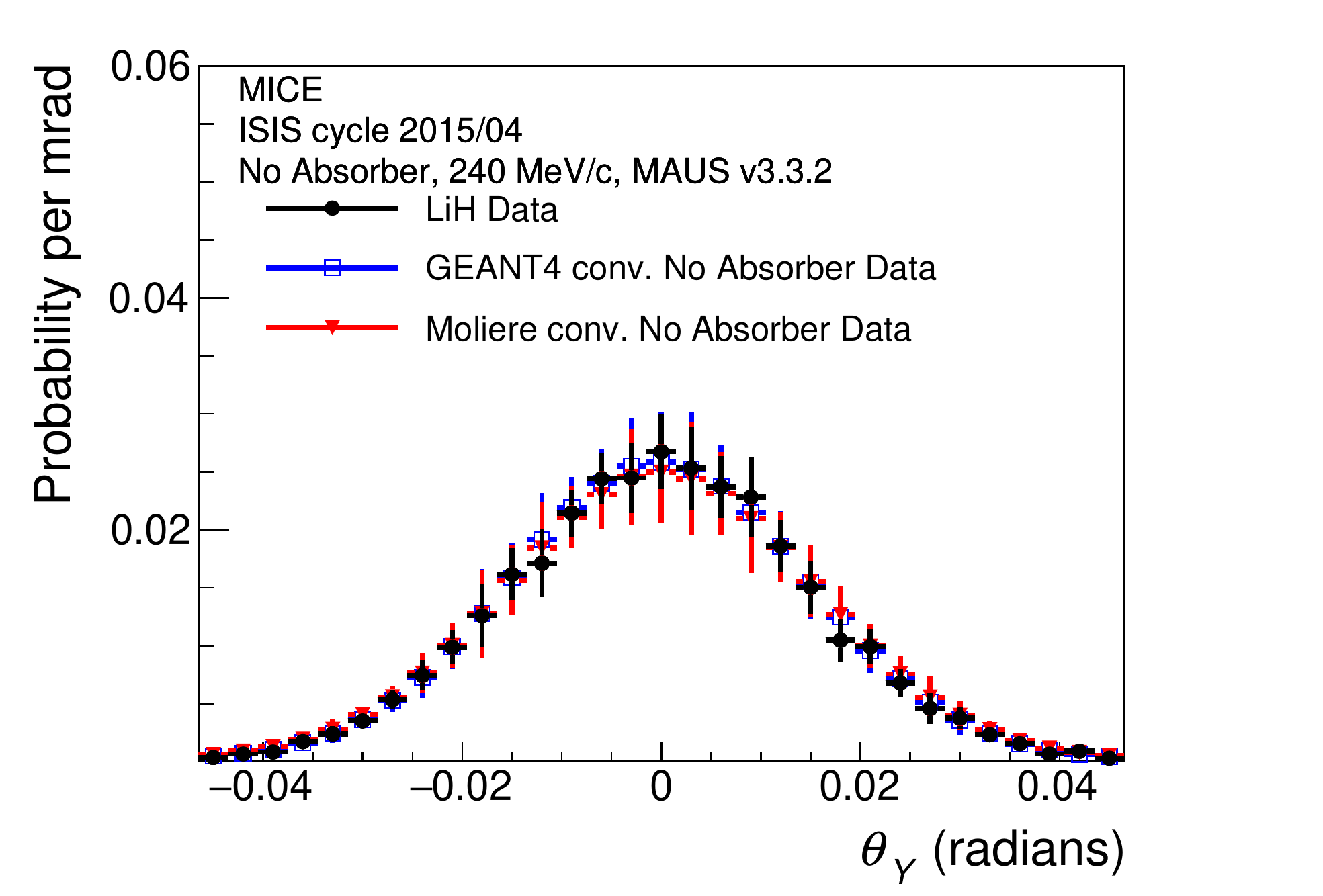}}
\caption{Scattering probability functions reconstructed from the 172, 200 and 240\,MeV/$c$ muon beams with the LiH absorber in place  (black dots) compared to the GEANT4 scattering model (blue dots) and the Moli\`ere model (red dots) in LiH convolved with the No Absorber distribution.}
\label{fig:raw_comp}
\end{center}
\end{figure*}

\begin{figure*} 
\begin{center}
\subfigure{\label{fig:con_resX_172}\includegraphics[width=0.48\textwidth]{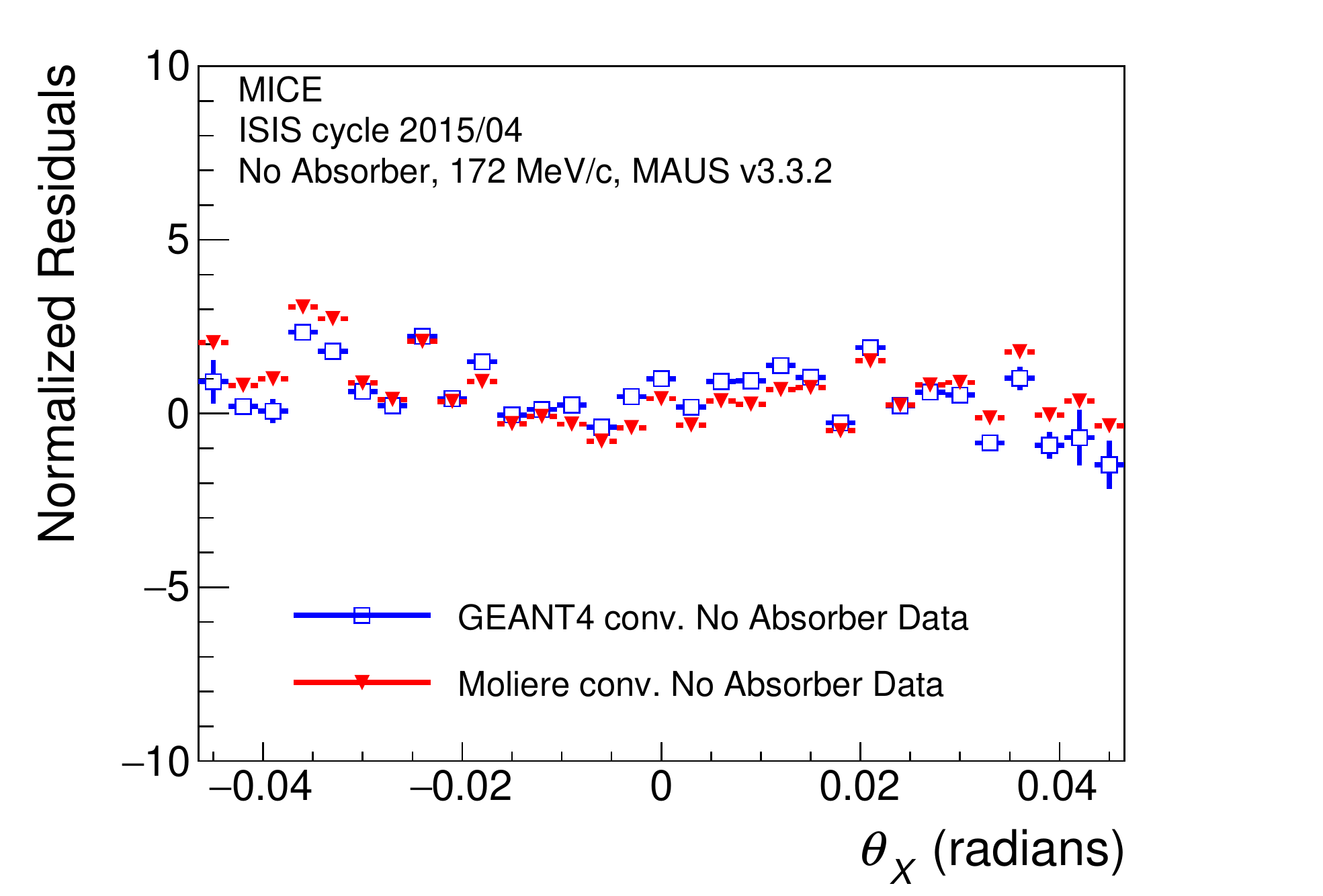}}
\subfigure{\label{fig:con_resY_172}\includegraphics[width=0.48\textwidth]{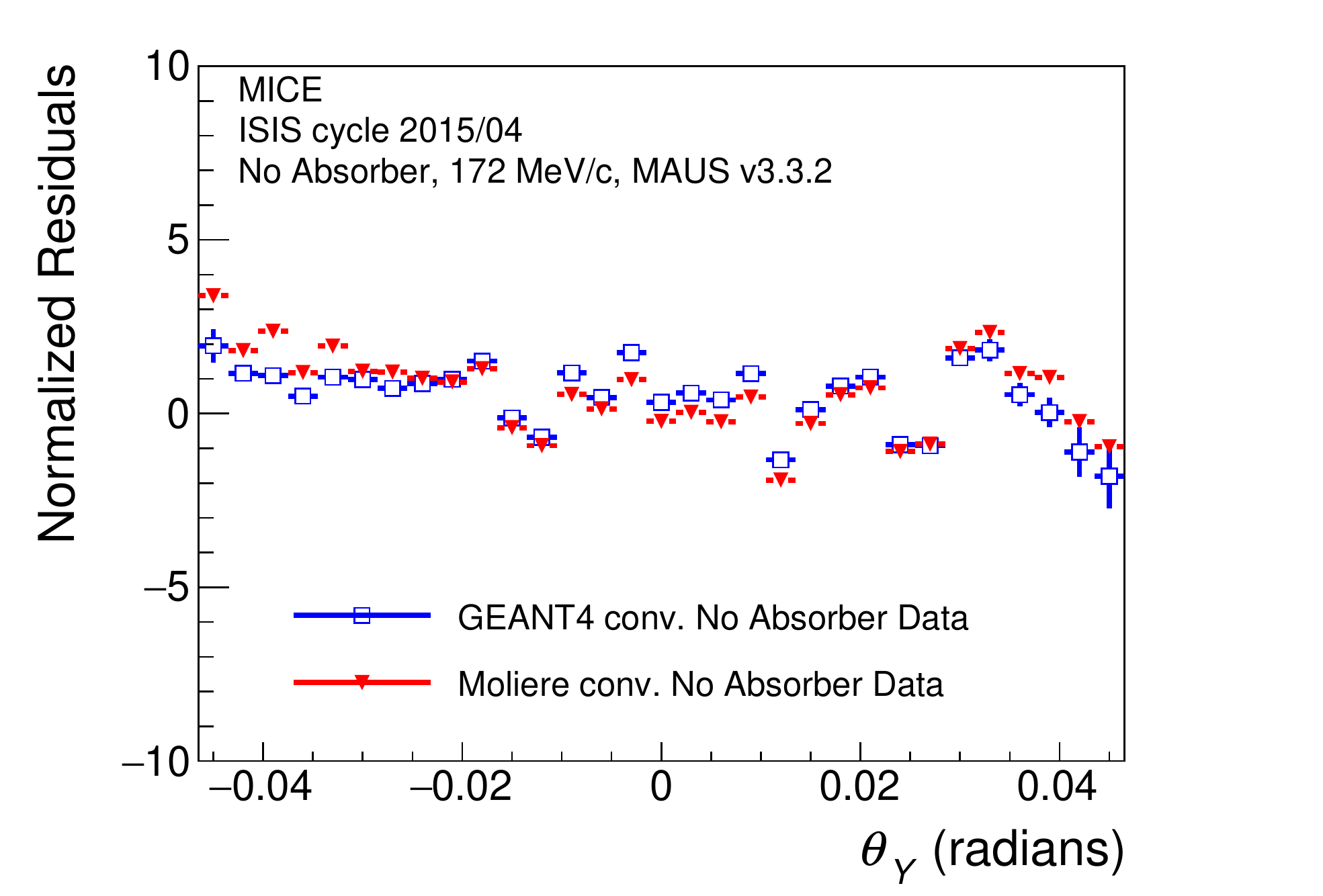}}
\subfigure{\label{fig:con_resX_200}\includegraphics[width=0.48\textwidth]{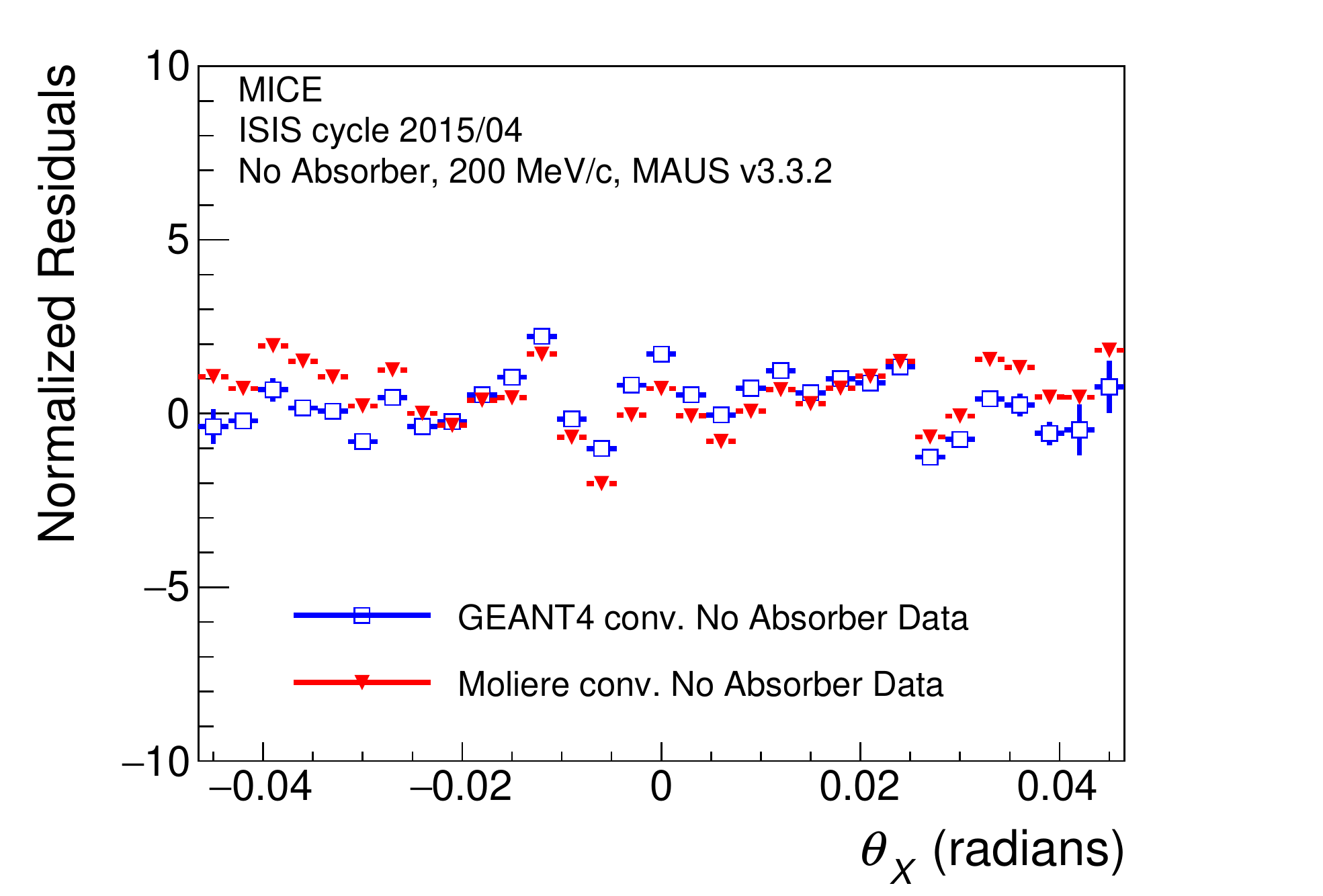}}
\subfigure{\label{fig:con_resY_200}\includegraphics[width=0.48\textwidth]{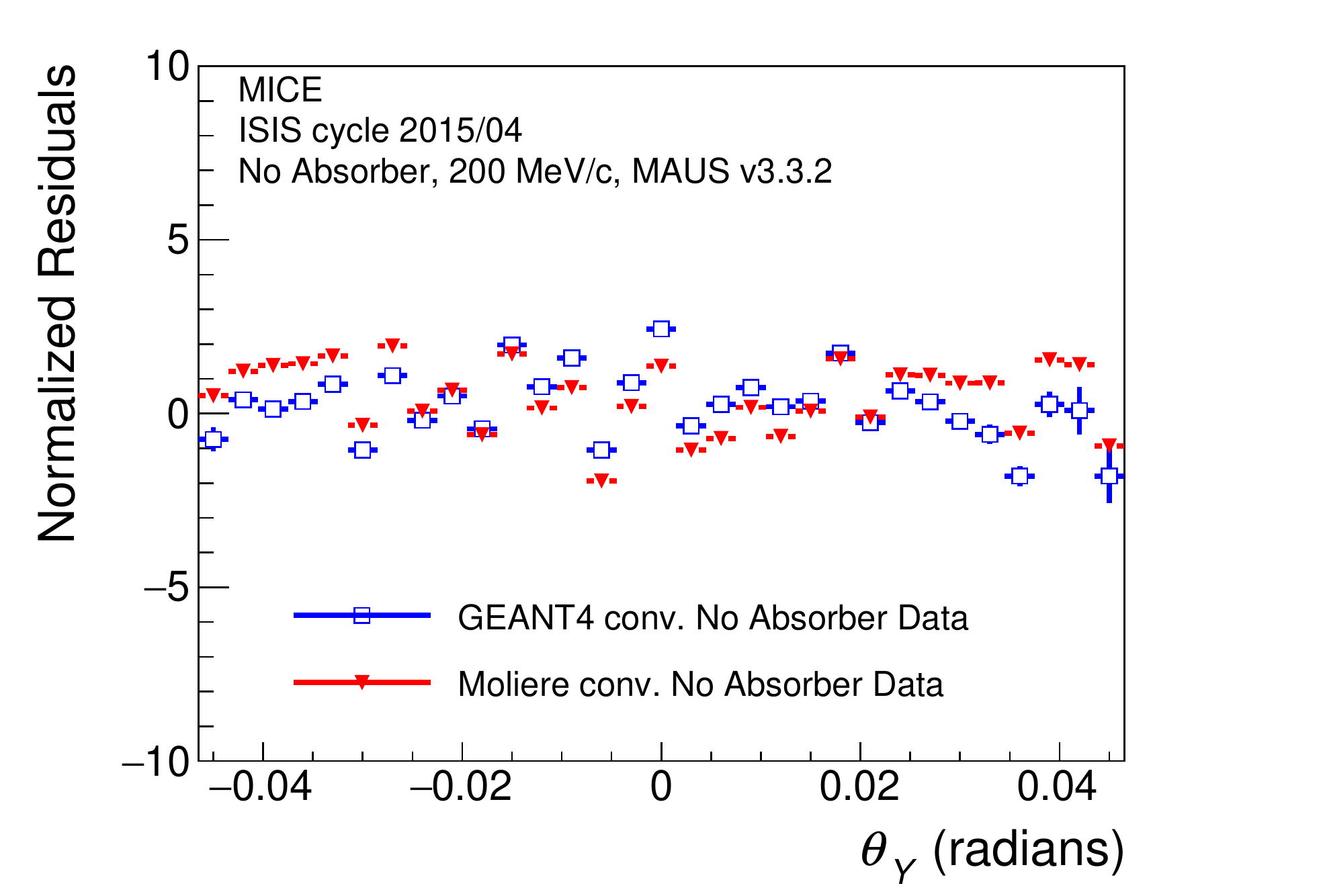}}
\subfigure{\label{fig:con_resX_240}\includegraphics[width=0.48\textwidth]{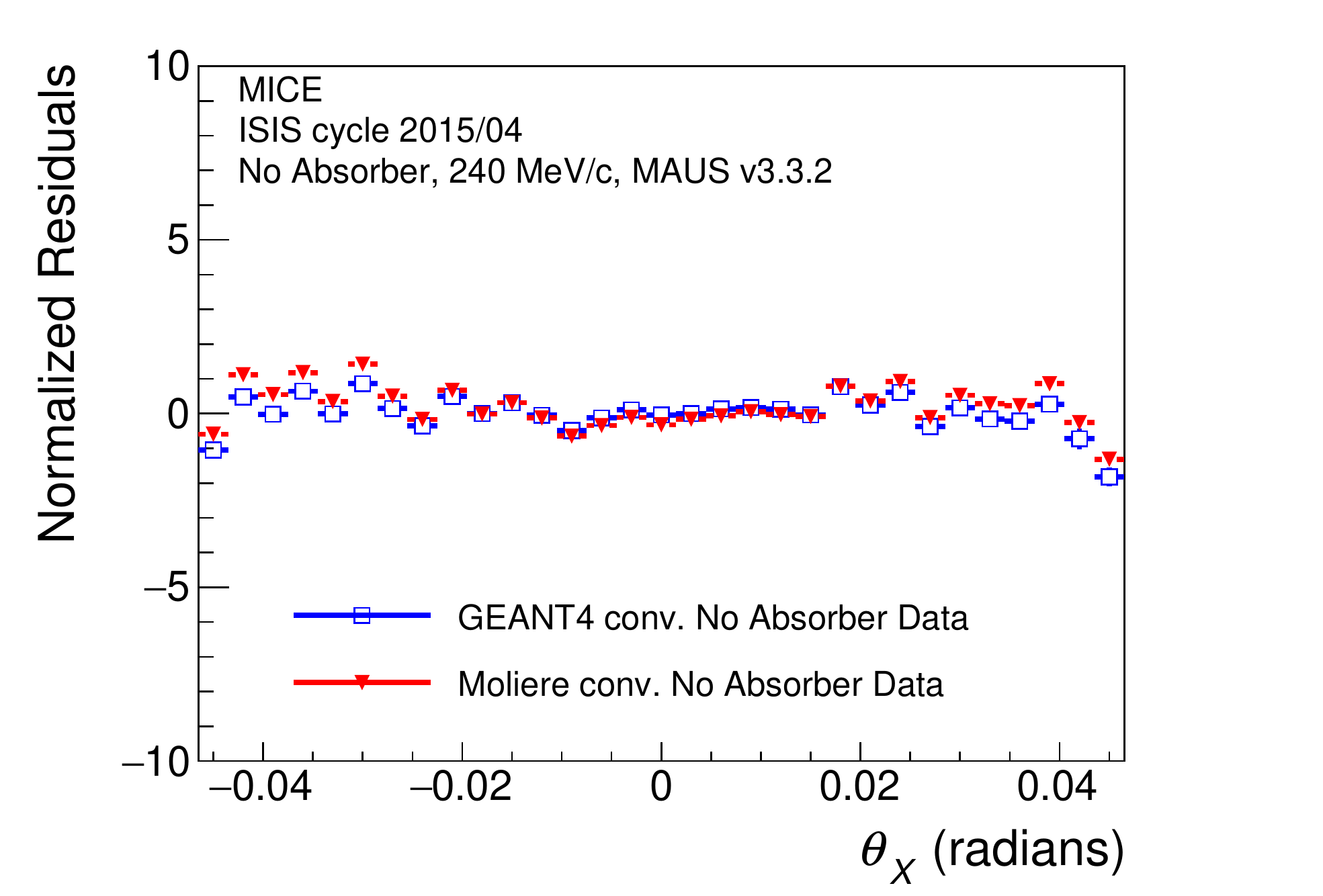}}
\subfigure{\label{fig:con_resY_240}\includegraphics[width=0.48\textwidth]{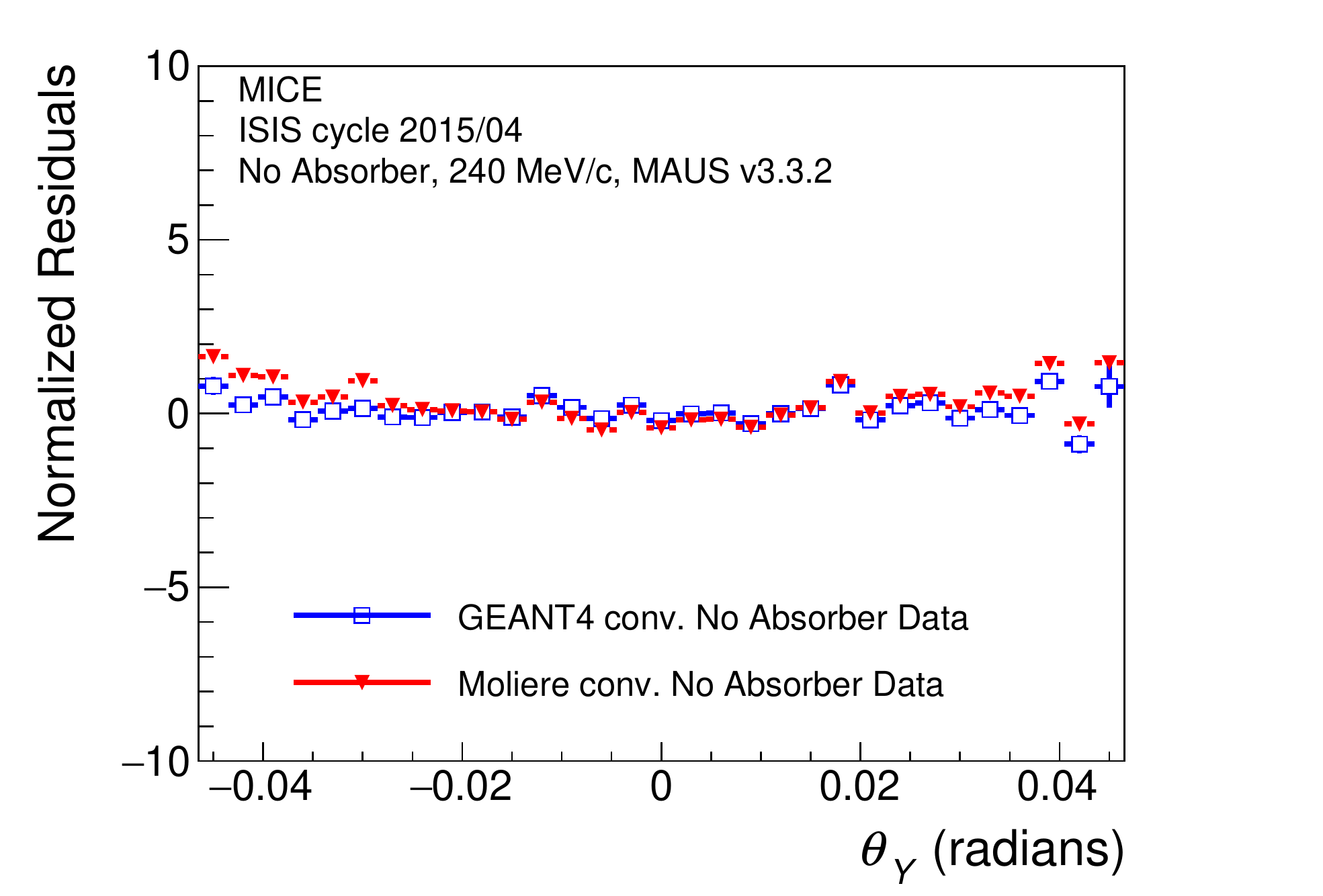}}
\caption{Scattering residuals between data with the LiH absorber and No Absorber data convolved with either GEANT4 or the Moli\`ere  scattering models in LiH for the 172, 200 and 240\,MeV/$c$ samples. The residuals are normalized to the estimated uncertainty in the data in each bin. The agreement improves at higher momentum where the scattering distributions are narrower.}
\label{fig:con_res}
\end{center}
\end{figure*}

The data collected with the absorber were compared to GEANT4 and the Moli\`ere scattering models by performing a convolution of the scattering model with No Absorber data. The convolution,

\begin{linenomath*}\begin{equation}
n_{\mathrm{conv}}(\theta)=n_{NA}(\theta)\ast n_{\mathrm{model}}(\theta),
\end{equation}\end{linenomath*}
where $n_{conv}(\theta)$ is the forward convolved distribution, $n_{NA}(\theta)$ is the scattering distribution measured with the No Absorber data and $n_{model}(\theta)$ is the scattering distribution predicted by the model, is performed by adding an angle sampled from the predicted scattering distribution in the absorber for a given model (GEANT4 or Moli\`ere) to the angle determined from a given trajectory selected from the No Absorber data. This takes into account scattering in the measurement system. The trajectories described by the sum of angles are extrapolated to the downstream tracker and if the track would not have been contained within the downstream tracker then it is not shown in the scattering distribution but the event is still counted in the normalisation. The net effect is a distribution, $n_{conv}(\theta)$, that is the convolution of the raw scattering model $n_{model}(\theta)$ with the detector effects given by the No Absorber distribution $n_{NA}(\theta)$. 
Plots of the Lithium Hydride Absorber data and the No Absorber data convolved with either the GEANT4 simulation or the Moli\`ere model are shown in Fig. \ref{fig:raw_comp}, with the residuals shown in Fig. \ref{fig:con_res}, and the results are summarised in Table \ref{tab:raw_comp}.

\begin{table*}
	\caption{Distribution widths of multiple scattering in lithium hydride data compared to No Absorber data convolved with two different models of scattering (Geant4 and Moli\`{e}re). The $\chi^{2}/N\!D\!F$ were calculated using the number of bins as the number of degrees of freedom. Statistical and systematic uncertainties are given for the data distributions.}
	\begin{ruledtabular}
	\begin{tabular}{cc|c|ccc|ccc}
	$p$ (MeV/$c$) & Angle & $\theta_{\mathrm{Data}}$ (mrad)& $\theta_{G4}$ (mrad) & $\chi^{2}/N\!D\!F$ & P-value & $\theta_{Mol}$ (mrad) &  $\chi^{2}/N\!D\!F$ & P-value \\
\hline
171.55& $\theta_X$ & 21.16$\pm$0.28$\pm$0.48 & 21.36$\pm$0.05 & 30.29 / 31 & 0.45 & 22.64$\pm$0.06 & 34.72 / 31 & 0.25\\
171.55& $\theta_Y$ & 20.97$\pm$0.27$\pm$0.48 & 21.32$\pm$0.05 & 29.10 / 31 & 0.51 & 22.58$\pm$0.06 & 41.14 / 31 & 0.08\\
\hline
199.93& $\theta_X$ & 18.38$\pm$0.18$\pm$0.33 & 18.09$\pm$0.03 & 21.78 / 31 & 0.86 & 19.00$\pm$0.04 & 28.04 / 31 & 0.57\\
199.93& $\theta_Y$ & 18.35$\pm$0.18$\pm$0.33 & 18.02$\pm$0.03 & 26.98 / 31 & 0.62 & 18.98$\pm$0.04 & 35.41 / 31 & 0.23\\
\hline
239.76& $\theta_X$ & 15.05$\pm$0.17$\pm$0.21 & 15.07$\pm$0.02 & 4.08 / 31 & 1.00 & 15.62$\pm$0.02 & 9.48 / 31 & 1.00\\
239.76& $\theta_Y$ & 15.03$\pm$0.16$\pm$0.21 & 15.11$\pm$0.02 & 3.44 / 31 & 1.00 & 15.70$\pm$0.02 & 8.62 / 31 & 1.00\\
	\end{tabular}
	\end{ruledtabular}
	\label{tab:raw_comp}
\end{table*}

The Moli\`ere distributions for the lithium hydride absorber were calculated using 
the procedure described by Gottschalk~\cite{Gottschalk:1992gg} for mixtures and compounds.
Pure $^6$LiH with a thickness of 4.498\,g\,cm$^{-2}$ was assumed. Distributions were calculated for
monoenergetic muons of 172, 200 and 240\,MeV/$c$. Because the muon energy loss is small -- about 11\,MeV -- 
the muon momentum was taken to be constant through the absorber. 

Fano's correction to the Moli\`ere distribution was used to account for the
scattering by atomic electrons. The values of the parameter $U_{\rm in}$, which appears in the correction, were 
$- U_{\rm in} = 3.6$ for hydrogen, as calculated exactly by Fano, and $- U_{\rm in} = 5.0$ for lithium as suggested
by Gottschalk for other materials. 

A cubic spline was used to interpolate between the tabulated points of the functions given by Moli\`ere and Bethe. 
Systematic errors in the calculation arising from, for example, the description of the absorber as pure $^6$LiH were 
estimated to be of the order of one percent.
 
The calculated widths, $\theta_m$, of the central Gaussian term of the projected Moli\`ere distributions are given in 
Table~\ref{Molwidtab}. If scattering by electrons is not included, i.e., Fano's electron correction is set to zero, the
distributions are approximately twenty percent narrower. We note that Bethe's 
{\it ansatz\/} $Z^2 \rightarrow Z(Z+1)$ to describe the electron contribution is inappropriate here because the maximum
kinematically allowed scattering angle of a 200\,MeV/$c$ muon by an electron is of the order of 4 milliradians, much less
than the width of the scattering distribution. The Moli\`ere predictions shown in Table~\ref{Molwidtab} differ from those shown in Table~\ref{tab:raw_comp} as these are the predictions solely from the Moli\`ere calculation not the Moli\`ere prediction convolved with MICE No Absorber scattering data.

\begin{table}[h]
\caption{Calculated widths, $\theta_m$, of the central Gaussian term of the projected Moli\`ere distribution for the 
lithium hydride absorber at each selected muon momentum.}
\begin{center}
\begin{ruledtabular}
\begin{tabular}{c|c}
Momentum & $\theta_m$\\
MeV/$c$  & milliradians\\
\hline
172 & 20.03 \\
200 & 16.87 \\
240 & 13.60 \\
\end{tabular}
\end{ruledtabular}
\end{center}
\label{Molwidtab}
\end{table}

\subsection{Deconvolution}
\label{subsec:deconvolution}
To determine the underlying scattering distribution in the absorber, the effects of scattering in non-absorber materials and the detector resolution must be deconvolved from the measured scattering distribution. The measured scattering distribution with the absorber in the MICE channel can be written

\begin{linenomath*}\begin{equation}
s'(i) = A(i) \sum_{k=0}^{k=31} s(k)( h(i-k) / A(i-k) ),
\end{equation}\end{linenomath*}
where $s'(i)$ is the number of events measured in the $i$th bin with the absorber in the channel, $s(k)$ is the scattering distribution due only to the absorber material without the detector, $h(i-k)$ is the No Absorber scattering distribution which includes the detector resolution and $A(i)$ is the acceptance function at bin $(i)$.  This system of linear equations can be written in matrix form as

\begin{linenomath*}\begin{equation}
\overrightarrow{s}'=\textbf{H}\overrightarrow{s}
\label{matrix}
\end{equation}\end{linenomath*}
where $\overrightarrow{s}'$ is the a vector where each entry is the number of events in a bin of the scattering distribution of all material in the channel. Similarly for $\overrightarrow{s}$ but for a scattering distribution of only the absorber and $\textbf{H}$ is a matrix which transforms one to the other. 
The unfolding step employs Gold's deconvolution algorithm to extract the true scattering distribution ($s$) solely due to the absorber material, as described in \cite{Morhac:1997} and implemented in the ROOT \cite{Brun:1997pa} TSpectrum class. The advantages of using the Gold deconvolution algorithm are that it does not rely on simulated data or scattering models and is a purely data-driven technique making use of all of the data collected. The output of the deconvolution is compared to the GEANT4 and Moli\`ere prediction in Fig. \ref{fig:scatt200G4}.

\subsection{Systematic uncertainties}
\label{subsec:systematics}

\begin{figure*}
\begin{center}
\subfigure{\label{fig:raw_thetaX_172}\includegraphics[width=0.48\textwidth]{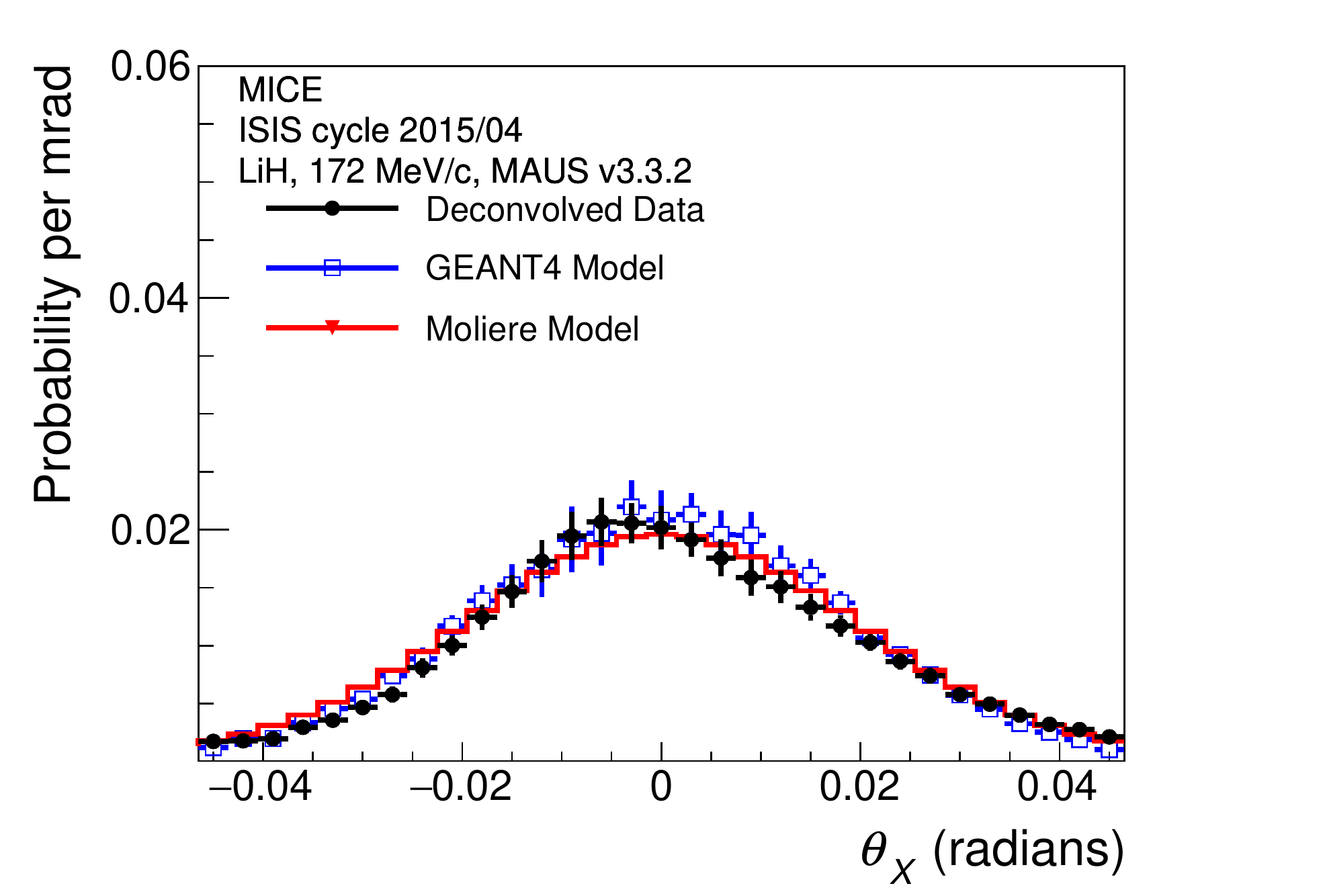}}
\subfigure{\label{fig:raw_thetaY_172}\includegraphics[width=0.48\textwidth]{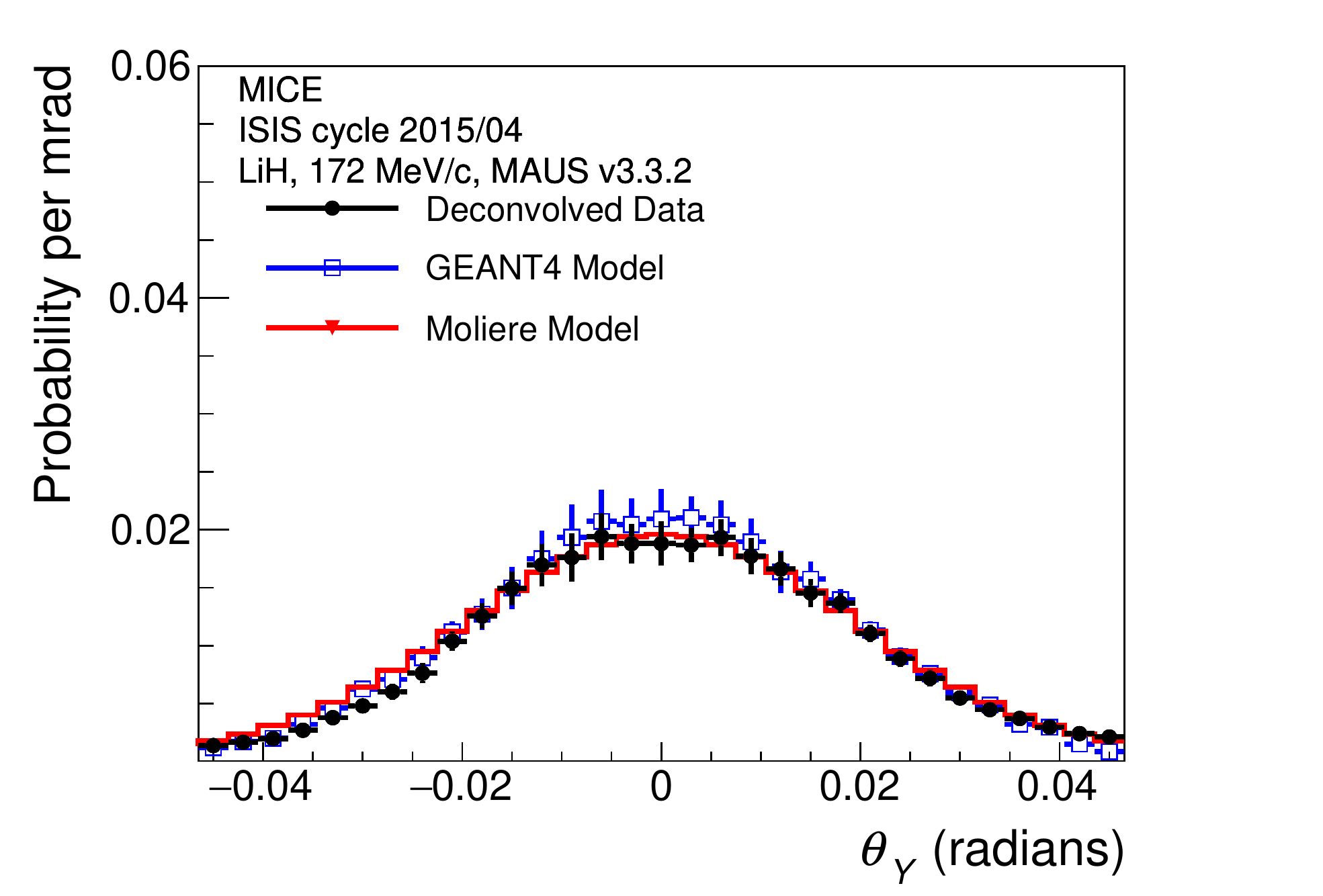}}
\subfigure{\label{fig:raw_thetaX_200}\includegraphics[width=0.48\textwidth]{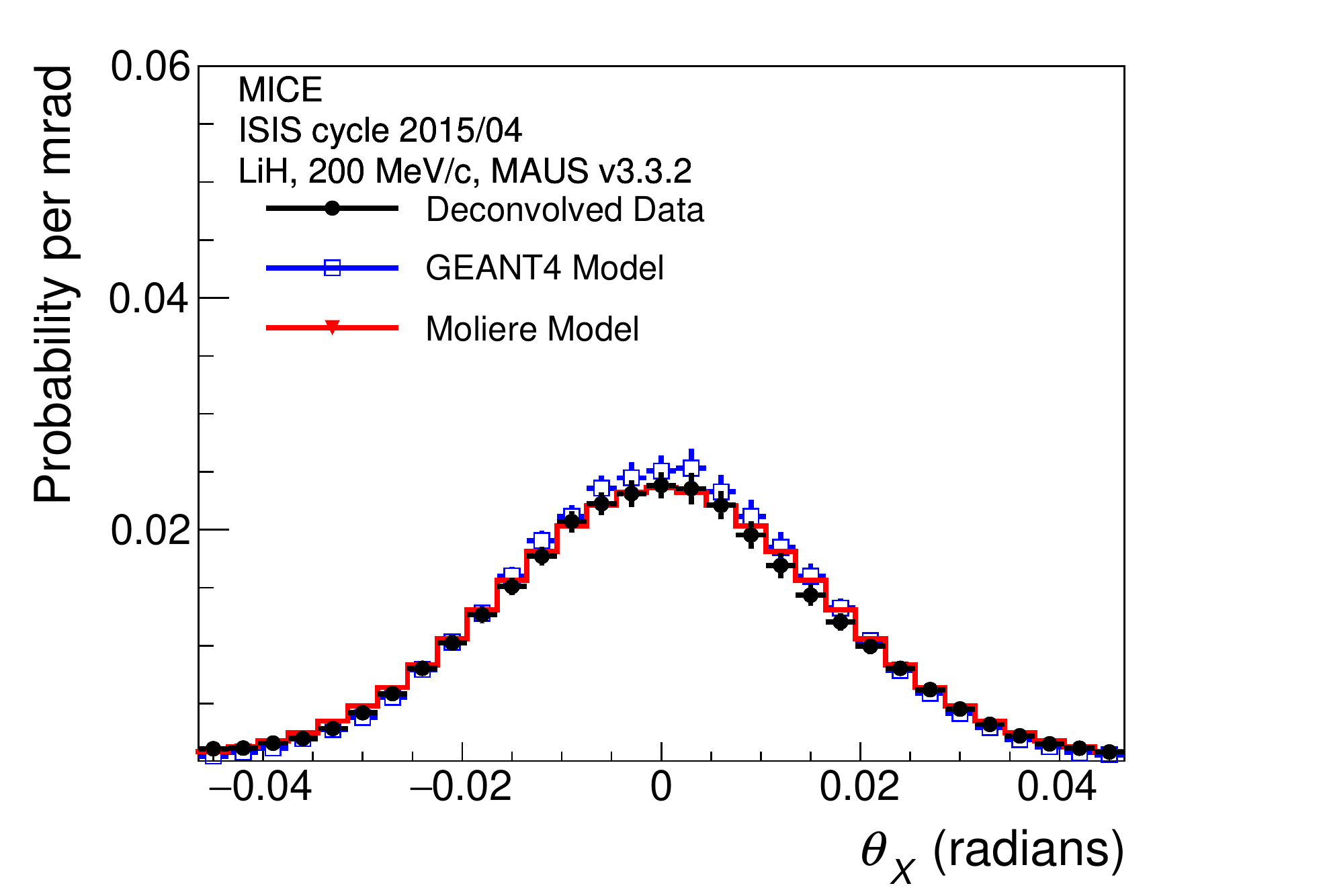}}
\subfigure{\label{fig:raw_thetaY_200}\includegraphics[width=0.48\textwidth]{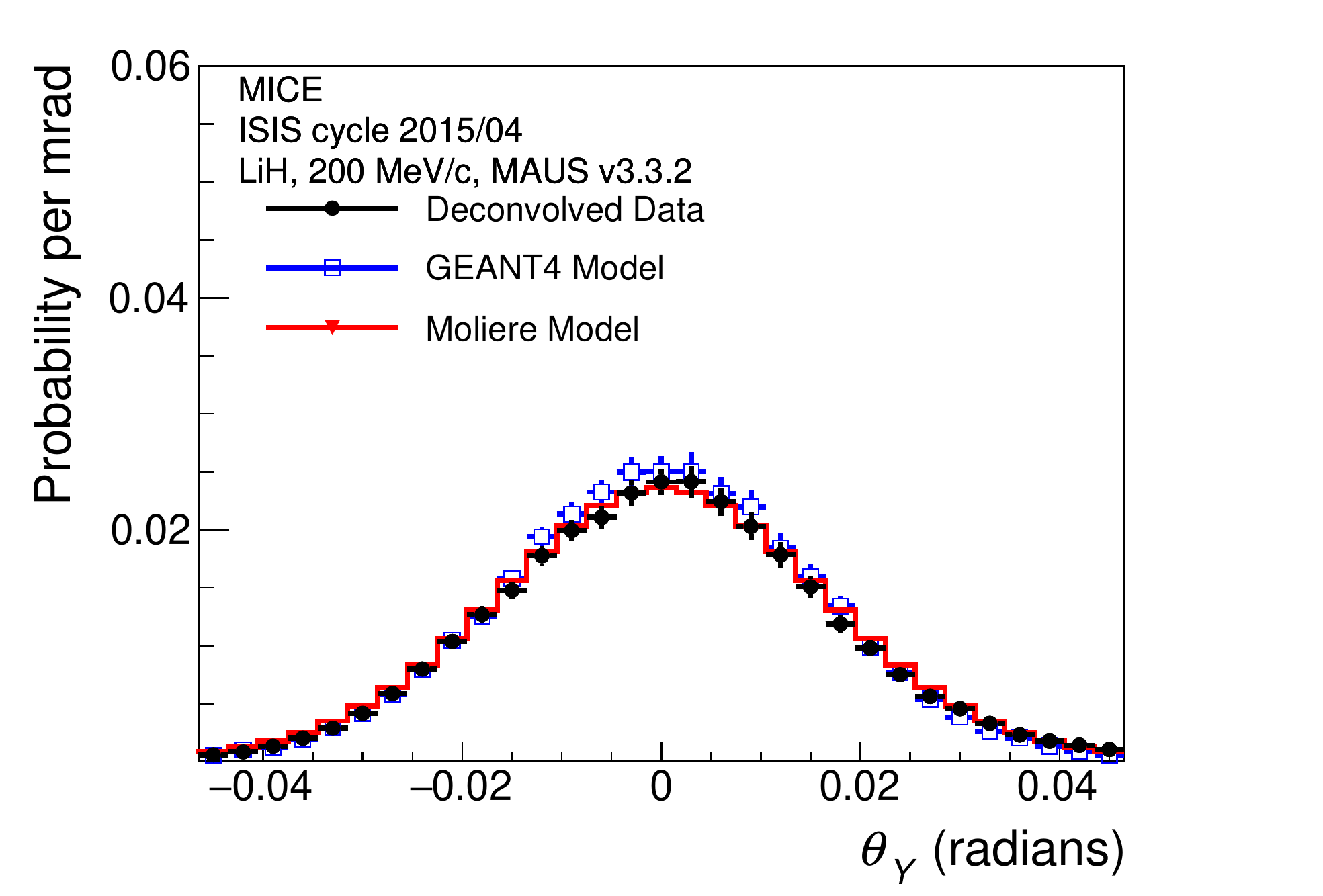}}
\subfigure{\label{fig:raw_thetaX_240}\includegraphics[width=0.48\textwidth]{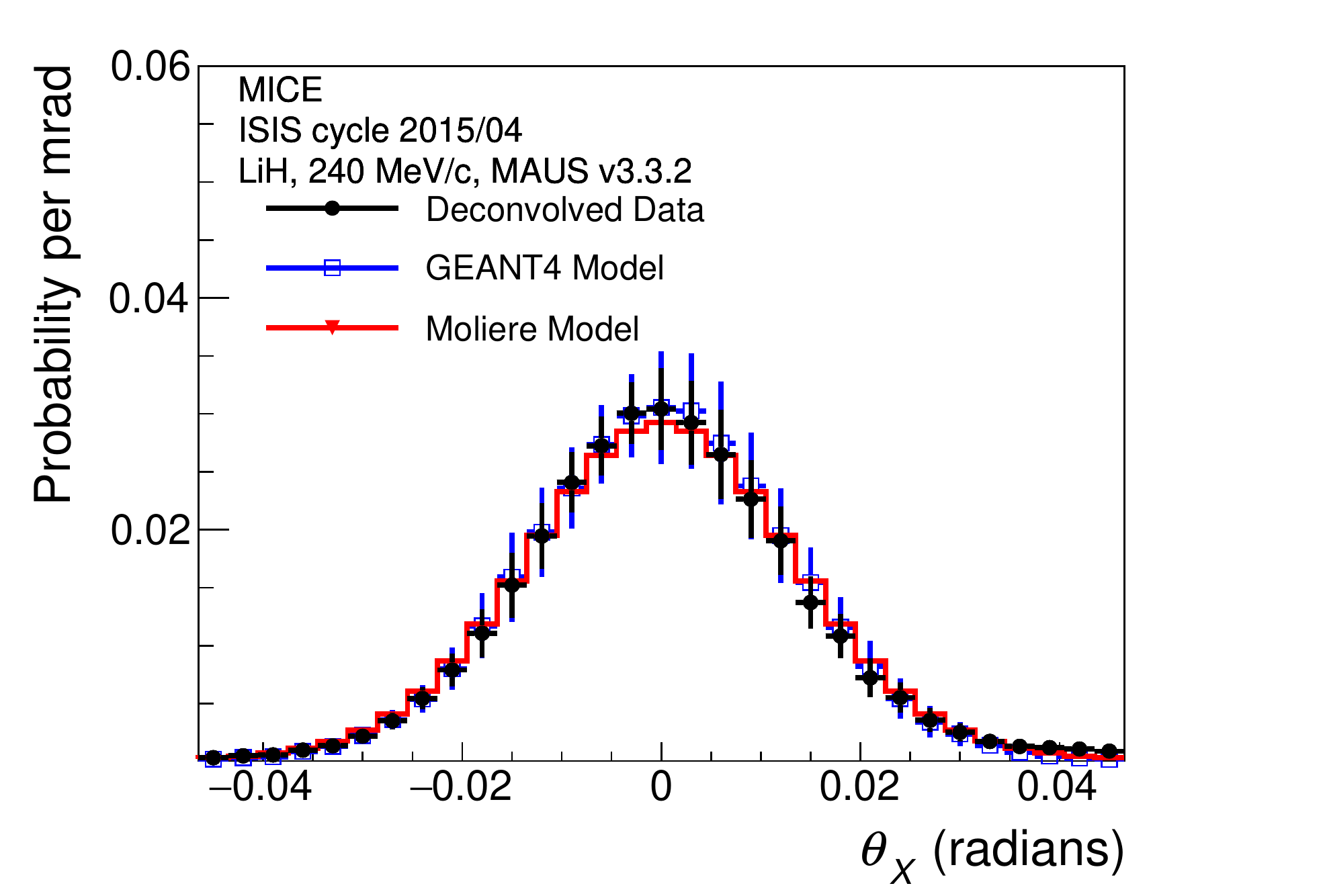}}
\subfigure{\label{fig:raw_thetaY_240}\includegraphics[width=0.48\textwidth]{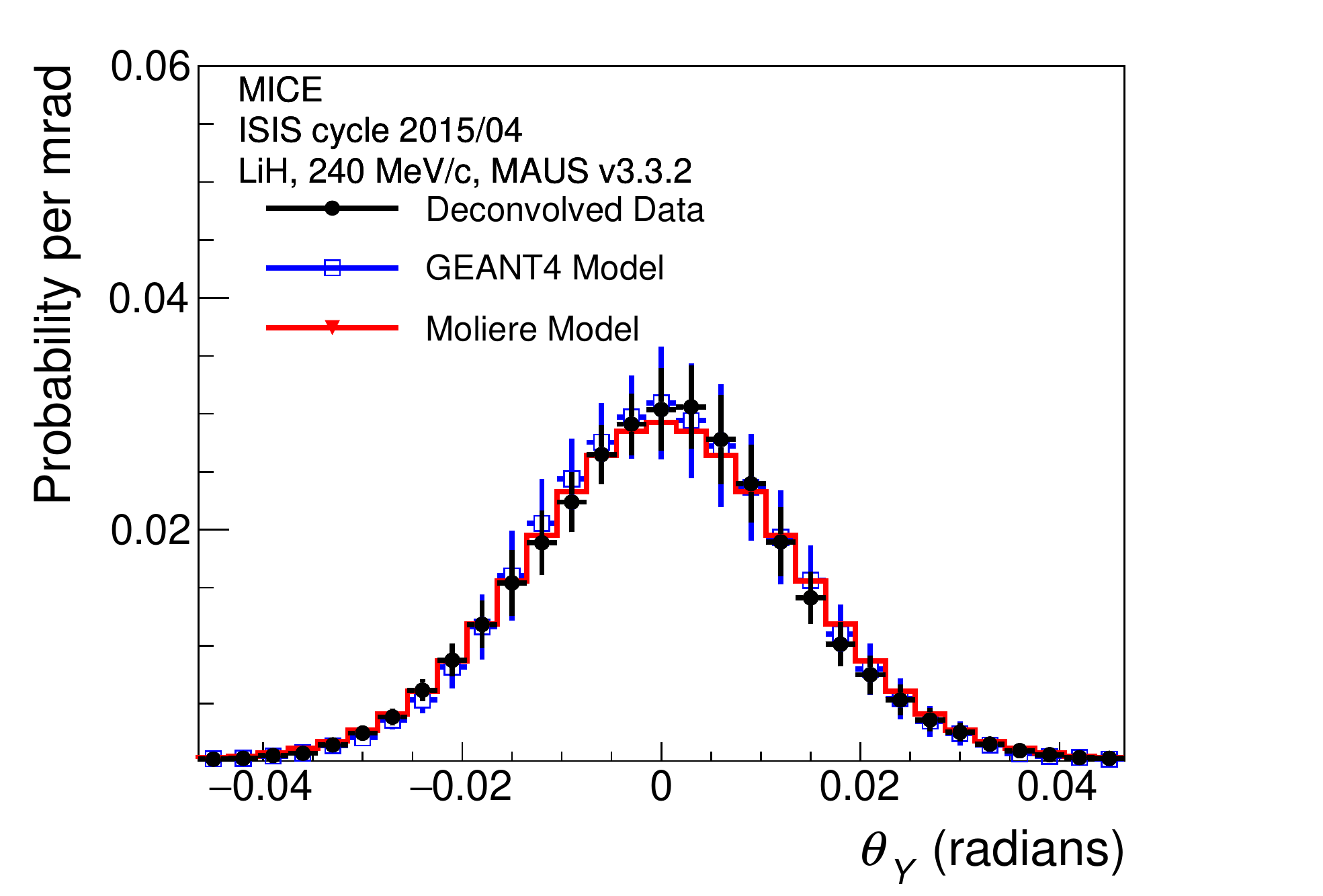}}
\caption{Projected $\theta_X$ and $\theta_Y$ multiple scattering probability functions at 172, 200 and 240\,MeV/$c$ after deconvolution. The GEANT4 and Moli\`ere scattering distributions in LiH are provided for comparison.}
\label{fig:scatt200G4}
\end{center}
\end{figure*}

Six contributions to the systematic uncertainty in the scattering distributions are considered here; uncertainties in the time of flight; measured alignment; fiducial radius; choice of plane in which to measure scattering; effect of pion contamination; and in the deconvolution procedure. To calculate the systematic uncertainty for the individual bins of the scattering plots shown in Figs. \ref{fig:MCdata_comp172}, \ref{fig:MCdata_comp200}, \ref{fig:MCdata_comp240}, \ref{fig:MCdata_comp2Scatt} and \ref{fig:scatt200G4} the numerical derivative is calculated with the expression

\begin{linenomath*}\begin{equation}
\sigma_{sys, i} = \frac{d n_{i}}{d\alpha}\sigma_{\alpha} \approx \frac{\Delta n_i}{\Delta \alpha} \sigma_\alpha\,,
\end{equation}\end{linenomath*}
where $\Delta n_i$ is the change in the number of entries in a bin that results from altering a parameter $\alpha$ with a known uncertainty $\sigma_{\alpha}$ in the analysis or simulation by an amount $\Delta\alpha$. The uncertainty in the measured width of the distribution is calculated in a similar way using
\begin{equation}
\sigma_{sys} \approx \frac{\Delta\theta_0}{\Delta\alpha}\sigma_{\alpha} \,  ,
\end{equation}
where $\Delta\theta_0$ is the change in the width of the scattering distribution when measured in either the $x$ or $y$ projection. The systematic uncertainties are reported for the RMS width of the $\theta_X$ distribution ($\theta_{0,X}$) and the width of the $\theta_Y$ distribution  ($\theta_{0,Y}$) separately.

A significant systematic uncertainty is due to the TOF selection criteria which directly impact the momentum range of the particles used in the scattering measurement. The scale is set using the measured 70~ps uncertainty on the time-of-flight measurement. 
The effect of particles incorrectly appearing inside or outside of the 200~ps bin selection window is determined by offsetting the No Absorber data by 200~ps and the change in the measured scattering width is treated as the systematic uncertainty.

Uncertainties in the alignment have a direct effect on the angles measured by the tracker. The alignment of the MICE trackers is characterized by offsets parallel to $x$ and $y$, with an uncertainty of 0.2~mm, and angles of rotation about the $x$ and $y$ axes, with an uncertainty of 0.07~mrad.  The uncertainties in the width of the scattering distributions were extracted from a number of pseudo-experiments, where the alignment parameters were varied in each iteration. 

The choice of the fiducial region may systematically affect the results. A scan over the possible values of the fiducial radius was completed and the variation in the width of the scattering distributions for samples adjacent to the selected value of 90~mm was used to set the uncertainty.

The definitions of the scattering angles are given in section \ref{Sect:Method} and the Appendix. In the definition of the projected scattering angles, $\theta_x$ and $\theta_y$, $\hat{\textbf{v}}$ is the unit vector mutually orthogonal
to the $y$ direction and the momentum vector and $\hat{\textbf{u}}$ is the unit vector parallel to the upstream
momentum vector. They are related via the formula
\begin{equation}
\hat{\textbf{v}}=\hat{\textbf{s}}\times\hat{\textbf{u}},
\end{equation}
 where $\hat{\vec{s}}$ is arbitrarily defined as $\hat{\textbf{s}}=(0,-1,0)$. This expression defines a direction perpendicular to a plane containing the upstream track. There are an infinite number of planes that contain this track, so we consider the uncertainty introduced by the definition of $\hat{\vec{s}}$ by rotating it between 0$^\circ$ and 180$^\circ$, in increments of 1$^\circ$, around the $x$-axis, with the analysis repeated after each increment. The resulting maximum change in measured scattering angle is included in the systematic uncertainties in Table \ref{tab:systematics}.

The MICE muon beam has pion contamination with an upper limit $f_\pi < 1.4\%$ at 90\% C.L. \cite{Adams:2015wxp}. To measure the effect of this contamination on the scattering measurement for muons, a Monte Carlo study was performed. The measurement was simulated with the MICE beam, including simulated impurities, and a pure muon sample, with the systematic error being the difference between the two results.

The difference between the deconvolved result and the true scattering distribution from a GEANT4 simulation was taken to be an additional source of systematic error. This accounts for any bias introduced by the Gold deconvolution procedure. The systematic uncertainties for the deconvolution procedure showed significant variation from bin to bin so a parabolic smoothing function was used to assign the systematic uncertainty to each bin. 

All systematic uncertainties, and their quadratic combination, for the three selected momenta of 172, 200 and 240\,MeV/$c$ are included in Table \ref{tab:systematics}. The dominant systematic uncertainties are those in the momentum scale of the TOF system and the deconvolution procedure.

\begin{table}[htp]
\begin{center}
{\normalsize
\caption{Systematic uncertainties associated with the width of the scattering distributions of $\theta_{0,X}$ and $\theta_{0,Y}$ in three representative momentum bins.
\vspace*{0.5cm}}
\label{tab:systematics}
\begin{ruledtabular}
\begin{tabular}{c|c|c|c}
$p$ (MeV/$c$)  & Type & $\Delta\theta_{0,X}$ & $\Delta\theta_{0,Y}$ \\
               &      & (mrad)               & (mrad)               \\
\hline
 & TOF selection & 0.64 & 0.64  \\
 & Alignment & $<0.01$ & 0.01 \\
 171.55  & Fiducial radius & $<0.01$ & $<0.01$  \\
 & $\theta$ angle definition  & $<0.01$ & $<0.01$ \\
 & $\pi$ contamination & $<0.01$ & $<0.01$  \\
 & Deconvolution & 1.25 & 1.19  \\
 \hline
  & Total sys. & 1.39 & 1.35 \\
 \hline
  & TOF selection & 0.29 & 0.29 \\
 & Alignment & 0.02 & $<0.01$  \\
199.93  & Fiducial radius & 0.01 & 0.01  \\
 & $\theta$ definition  & $<0.01$ & $<0.01$  \\
 & $\pi$ contamination   &  $<0.01$ &  $<0.01$\\
 & Deconvolution & 0.70 & 0.47  \\
 \hline
 & Total sys. & 0.73 & 0.54  \\
\hline
  & TOF selection & 0.27 & 0.27 \\
 & Alignment & $<0.01$ & $<0.01$  \\
239.76  & Fiducial radius & 0.01 & 0.01 \\
 & $\theta$ definition & $<0.01$ & $<0.01$  \\
 & $\pi$ contamination  &  0.01 & 0.01 \\
 & Deconvolution & 0.27 & 0.41  \\
 \hline
 & Total sys. & 0.36 & 0.49   \\
\end{tabular}
\end{ruledtabular}
}
\end{center}
\end{table}%

\subsection{Model comparisons}\label{s:Results}

The residual between the scattering distribution in data and that predicted by the models is used to quantify the level of agreement between data and simulation. The normalised residual is defined as 
\begin{equation}
\text{residual} = \frac{p_{\mathrm{data}}(\theta_{i})-p_{\mathrm{simulation}}(\theta_{i})}
			{\sqrt{\sigma^2_{\mathrm{stat}}+\sum\sigma^{2}_{\mathrm{sys},i}}}	
\end{equation}
 where $p_{\mathrm{data}}(\theta_{i})$ is the probability of scattering at angle $\theta_i$ measured with the MICE data and $p_{\mathrm{simulation}}(\theta_{i})$ is the probability of scattering predicted by the corresponding model. The systematic uncertainties $\sigma^{2}_{\mathrm{sys},i}$, discussed in Sec. \ref{subsec:systematics}, are calculated and summed in quadrature on a bin by bin level. 
 The $\chi^{2}$ derived from these residuals appears in Table \ref{tab:dataMC}. The $\chi^{2}$ between the scattering distribution from the data and that predicted by the model is calculated using
\begin{equation}\label{eq:chi2}
\chi^{2} = \sum_{i=0}^{N}\frac{\left(p_{\mathrm{data}}(\theta_{i})-p_{\mathrm{simulation}}(\theta_{i})\right)^2}{{\sigma_{\mathrm{stat}}^2 + \sum_{\mathrm{sys}}\sigma^{2}_{\mathrm{sys},i}}}
\end{equation}
where $N$ is the number of bins and $\mathrm{sys}$ is the number of systematic errors.
The $\chi^{2}$ was calculated using 31 data points and demonstrates good agreement between data and MC.
The $\chi^2$ calculation in Eqn. \ref{eq:chi2} was repeated for both the forward convolution comparison to real data and for the comparison between the deconvolved data and the GEANT4 and Moli\`ere models. The systematic uncertainties are added on a bin by bin basis in the calculation of the $\chi^{2}$ in
Eq. \ref{eq:chi2}. 

There is very little difference between the GEANT4 simulation, the Moli\`ere calculations and the deconvolved data.  The deconvolved $\theta_X$ and $\theta_Y$ multiple scattering distributions on lithium hydride for the 172, 200 and 240\,MeV/$c$  muon samples are shown in Fig. \ref{fig:scatt200G4}, and these are compared with a GEANT4 LiH simulation and the Moli\`ere calculation. 

The distributions of the projections in $\theta_X$ and $\theta_Y$ were characterized using a Gaussian fit within a $\pm 45$~mrad range, with the results shown in Table \ref{tab:G4gausfit} for deconvolved data using the Gold deconvolution algorithm and the true distributions extracted from the GEANT4 simulation and the Moli\`ere model calculation. The table shows that the deconvolved $\theta_X$ and $\theta_Y$ projections of the scattering distributions are approximately consistent with the GEANT4 and Moli\`ere distributions, but the  Moli\`ere distribution is systematically wider than the rest and significantly wider than that given by GEANT4. 

\begin{table}
\caption{Widths of best fit Gaussian fitted to central $\pm45$ mrad of scattering distributions after deconvolution compared to GEANT4 and Moli\`ere models. Statistical and systematic uncertainties are given for the data distributions. Only statistical uncertainties are given for the GEANT4 model.}
\vspace{3mm}
\begin{center}
\begin{ruledtabular}
	\begin{tabular}{cc|c|cc}
	$p$ (MeV/$c$) & Angle & $\theta^{\mathrm{meas}}_{\mathrm{Gold}}$ (mrad)& $\theta^{\mathrm{true}}_{G4}$ & $\theta^{\mathrm{true}}_{\mathrm{Moli\grave{e}re}}$ \\
	  &            &                         &(mrad)          & (mrad) \\
	\hline
171.55& $\theta_X$ & 19.03$\pm$0.26$\pm$1.39 & 18.62$\pm$0.13 & 20.03 \\
171.55& $\theta_Y$ & 18.95$\pm$0.24$\pm$1.35 & 18.59$\pm$0.12 & 20.03 \\
\hline
199.93& $\theta_X$ & 16.59$\pm$0.17$\pm$0.73 & 15.82$\pm$0.05 & 16.87 \\
199.93& $\theta_Y$ & 16.36$\pm$0.17$\pm$0.55 & 15.82$\pm$0.05 & 16.87 \\
\hline
239.76& $\theta_X$ & 13.29$\pm$0.17$\pm$0.37 & 13.16$\pm$0.04 & 13.60 \\
239.76& $\theta_Y$ & 13.21$\pm$0.16$\pm$0.49 & 13.10$\pm$0.04 & 13.60 \\
	\end{tabular}
	\end{ruledtabular}
	\label{tab:G4gausfit}
\end{center}
\end{table}

\subsection{Momentum-dependent measurements}

\begin{figure}[htp]
\subfigure{\label{fig:thetaXvp}\includegraphics[width=0.48\textwidth]{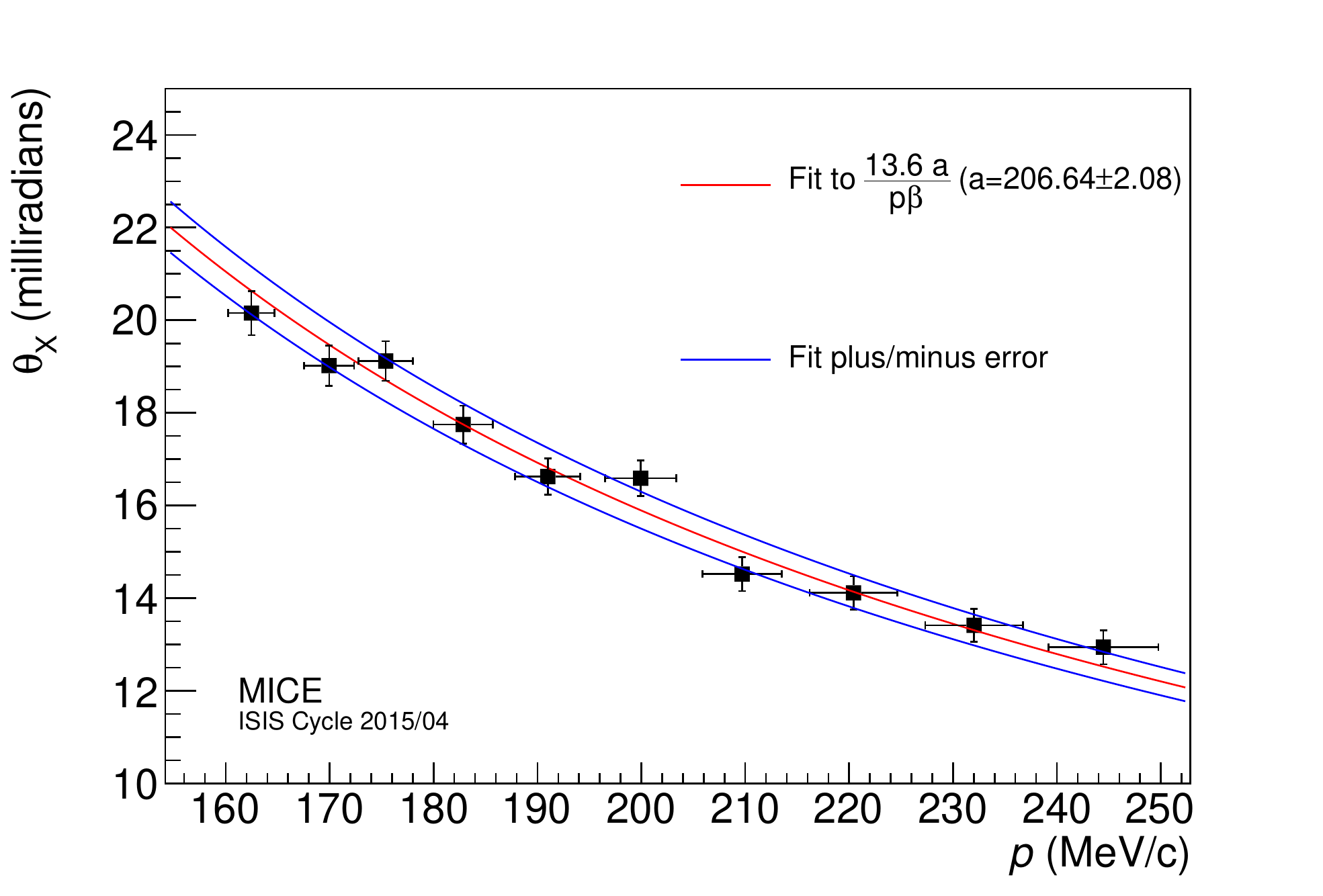}}
\subfigure{\label{fig:thetaYvp}\includegraphics[width=0.48\textwidth]{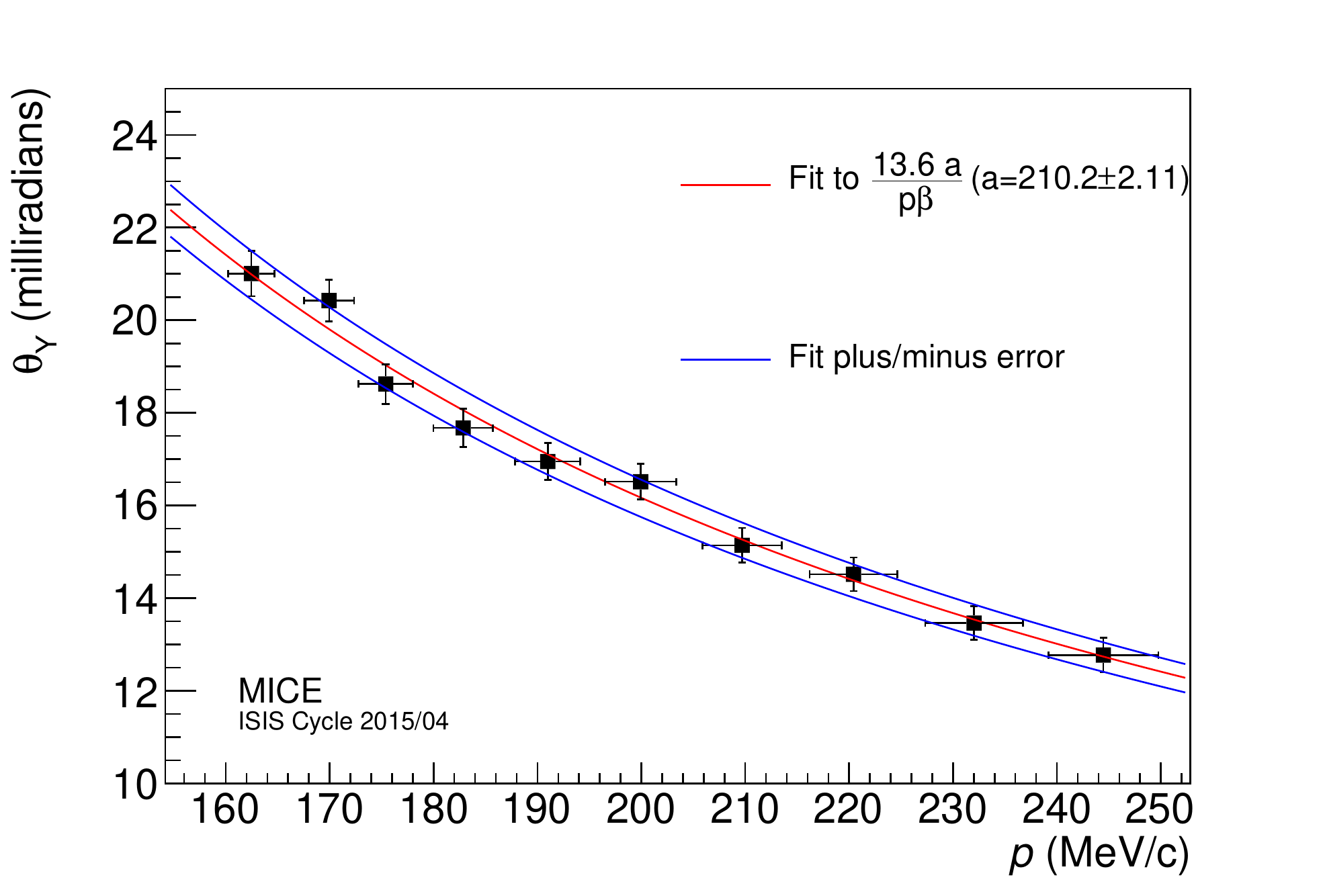}}
\caption{The results of the scattering analysis using data in a number of momentum bins. Scattering widths are reported after application of the Gold deconvolution.}
\label{fig:thetavp}
\end{figure}

The selected samples are plotted as a function of mean momentum for each sample, to confirm the dependence of the widths of the scattering distributions on momentum.  The number of events contained in each TOF bin is between 3500 and 9000 events. The deconvolved scattering widths as a function of momentum are shown in Fig. \ref{fig:thetavp}. The widths, $\theta_0$, are fitted to
\begin{equation}
\label{eq:fit}
 \theta_0=\frac{13.6\,[\mathrm{MeV}/c]a}{p \beta } ,
\end{equation}
where $a$ is a fit coefficient, motivated by Eq.~\ref{eq:pdg}, where the $\beta$ dependence of the log term is negligible, changing the calculated value by less than 1\%. 

 The coefficient, $a$, is compared with the prediction from the PDG formula in Eq.~\ref{eq:pdg}. The values of the coefficients, $a$, determined from the fits to the $\theta_{0,X}$ and $\theta_{0,Y}$ distributions are shown in Table~\ref{tab:fitparams}. The numerical derivative of the momentum with respect to TOF of the sample was calculated and used to assess the systematic uncertainty associated with the measurement. 

Measurements using the projected angles are systematically smaller than the PDG prediction. The average of the two fits to the $\theta_{0,X}$ and $\theta_{0,Y}$ muon scattering widths as a function of momentum yields $a = 208.1 \pm 1.5$~mrad, which is 9\% smaller than the value proposed by the PDG formula, $a = 226.7$~mrad, but still within the uncertainties of that approximate formula,   Eq. \ref{eq:pdg}, which is quoted as accurate to 11\%.

\begin{table}[htp]
\caption{Results of the fit to the scattering widths as a function of momentum, given by Eq. \ref{eq:fit}. The value predicted by the PDG is also shown.}
	\begin{center}
	\begin{ruledtabular}
		\begin{tabular}{c|c}
		Angle & $a$ (mrad) \\
		\hline
		$\theta_{0,X}$ & 206.6$\pm$2.1  \\
		$\theta_{0,Y}$ & 210.2$\pm$2.1  \\
		PDG & 226.7 \\
		\end{tabular}
		\end{ruledtabular}
	\end{center}
\label{tab:fitparams}
\end{table}%

\section{\label{Sect:Conclusions}CONCLUSIONS}

Presented here is an analysis of the LiH multiple Coulomb scattering data taken during ISIS user run 2015/04 using MICE. These data were compared to the GEANT4 (v9.6) default scattering model \cite{Agostinelli:2002hh} and the full Moli\`ere calculation \cite{Moliere:1947zza,Moliere:1948zz}. A $\chi^{2}$ statistic was used to make quantitative statements about the validity of the proposed models. Three approaches are taken; the measured LiH and No Absorber scattering distributions were compared to GEANT4, the forward convolution using the No Absorber data was compared to both GEANT4 and the Moli\`ere model and the deconvolution of the LiH scattering data using the No Absorber data was compared to both GEANT4 and the Moli\`ere model. In all cases the GEANT4 scattering widths agreed with the measured data at each of the nominal momenta, but the Moli\`ere model produces systematically wider distributions. 

The momentum dependence of scattering was examined by selecting 200~ps time of flight samples from the muon beam data. The momentum dependence from 160 to 245\,MeV/$c$ was compared to the dependence in Eq. \ref{eq:pdg}, from the PDG \cite{Zyla:2020zbs}, and it was found that the measured RMS scattering width is about 9\% smaller than the approximate PDG estimation, but within the latter's stated uncertainty.

\newpage
\appendix*

\section{\label{Sect:Appendix}Definition of Scattering Angles}

The projections of the scattering angle
onto the $y$-$z$ or $x$-$z$ plane, angles $\theta_X$ and $\theta_Y$, are defined by
considering the inner product of the downstream momentum $\textbf{p}_{DS}$ with the
component of the upstream momentum vector $\textbf{p}_{US}$, perpendicular to the projection plane. The scattering projection into
the plane defined by the momentum vector and the $y$-axis is
\begin{equation}
\label{eqn:theta_y}
\theta_{Y} = \arctan \left(
\frac{\textbf{p}_{DS}\cdot\hat{\textbf{v}}}{\textbf{p}_{DS}\cdot\hat{\textbf{u}}} \right)= \arctan \left(
\frac{\textbf{p}_{DS}\cdot(\hat{\textbf{y}}\times\textbf{p}_{US})|\textbf{p}_{US}|}
     {(\textbf{p}_{DS}\cdot\textbf{p}_{US})|\hat{\textbf{y}}\times\textbf{p}_{US}|}\right),
\end{equation}
where $\hat{\textbf{y}}$ is the unit vector in the $y$ direction,
$\hat{\textbf{v}} = \hat{\textbf{y}}\times \textbf{p}_{US} /
|\hat{\textbf{y}}\times\textbf{p}_{US}|$ is the unit vector mutually orthogonal to the $y$
direction and the momentum vector and
$\hat{\textbf{u}}=\textbf{p}_{US}/|\textbf{p}_{US}|$ is the unit
vector parallel to the upstream momentum vector. A
scattering angle in the perpendicular plane must then be defined as
\begin{equation}
\label{eqn:theta_x}
\theta_{X} = \arctan \left( |\textbf{p}_{US}|
		\frac{\textbf{p}_{DS}\cdot(\textbf{p}_{US}\times(\hat{\textbf{y}}\times\textbf{p}_{US}))}{|\textbf{p}_{US}\times(\hat{\textbf{y}}\times\textbf{p}_{US})|\textbf{p}_{DS}\cdot\textbf{p}_{US}}\right),
\end{equation}
where the downstream vector is now projected onto the unit vector
$\hat{\textbf{v}} = \textbf{p}_{US}\times(\hat{\textbf{y}}\times \textbf{p}_{US}) /
|\textbf{p}_{US}\times(\hat{\textbf{y}}\times \textbf{p}_{US})|$. These
two expressions can be expressed in terms of the gradients of the muon
tracks before and after the scatters,
\begin{widetext}
\begin{eqnarray}
\theta_{Y} &=& \arctan \left\{\frac{\sqrt{1 + \left(\frac{dx}{dz}\right)_{US}^{2} + \left(\frac{dy}{dz}\right)_{US}^{2}}}{\sqrt{1 + \left(\frac{dx}{dz}\right)_{US}^{2}}} 
		\left(\frac{\left(\frac{dx}{dz}\right)_{DS} - \left(\frac{dx}{dz}\right)_{US}}{1+\left(\frac{dx}{dz}\right)_{US}\left(\frac{dx}{dz}\right)_{DS} + \left(\frac{dy}{dz}\right)_{US}\left(\frac{dy}{dz}\right)_{DS}}\right)\right\}, 
\label{eq:thetay}
\end{eqnarray}
\begin{equation}
\theta_{X} = \arctan \left\{
	\sqrt{\frac{1 + \left(\frac{dx}{dz}\right)_{US}^{2} + \left(\frac{dy}{dz}\right)_{US}^{2}}
			{\left(1+\left(\frac{dx}{dz}\right)_{US}^{2} + \left(\frac{dy}{dz}\right)_{US}^{2}\right)\left(1 + \left(\frac{dx}{dz}\right)^{2}_{US}\right)}}\right. 
	\times \left. \left(\frac{\left(\frac{dy}{dz}\right)_{DS}\left(1 + \left(\frac{dx}{dz}\right)_{US}^{2}\right) + 
		\left(\left(\frac{dx}{dz}\right)_{DS}\left(\frac{dx}{dz}\right)_{US} - 1\right)\left(\frac{dy}{dz}\right)_{US}}
		{1 + \left(\frac{dx}{dz}\right)_{US}\left(\frac{dx}{dz}\right)_{DS} + \left(\frac{dy}{dz}\right)_{US}\left(\frac{dy}{dz}\right)_{DS}}\right)
\right\}.
\label{eq:thetax}
\end{equation}
\end{widetext}
In the approximation of small angles
(i.e. $\frac{dx}{dz}\approx\frac{dy}{dz} \ll 1$) these produce the
more familiar forms
\begin{equation}
\theta_X=\left(\frac{dy}{dz}\right)_{DS} - \left(\frac{dy}{dz}\right)_{US}
\end{equation}
for scattering about the $x$-axis or
\begin{equation}
\theta_Y=\left(\frac{dx}{dz}\right)_{DS} - \left(\frac{dx}{dz}\right)_{US}
\end{equation}
for scattering about the $y$-axis. The more exact expressions, equations \ref{eq:thetay} and \ref{eq:thetax}, are
used throughout for this analysis.

\begin{acknowledgments}
The work described here was made possible by grants from the Science and Technology Facilities Council (UK), the Department of Energy and the National Science Foundation (USA), the Istituto Nazionale di Fisica Nucleare (Italy), the European Union under the European Union’s Framework Programme 7 (AIDA project, grant agreement number 262025; TIARA project, grant agreement number 261905; and EuCARD), the Japan Society for the Promotion of Science, the National Research Foundation of Korea (number NRF-2016R1A5A1013277), the Ministry of Education, Science and Technological Development of the Republic of Serbia, the Institute of High Energy Physics/Chinese Academy of Sciences fund for collaboration between the People’s Republic of China and the USA, and the Swiss National Science Foundation in the framework of the SCOPES programme. We gratefully acknowledge all sources of support. We are grateful for the support given to us by the staff of the STFC Rutherford Appleton and Daresbury laboratories. We acknowledge the use of Grid computing resources deployed and operated by GridPP in the UK, http://www.gridpp.ac.uk/.

\end{acknowledgments}

\appendix

\bibliography{apssamp}

\end{document}